\documentclass[prx, twocolumn,superscriptaddress, notitlepage, showpacs, floatfix]{revtex4-1}

\usepackage{amsmath,bm}
\usepackage{graphicx}
\usepackage{amsfonts}
\usepackage{amssymb}
\usepackage{epstopdf}
\usepackage{hyperref}
\usepackage{color}
\usepackage{color,tikz,float}
\usepackage{dsfont}

\usetikzlibrary{shapes.misc}

\newcommand{\ket}[1]{| #1 \rangle}
\newcommand{\bra}[1]{\langle #1 |}

\newcommand{\dd}{\mathrm{d}}
\newcommand{\ii}{\mathrm{i}}

\newcommand{\tr}{\mathrm{Tr}}
\renewcommand{\t}[1]{\mathrm{#1}}
\newcommand{\be}{\begin{equation}}
\newcommand{\ee}{\end{equation}}

\begin{document}
\title{Tensor-Network Approach for Quantum Metrology in Many-Body Quantum Systems}
\author{Krzysztof Chabuda}
\affiliation{Faculty of Physics, University of Warsaw,  ul. Pasteura 5, PL-02-093 Warszawa, Poland}
\author{Jacek Dziarmaga}
\affiliation{Institute of Physics, Jagiellonian University, {\L}ojasiewicza 11, PL-30348 Krak{\'o}w, Poland}
\author{Tobias J.\ Osborne}
\affiliation{Institut f{\"u}r Theoretische Physik, Leibniz Universit{\"a}t Hannover, Appelstr. 2, 30167 Hannover, Germany}
\author{Rafa{\l} Demkowicz-Dobrza{\'n}ski}
\affiliation{Faculty of Physics, University of Warsaw,  ul. Pasteura 5, PL-02-093 Warszawa, Poland}

\begin{abstract}
Identification of the optimal quantum metrological protocols in realistic many particle quantum models is in general a challenge that cannot be efficiently addressed by the state-of-the-art numerical and analytical methods. Here we provide a comprehensive framework exploiting matrix product operators (MPO) type tensor networks for quantum metrological problems. Thanks to the fact that the MPO formalism allows for an efficient description of short-range spatial and temporal noise correlations, the maximal achievable estimation precision in such models, as well as the optimal probe states in previously inaccessible regimes can be identified. Moreover, the application of infinite MPO (iMPO) techniques allows for a direct and efficient determination of the asymptotic precision of optimal protocols in the limit of infinite particle numbers. We illustrate the potential of our framework in terms of an atomic clock stabilization (temporal noise correlation) example as well as for magnetic field sensing in the presence of locally correlated magnetic field fluctuations (spatial noise correlations). As a byproduct, the developed methods for calculating the quantum Fisher information via MPOs may be used to calculate the fidelity susceptibility---a parameter widely used in many-body physics to study phase transitions.
\end{abstract}

\maketitle

\section{Introduction} \label{sec:intro}
Quantum metrology \cite{Giovannetti2006, Paris2009, tothQuantumMetrologyQuantum2014a, Demkowicz2015, Schnabel2016, degen2017quantum, Pezze2018, Pirandola2018} is plagued by the same computational difficulties afflicting all quantum information processing technologies, namely, the exponential growth of the dimension of many particle Hilbert space \cite{feynmanSimulatingPhysicsComputers1981}. While small-scale problems are feasible via direct numerical study, even a slight increase in the number of elementary objects quickly makes such an approach intractable. This presents a serious conceptual roadblock for the development of quantum technologies---once inside the quantum enhanced regime one can no longer validate the performance of quantum devices using naive classical simulation \cite{Aolita2015}. For example, just storing the tomographic reconstruction of a multiqubit quantum state quickly becomes unfeasible for more than 40 particles \cite{Cramer2010}.

Despite the curse of dimensionality there has nevertheless been considerable progress in quantum metrology. This is because in the noiseless case \cite{Giovannetti2006}, as well as in some noisy models including e.g.\ perfectly correlated dephasing \cite{Dorner2012, Macieszczak2014}, the description of metrological phase/frequency estimation protocols may, without loss of generality, be restricted to the fully symmetric subspace. Since the dimension of the symmetric subspace scales \emph{linearly} with particle number it is possible to perform direct and efficient calculations in the large particle limit. Apart from that, a number of powerful methods have been designed to obtain fundamental precision bounds for uncorrelated noise models, circumventing the issue that the states involved in the protocols are no longer restricted to the fully symmetric subspace \cite{Escher2011, Demkowicz2012, Kolodynski2013, Knysh2014, demkowicz2014using, sekatski2017quantum, demkowicz2017adaptive, zhou2018achieving}. However, in cases when one deals with partially correlated noise metrological models, or whenever one wishes to study the performance of states outside the fully symmetric subspace, there are no efficient methods that can be applied and one is forced to restrict considerations to small-scale problems.

There are several ways in which correlated noise manifests itself in metrological problems. The first is when environmental memory effects are significant, leading to probe dynamics with a time-correlated noise. The simplest example here is that of the atomic clock stabilization problem, where the effective dephasing process of atoms is temporally correlated as a result of correlations of the frequency fluctuations of the local oscillator \cite{Ludlow2015}. Because of this feature, identification of the optimally quantum clock stabilization strategies taking into account all the possible states of the atoms as well as quantum measurement and feedback strategies is a highly non-trivial task \cite{Andre2004, Borregaard2013b, Kessler2014}. One of the approaches, addressing the optimality of the protocols to stabilise atomic clocks, was based on the concept of the \emph{quantum Allan variance} (QAVAR) \cite{Chabuda2016}. Unfortunately the computation of the QAVAR for even small systems becomes quickly unfeasible, as its computational complexity grows exponentially with the number of atomic clock interrogations. An even more challenging case involves models where time-correlated noise cannot be effectively described via some classical stochastic process  \cite{Clerk2010, Szankowski2017} and as such manifests non-Markovian features of quantum dynamics \cite{Chin2012}, as e.g.\ in NV-center sensing models \cite{Rondin2014, Paz2017, Kwiatkowski2018}. A second natural setting exhibiting nontrivial noise correlations is that of many-body systems such as, e.g., spin chains. Here, typically, \emph{spatially} correlated noise emerges. This is of crucial relevance for any models where the effective signal is obtained from spatially distributed probes in the mean field estimation \cite{Jeske2014} or e.g.\ in \emph{field-gradient metrology} \cite{Altenburg2017, Appelaniz2018}.

Although realistic quantum noise is unlikely to be strictly uncorrelated, it is not going to be arbitrarily complicated. On the contrary, temporal noise correlations usually decay rapidly. Similarly, in the spatially correlated case one expects on dimensional and energetic grounds that noise will be short-range correlated. For typical short-range correlated noise processes the optimal performance of a metrological protocol is expected to be attainable with input probe states exhibiting entanglement within groups of a finite number of particles similarly as in the uncorrelated noise models \cite{Jarzyna2013a}. This key physical insight is a strong hint for how to go beyond the extant methods for uncorrelated and Markovian noise: one needs a way to represent short-range correlations in many body systems.

The most successful approach for classically simulating short-range correlated many body systems is via the variational \emph{tensor-network state} (TNS) ansatz (see, e.g., \cite{bridgemanHandwavingInterpretiveDance2017} for a recent review). Expressive ansatz classes such as the \emph{matrix product states} (MPS) \cite{fannesFinitelyCorrelatedStates1992,ostlundThermodynamicLimitDensity1995}, projected entangled-pair states (PEPS) \cite{verstraeteRenormalizationAlgorithmsQuantumMany2004}, or the multiscale entanglement renormalisation ansatz (MERA) \cite{vidalEntanglementRenormalization2007}, have led to unparalleled insights into the physics of quantum many body systems ranging from the description of quantum phase transitions to new phases of matter such as topological order. Most relevant for the present work is the application of tensor networks to the study of dissipative physics via the \emph{matrix product operator} (MPO) ansatz for density operators \cite{zwolakMixedStateDynamicsOneDimensional2003,verstraeteMatrixProductDensity2004} and also their infinite particle limit known as \emph{infinite} MPO (iMPO) \cite{TopoCincioVidal}.

With a few notable examples, the application of tensor-network methods to quantum metrology is still in its infancy. At the present time calculations of the QFI for mixed tensor-network states has been via approximate upper bounds, effectively amounting to a weighted sum of the QFI calculated on pure states \cite{Jarzyna2013a}---~a task equivalent to calculating the variance of a local observable. A further problem is that quantities such as the Quantum Fisher Information (QFI) or the Bayesian cost are not familiar in the tensor-network literature and, until now, it was far from clear whether they can be easily calculated at all in the mixed-state setting.

In this Paper we develop tensor-network based methods to address quantum metrological problems, in particular that involving short-range quantum noise. We demonstrate how to directly calculate relevant metrological quantities, such as the QFI and Bayesian-type cost, using the MPO formalism. Moreover, we also show how to optimize input probe states for metrological problems using the same formalism. As a result, we provide an efficient iterative procedure to design optimal metrological protocols that remains within the MPO formalism and does not suffer from the curse of dimensionality.

The Paper is extensive as its aim is also to bridge the gap between the tensor network and quantum metrology communities and the outline is as follows. In Sec.~\ref{sec:qmetrology} we review the main concepts of quantum metrology, including, the Cram{\'e}r-Rao bound and the quantum Fisher information, and specify the optimization algorithm that leads to the identification of the optimal metrological protocols. We also define the class of metrological models with correlated (but short range) noise for which the developed MPO methods are expected to be efficient. In Sec.~\ref{sec:mpo} we review the tensor networks formalism with a particular focus on the MPO and MPS construction and describe the whole metrological optimization algorithm in the language of MPO operations. We also present the iMPO approach where asymptotic performance of metrological protocols in the limit of large number of particles can be obtained in a direct way. This is followed, in Sec.~\ref{sec:example}, with a series of applications including: magnetic field sensing in presence of spatially correlated fluctuating field, calculation of the QAVAR for the atomic clock stabilisation problem taking into account temporal correlations of the local oscillator fluctuations, and finally demonstrate how the developed tools may be applied to problems outside the field of quantum metrology, namely, the calculation of the fidelity susceptibility of a finite-temperature thermal state of a many-body spin system. Finally, in Sec.~\ref{sec:conclusions} we conclude and provide future directions.

\section{Optimal quantum metrological protocols} \label{sec:qmetrology}
\subsection{General models} \label{subsec:qmetrology intro}
A typical problem in quantum metrology may be formulated as follows:
\begin{center}
	\begin{tikzpicture}[scale=0.5]
		\draw (-2+0.5, 0.5) -- (0, 0.5);
		\draw (-2, 0.5) circle[radius=0.5] node {$\rho_0$};
		\draw (0, 0) rectangle (1,1);
		\draw (0.5, 0.5) node {$\Lambda_\varphi$};
		\draw (1, 0.5) -- (2.5, 0.5);
		\draw (2.5, 0) -- (2.5, 1);
		\draw (3, 0.5) node {$\Pi_x$};
		\draw (2.5,1) .. controls (4,1) and (4,0) .. (2.5,0);
		\draw[dashed] (3.625,0.5) -- node[above] {$x$} (5,0.5);
		\draw (5, 0) rectangle (6.5,1);
		\draw (5.75,0.5) node {$\widetilde{\varphi}(x)$};
	\end{tikzpicture}.
\end{center}
Here a probe state $\rho_0$ is subject to a parameter-dependent quantum evolution, mathematically represented by a parameter-dependent quantum channel $\Lambda_{\varphi}$. A POVM type measurement $\{{\Pi}_x\}_x$ \cite{Nielsen2000} is then carried yielding a conditional probability distribution
\begin{equation}
	p(x|\varphi) = \tr({\rho}_{\varphi}{\Pi}_x),
\end{equation}
where
\begin{equation}
	\rho_\varphi=\Lambda_{\varphi}\left(\rho_{0}\right).
\end{equation}
Given the conditional probability distribution $p(x|\varphi)$ the objective is to estimate the value of the unknown parameter $\varphi$. To this end one employs an estimator function $\widetilde{\varphi}(x)$ to produce a given estimate for $\varphi$ given the measurement outcome $x$. The performance of an estimator function is limited by the average uncertainty
\begin{equation} \label{eq:var tilde phi}
	\Delta^2 \widetilde{\varphi} = \langle(\widetilde{\varphi}-\varphi)^2 \rangle,
\end{equation}
where the expectation is over all measurement results $x$. The central goal of quantum metrology is to find the best input probe state $\rho_0$, the best measurement and estimator so as to minimize $\Delta^2 \widetilde{\varphi}$. This is a challenging problem in general as both the optimal probe state and the measurement depend in a deeply nontrivial way on the channel $\Lambda_\varphi$.

Progress can be made by exploiting a fundamental result in quantum metrology, namely, the \emph{Cram{\'e}r-Rao} inequality \cite{Helstrom1976, Holevo1982, Braunstein1994}. This crucial result lower-bounds the average uncertainty of the best possible estimator $\widetilde{\varphi}(x)$ in terms of a quantity known as the QFI:
\begin{equation}
	\Delta^2 \widetilde{\varphi} \ge \frac{1}{F(\rho_0)},
\end{equation}
where
\begin{equation} \label{eq:qfi}
	F(\rho_0) = \tr(\rho_\varphi L^2),
\end{equation}
and $L$ is the \emph{symmetric logarithmic derivative} (SLD) defined implicitly via
\begin{equation} \label{eq:sld}
	\rho'_\varphi = \frac{1}{2}\left(L\rho_\varphi + \rho_\varphi L\right),
\end{equation}
where $\rho^\prime_\varphi$ is derivative of $\rho_\varphi$ with respect to $\varphi$. Instead of directly minimising $\Delta^2 \widetilde{\varphi}$ from Eq.~\eqref{eq:var tilde phi} one can instead maximise the QFI over input states $\rho_0$, an easier task in general as the optimization of the measurement $\Pi_x$ and the estimator is no longer required.

A calculation of the QFI requires solving  Eq.~\eqref{eq:sld} for the SLD operator $L$. This is a linear equation; its solution generally requires finding the eigenvalues and eigenvectors of $\rho_\varphi = \sum_j \lambda_j |\lambda_j\rangle\langle\lambda_j|$ \cite{Helstrom1976, Holevo1982, Braunstein1994}:
\begin{equation} \label{eq:L}
	L = \sum_{j,k} \frac{2 \langle \lambda_j|\rho'_\varphi|\lambda_k\rangle}{\lambda_j+\lambda_k} |\lambda_j\rangle\langle\lambda_k|.
\end{equation}
Performing a full eigendecomposition is  generally unfeasible for a system comprised of more than a small number of particles, and cannot be easily implemented using the efficient MPO description advocated in this paper. For this reason we do not use the above formula. Instead, it is much more numerically efficient to solve the linear equation \eqref{eq:sld} directly using standard linear equation solving methods in order to find $L$. Alternatively, one may reformulate the calculation of the QFI as the following maximization problem \cite{Macieszczak2013a, Macieszczak2014}:
\begin{equation} \label{eq:qfimax}
	F(\rho_0)= \sup_{L} F(\rho_0,L), \quad F(\rho_0,L) = 2\tr\left(\rho'_\varphi L \right)-\tr\left(\rho_\varphi L^2\right),
\end{equation}
where we will refer to $F(\rho_0,L)$ as the figure of merit (FoM) for the problem. To see the equivalence of the definition Eq.~\eqref{eq:qfi} with Eq.~\eqref{eq:qfimax} note that the above formulation is the supremum of a quadratic function of a Hermitian operator $L$. This optimization may be solved by formally taking the derivative with respect to $L$ and setting it to zero. The resulting extremum condition yields the equation for the SLD and hence the formula Eq.~\eqref{eq:qfi} for the QFI.

The QFI formula~\eqref{eq:qfimax} has an advantage over the original \eqref{eq:qfi}, when one wants to additionally perform optimization of the QFI over the input states $\rho_0$ in order to find the optimal quantum metrological protocol. Using \eqref{eq:qfimax} this problem can  be written as a double maximization problem:
\begin{equation}
	F = \sup_{\rho_0} F(\rho_0) = \sup_{\rho_0, L} F(\rho_0,L),
\end{equation}
where the FoM is linear in $\rho_0$ and quadratic in $L$. This formulation leads to an extremely efficient iterative numerical procedure for determining the optimal input probe state: start with some random (or an educated guess for an) input state, determine the corresponding optimal $L$ by performing the relevant optimization. Then, for the $L$ just found, reverse the procedure and look for the optimal input state. This procedure converges very quickly and yields the optimal input probe state as well as the corresponding QFI. This approach was first proposed in \cite{Demkowicz2011,Macieszczak2014} in the Bayesian estimation context, and then applied to the QFI FoM in \cite{Macieszczak2013a} (recently the method has been rediscovered in a slightly modified incarnation in \cite{Toth2018} and proved useful in studying metrological properties of PPT states). As we will see, each of these iterative steps may be performed efficiently using tensor networks.

Even though, in the above formulation, we focused solely on the QFI based approach, the above considerations are valid whenever the quantity to be optimized is given in the form \eqref{eq:qfimax}: $\rho^\prime_\varphi$ need not necessarily be the derivative of the state with respect to the estimated parameter. As such, this procedure is applicable in the Bayesian approach, and also in case of less trivial FoMs such as the QAVAR as discussed in Sec.~\ref{subsec:example allan}.

\subsection{Many-particle models with local parameter encoding and locally correlated noise} \label{subsec:qmetrology local}
In this subsection we describe the class of metrological models currently challenging for state-of-the-art methods. These are the main motivation for the development of MPO-based techniques. We focus on systems comprised of $N$ distinguishable $d$-dimensional particles, so that the total Hilbert space is $\mathcal{H} = \bigotimes_{n=1}^N \mathbb{C}^d$. We assume that the parameter $\varphi$ is \emph{unitarily} encoded in the output state $\rho_\varphi$ according to a product of unitaries given by the exponential of local generators (or \emph{Hamiltonians}):
\begin{equation} \label{eq:rhophi}
	\rho_\varphi = \Lambda_\varphi(\rho_0) = e^{ - \mathrm{i} H \varphi }\Lambda(\rho_0) e^{\mathrm{i} H \varphi}, \quad H = \sum_{n=1}^N h^{[n]},
\end{equation}
where $h^{[n]}$ is the generator acting on the $n$th particle.

Most importantly for this paper, the noise, represented above by the operator $\Lambda$, is \emph{not} assumed to be local, which makes the problem particularly challenging. Powerful methods capable of yielding fundamental metrological bounds in the large particle number regime effectively work only in case of uncorrelated noise models \cite{Escher2011, Demkowicz2012, Kolodynski2013, Knysh2014, demkowicz2014using, demkowicz2017adaptive, zhou2018achieving} and cannot be directly used to study the effects of noise correlations. In many physically realistic situations, however, correlated noise tends to be only locally correlated, which gives one hope that more efficient methods to deal with such problems than simple brute force numerical optimization may be found.

We assume that $\Lambda$ may be effectively approximated as a product of single ($\Lambda^{[n]}$), two- ($\Lambda^{[n,n+1]}$), three- ($\Lambda^{[n,n+1,n+2]}$), etc.\ particle maps up to some cut-off point after which to is assumed that noise correlations do not extend beyond $r$ neighbouring particles. From a more physical perspective, consider a time-independent quantum master equation \cite{breuer2002theory} describing the noisy part of the evolution of an $N$-body quantum system:
\begin{align}
\label{eq:master}
    \frac{\dd \rho}{\dd t} &= \sum_{k=1}^{r}\sum_{n=1}^N  \mathcal{L}^{(k,n)}(\rho), \\ \nonumber
    \mathcal{L}^{(k,n)}(\rho) &= \sum_{j} \mathcal{D}\left[ L^{[n,\dots,n+k-1]}_j\right] (\rho),
\end{align}
where
\begin{equation}
\label{eq:master2}
	\mathcal{D}[c](\rho) =  c\rho c^\dagger -
	\frac{1}{2}\left( \rho  c^{\dagger}  c  + c^{\dagger} c \rho \right).
\end{equation}
In the above, the $L^{[n,\dots,n+k-1]}_j$ are noise operators acting on $k$ neighbouring particles (not to be confused with the SLD), and the combined effect of the $k$-particle noise acting on the subset $[n,\dots,n+k-1]$ of particles is represented by the $\mathcal{L}^{(k,n)}$ operator, where we neglect terms with higher range than $r$---note that a three-particle term may in particular represent a two-body interaction between next-nearest neighbours. The channel $\Lambda$ can now be obtained by integrating the evolution over some fixed time $t$:
\begin{equation}\label{eq:localnoisechannel}
	\Lambda = \exp\left(\sum_{k=1}^r\sum_{n=1}^N \mathcal{L}^{(k,n)} t\right).
\end{equation}
If all the operators in the above exponent commute we can immediately write $\Lambda$ as a product of single, two-, three- etc.\ maps acting on different subsets of particles, which will lead us immediately to an efficient MPO description of the dynamics, see Sec.~\ref{sec:mpo}. Otherwise, one may approximate the evolution for a time $t$ as a product of short time steps, where in each time step we perform the Suzuki-Trotter decomposition \cite{Trotter_59,Suzuki_66,Suzuki_76}.

In what follows we often use a vectorized density matrix notation $\rho \rightarrow |\rho\rangle$, which is useful in the MPO approach, and where clear from context, we switch between the cases where $\Lambda$ is understood as acting on $\rho$ or $\ket{\rho}$. For example, $|\Lambda(\rho_0)\rangle = \Lambda |\rho_0\rangle$.

For definiteness, we assumed above that the noise acts \emph{before} the unitary encoding. This entails no loss of generality if the noise commutes with the encoding, as is the case in the most popular metrological models of phase/frequency estimation in presence of dephasing or loss \cite{Dorner2008, Escher2011, Demkowicz2012}. If needed, one can extend our formalism to the case where the parameter dependence no longer commutes. This comes at the expense of a slightly higher complexity of formulas and numerics, since we are no longer able to write the derivative of $\rho_\varphi$ over the parameter as a commutator with the Hamiltonian. Instead, the entire channel structure $\Lambda_\varphi$ determines the derivative.

Note that, thanks to the assumptions of the model, if the input probe state $\rho_0$ is only locally correlated, which is a sufficient condition for an efficient MPO description, then it remains so under the above evolution. Moreover, the derivative of the output state with respect to the parameter $\varphi$---required for calculations of the QFI---reads $\rho'_\varphi = \ii \left[\rho_\varphi,H\right]$ and, since $H$ is the sum of local Hamiltonians, is also be efficiently describable using MPO.

Most importantly, in essentially all realistic metrological protocols, the maximal achievable QFI scales linearly with $N$ in the limit of large particle numbers \cite{Escher2011, Demkowicz2012}, and the quantum-enhancement advantage appears in the form of a constant factor. This further  implies that it is enough to consider input states $\rho_0$ with finite-range correlations to achieve almost optimal metrological performance \cite{Jarzyna2013a}. As a result the MPO formalism is ideally suited to tackle this class of metrological problems and guarantees that the optimal values of QFI as well as the optimal probe states will be found via this approach.

\section{Matrix product operator approach} \label{sec:mpo}
In this section we present the formalism of MPO (and also closely connected MPS) type tensor networks, adapted to quantum metrological problems, and show effective ways to use it in order to solve the optimizations formulated in the previous section.

\subsection{Review of tensor networks} \label{subsec:mpo review}
First we give a short review of the tensor-network formalism (for a more comprehensive recent review see \cite{bridgemanHandwavingInterpretiveDance2017}). To succinctly describe tensor networks we represent tensors diagrammatically: suppose that $T$ is a tensor with $N$ indices, each ranging from $0$ to $d-1$, elements of this tensor we denote as $T_{j_1j_2\dots j_N}$. We depict such a tensor as a circle with $N$ legs, each labelled with an index:
\begin{center}
	\begin{tikzpicture}[scale=0.75]
		\draw (-3.5,0) node {$T_{j_1j_2\dots j_N} =$};
		\foreach \x / \myn in {0/1, 60/2, 120/3, 180/4, 240/5}
		{
			\draw (0,0) -- +(\x:1.2);
			\draw +(\x:1.4) node {$j_{\myn}$};
		}
		\draw (0,0) -- +(300:1.2);
		\draw +(300:1.4) node {$j_N$};
		\draw[fill=white] circle (0.4);
		\draw (0,0) node {$T$};
		\draw +(270:1.2) node {$\dots$};
	\end{tikzpicture}.
\end{center}
Tensors in this paper are nothing more than lists of numbers; we don't assume any particular transformation properties for our tensors. Some simple special cases include kets $|v \rangle$, bras $\langle v|$ ($N=1$) and operators $\mathcal{M}$ ($N=2$):
\begin{center}
	\begin{tikzpicture}[scale=0.75]
		\draw (-1,0) -- (0,0);
		\draw[fill=white] circle (0.4);
		\draw (0,0) node {$v$};
	\end{tikzpicture},\hspace{1cm}
	\begin{tikzpicture}[scale=0.75]
		\draw (0,0) -- (1,0);
		\draw[fill=white] circle (0.4);
		\draw (0,0) node {$\overline{v}$};
	\end{tikzpicture},\hspace{1cm}
	\begin{tikzpicture}[scale=0.75]
		\draw (-1,0) -- (1,0);
		\draw[fill=white] circle (0.4);
		\draw (0,0) node {$\mathcal{M}$};
	\end{tikzpicture},
\end{center}
where the overline denotes complex conjugation.

\emph{Tensor contraction}, whereby the components of two tensors $T$ and $W$ are multiplied and summed over repeated indices, is depicted by connecting the legs, e.g.,
\begin{center}
	\begin{tikzpicture}[scale=0.75]
		\foreach \x / \myn in {0, 60, 120, 180, 240}
		{
			\draw (0,0) -- +(\x:1.2);
		}
		\foreach \x / \myn in {60/2, 120/3, 180/4, 240/5}
		{
			\draw +(\x:1.4) node {$j_{\myn}$};
		}
		\draw (0,0) -- +(300:1.2);
		\draw +(300:1.4) node {$j_N$};
		\draw[fill=white] circle (0.4);
		\draw (0,0) node {$T$};
		\draw +(270:1.2) node {$\dots$};
		\foreach \x / \myn in {0/1, 60/2, 120/3, 180/4, 240/5}
		{
			\draw (2.4,0) -- +(\x:1.2);
		}
		\foreach \x / \myn in {0/1, 60/2, 120/3, 240/5}
		{
			\draw (2.4,0)+(\x:1.4) node {$k_{\myn}$};
		}
		\draw (2.4,0) -- +(300:1.2);
		\draw (2.4,0)+(300:1.4) node {$k_M$};
		\draw[fill=white] (2.4,0) circle (0.4);
		\draw (2.4,0) node {$W$};
		\draw (2.4,0)+(270:1.2) node {$\dots$};
	\end{tikzpicture}.
\end{center}
Here this \emph{tensor network} depicts the contraction
\begin{equation}
	\sum_{l=0}^{d-1} T_{lj_2\dots j_N} W_{k_1\dots k_3lk_5 \dots k_M},
\end{equation}
which is a tensor with $N+M-2$ legs. By combining tensors with three or more legs via tensor contraction we can build networks of arbitrary complexity. (Tensor networks involving tensors with one or two legs are necessarily a combination of lines and cycles.)

Among various tensor network classes, the most important for this paper are the \emph{matrix product state} (MPS) and \emph{matrix product operator} (MPO) tensor networks. MPS/MPO are a natural compact representation for states/operators with finite correlations so we expect that they form natural language to study quantum metrology problems with locally correlated noise. An MPS with \emph{open boundary conditions} (OBC) is a state of the form
\begin{center}
	\begin{tikzpicture}[scale=0.75,every node/.style={scale=0.75}]
		\draw (-1.5,0) node[scale=1.33] {$|\psi\rangle =$};
		\draw (0,0) -- (0,1.5);	
		\foreach \x in {1.5, 3.0, ..., 7}
		{
			\draw (\x-1.5,0) -- (\x,0);
			\draw[line width=2mm, white] (\x,0) -- (\x,1.5);
			\draw (\x,0) -- (\x,1.5);
		}
		\draw (6,0) -- (6.45,0);
		\draw (6.8,0) node {$\dots$};
		\draw (7.05,0) -- (7.5,0);
		\draw[line width=2mm, white] (7.5,0) -- (7.5,1.5);
		\draw (7.5,0) -- (7.5,1.5);
		\foreach \x/\c in {0/{1}, 1.5/{2}, 3/{3}, 4.5/{4}, 6/{5}}
		{
			\draw[fill=white] (\x,0) circle (0.4);
			\draw (\x,0) node {$A[\c]$};
		}
		\draw[fill=white] (7.5,0) circle (0.4);
		\draw (7.5,0) node {$A[N]$};
	\end{tikzpicture}.
\end{center}
Matrix product states with OBCs are now known to provide an excellent model for the ground states of one-dimensional quantum spin chains with (constant) spectral gap \cite{hastings_area_2007}.

We can accommodate \emph{periodic boundary conditions} (PBCs) by joining the last tensor to the first via an additional ``horizontal leg'':
\begin{center}
	\begin{tikzpicture}[scale=0.75,every node/.style={scale=0.75}]
		\draw (-1.5,0) node[scale=1.33] {$|\psi\rangle =$};
		\draw (-0.3,0) arc (-90:-270:0.5);
		\draw (-0.3,1) -- (7.8,1);
		\draw[line width=2mm, white] (0,0) -- (0,1.5);
		\draw (0,0) -- (0,1.5);	
		\foreach \x in {1.5, 3.0, ..., 7}
		{
			\draw (\x-1.5,0) -- (\x,0);
			\draw[line width=2mm, white] (\x,0) -- (\x,1.5);
			\draw (\x,0) -- (\x,1.5);
		}
		\draw (6,0) -- (6.45,0);
		\draw (6.8,0) node {$\dots$};
		\draw (7.05,0) -- (7.5,0);
		\draw (7.8,0) arc (-90:90:0.5);
		\draw[line width=2mm, white] (7.5,0) -- (7.5,1.5);
		\draw (7.5,0) -- (7.5,1.5);
		\foreach \x/\c in {0/{1}, 1.5/{2}, 3/{3}, 4.5/{4}, 6/{5}}
		{
			\draw[fill=white] (\x,0) circle (0.4);
			\draw (\x,0) node {$A[\c]$};
		}
		\draw[fill=white] (7.5,0) circle (0.4);
		\draw (7.5,0) node {$A[N]$};
	\end{tikzpicture}.
\end{center}
Such MPS with PBCs offer some numerical advantages for quantum spin systems on rings \cite{verstraete_density_2004}.

A \emph{matrix product operator} (MPO) is a tensor network which is, in the PBCs case, a linear operator from $\left(\mathbb{C}^d\right)^{\otimes N}$ to itself parametrised according to
\begin{center}
	\begin{tikzpicture}[scale=0.75, every node/.style={scale=0.75}]
		\draw (-0.3,0) arc (-90:-270:0.5);
		\draw (-0.3,1) -- (7.8,1);
		\draw[line width=2mm, white] (0,-1.5) -- (0,1.5);
		\draw (0,-1.5) -- (0,1.5);	
		\foreach \x in {1.5, 3.0, ..., 7}
		{
			\draw (\x-1.5,0) -- (\x,0);
			\draw[line width=2mm, white] (\x,-1.5) -- (\x,1.5);
			\draw (\x,-1.5) -- (\x,1.5);
		}
		\draw (6,0) -- (6.45,0);
		\draw (6.8,0) node {$\dots$};
		\draw (7.05,0) -- (7.5,0);
		\draw (7.8,0) arc (-90:90:0.5);
		\draw[line width=2mm, white] (7.5,-1.5) -- (7.5,1.5);
		\draw (7.5,-1.5) -- (7.5,1.5);
		\foreach \x/\c in {0/{1}, 1.5/{2}, 3/{3}, 4.5/{4}, 6/{5}}
		{
			\draw[fill=white] (\x,0) circle (0.4);
			\draw (\x,0) node {$A[\c]$};
		}
		\draw[fill=white] (7.5,0) circle (0.4);
		\draw (7.5,0) node {$A[N]$};
	\end{tikzpicture},
\end{center}
where the last horizontal leg is contracted with the first. Here the tensor $A[n]$ has four legs:
\begin{center}
	\begin{tikzpicture}[scale=0.75, every node/.style={scale=0.75}]
		\draw (0,-1) -- (0,1);
		\draw (-1,0) -- (1,0);
		\draw[fill=white] (0,0) circle (0.4);
		\draw (0,0) node {$A[n]$};
		\draw (-1.25,0) node {$\alpha$};
		\draw (1.25,0) node {$\beta$};
		\draw (0,-1.25) node {$k$};
		\draw (0,1.25) node {$j$};
	\end{tikzpicture}.
\end{center}
The horizontal-leg indices (also called the virtual indices) $\alpha$ and $\beta$ range from $1,2,\dots, D$, where the parameter $D$ is called the \emph{bond dimension} and the vertical leg indices (also called the physical indices) $j$ and $k$ range from $0,1,\dots, d-1$, where $d$ is the \emph{physical dimension}. Thus an MPO $\mu$ with PBCs is the following operator
\begin{multline}
	\mu = \sum_{\substack{j_1, \dots, j_N,\\ k_1, \dots, k_N = 0}}^{d-1}\tr\left(A[1]^{j_1}_{k_1}A[2]^{j_2}_{k_2} \dots A[N]^{j_N}_{k_N}\right)\times \\
	|j_1,j_2,\dots, j_N\rangle\langle k_1,k_2,\dots, k_N|.
\end{multline}

Exploiting channel/state duality we can bend the vertical legs upward ($\ket{j}\bra{k} \rightarrow \ket{j,k}$) to write an MPO $\mu$ as an MPS $|\mu\rangle$ of $2N$ particles:
\begin{center}
	\begin{tikzpicture}[scale=0.75, every node/.style={scale=0.75}]
		\draw (-0.3,0) arc (90:270:0.5);
		\draw (-0.3,-1) -- (7.8,-1);
		\draw[line width=2mm, white] (0,0) -- (0,1.5);
		\draw (0,0) -- (0,1.5);	
		\foreach \x in {1.5, 3.0, ..., 7}
		{
			\draw (\x-1.5,0) -- (\x,0);
			\draw[line width=2mm, white] (\x,0) -- (\x,1.5);
			\draw (\x,0) -- (\x,1.5);
		}
		\draw (6,0) -- (6.45,0);
		\draw (6.8,0) node {$\dots$};
		\draw (7.05,0) -- (7.5,0);
		\draw (7.8,0) arc (90:-90:0.5);
		\draw[line width=2mm, white] (7.5,0) -- (7.5,1.5);
		\draw (7.5,0) -- (7.5,1.5);
		\foreach \x/\c in {0/{1}, 1.5/{2}, 3/{3}, 4.5/{4}, 6/{5}}
		{
			\draw[fill=white] (\x,0) circle (0.4);
			\draw (\x,0) node {$A[\c]$};
		}
		\draw[fill=white] (7.5,0) circle (0.4);
		\draw (7.5,0) node {$A[N]$};
		\foreach \x in {0, 1.5, ..., 8.5}
		{
			\draw[line width=2mm, white] (\x+0.6, -0.4) -- (\x+0.6, 1.5);
			\draw (\x,-0.4) arc (-180:0:0.3);
			\draw (\x+0.6, -0.4) -- (\x+0.6, 1.5);
		}
	\end{tikzpicture}.
\end{center}
Defining a new vertical line \raisebox{-3pt}{
\begin{tikzpicture}[scale=0.75]
	\draw[line width=0.5mm] (-0.2,0) -- (-0.2,0.5);
	\draw (0.2,0.25) node {$=$};
	\draw (0.6,0) -- (0.6,0.5);
	\draw (0.8,0) -- (0.8,0.5);
\end{tikzpicture}
} to range over a doubled index $(j,k)$, with $j,k=0,1, \dots, d-1$, we arrive at the equivalent tensor network for $|\mu\rangle$:
\begin{center}
	\begin{tikzpicture}[scale=0.75, every node/.style={scale=0.75}]
		\draw (-0.3,0) arc (90:270:0.5);
		\draw (-0.3,-1) -- (7.8,-1);
		\draw[line width=0.5mm] (0,0) -- (0,1.5);	
		\foreach \x in {1.5, 3.0, ..., 7}
		{
			\draw (\x-1.5,0) -- (\x,0);
			\draw[line width=0.5mm] (\x,0) -- (\x,1.5);
		}
		\draw (6,0) -- (6.45,0);
		\draw (6.8,0) node {$\dots$};
		\draw (7.05,0) -- (7.5,0);
		\draw (7.8,0) arc (90:-90:0.5);
		\draw[line width=2mm, white] (7.5,0) -- (7.5,1.5);
		\draw[line width=0.5mm] (7.5,0) -- (7.5,1.5);
		\foreach \x/\c in {0/{1}, 1.5/{2}, 3/{3}, 4.5/{4}, 6/{5}}
		{
			\draw[fill=white] (\x,0) circle (0.4);
			\draw (\x,0) node {$A[\c]$};
		}
		\draw[fill=white] (7.5,0) circle (0.4);
		\draw (7.5,0) node {$A[N]$};
	\end{tikzpicture}.
\end{center}

The multiplication of two MPOs $\mu$ and $\nu$, with bond dimensions $D_1$ and $D_2$, respectively, is given by
\begin{center}
	\begin{tikzpicture}[scale=0.75, every node/.style={scale=0.75}]
		\draw (-0.3,-1.5) arc (-90:-270:0.5);
		\draw (-0.3,1-1.5) -- (7.8,1-1.5);
		\draw[line width=2mm, white] (0,-1.5-1.5) -- (0,1.5-1.5);
		\draw (0,-1.5-1.5) -- (0,1.5-1.5);	
		\foreach \x in {1.5, 3.0, ..., 7}
		{
			\draw (\x-1.5,0-1.5) -- (\x,0-1.5);
			\draw[line width=2mm, white] (\x,-1.5-1.5) -- (\x,1.5-1.5);
			\draw (\x,-1.5-1.5) -- (\x,1.5-1.5);
		}
		\draw (6,0-1.5) -- (6.45,0-1.5);
		\draw (6.8,0-1.5) node {$\dots$};
		\draw (7.05,0-1.5) -- (7.5,0-1.5);
		\draw (7.8,0-1.5) arc (-90:90:0.5);
		\draw[line width=2mm, white] (7.5,-1.5-1.5) -- (7.5,1.5-1.5);
		\draw (7.5,-1.5-1.5) -- (7.5,1.5-1.5);
		\foreach \x/\c in {0/{1}, 1.5/{2}, 3/{3}, 4.5/{4}, 6/{5}}
		{
			\draw[fill=white] (\x,0-1.5) circle (0.4);
			\draw (\x,0-1.5) node {$B[\c]$};
		}
		\draw[fill=white] (7.5,0-1.5) circle (0.4);
		\draw (7.5,0-1.5) node {$B[N]$};
		\draw (-0.3,0) arc (-90:-270:0.5);
		\draw (-0.3,1) -- (7.8,1);
		\draw[line width=2mm, white] (0,0) -- (0,1.5);
		\draw (0,0) -- (0,1.5);	
		\foreach \x in {1.5, 3.0, ..., 7}
		{
			\draw (\x-1.5,0) -- (\x,0);
			\draw[line width=2mm, white] (\x,0) -- (\x,1.5);
			\draw (\x,0) -- (\x,1.5);
		}
		\draw (6,0) -- (6.45,0);
		\draw (6.8,0) node {$\dots$};
		\draw (7.05,0) -- (7.5,0);
		\draw (7.8,0) arc (-90:90:0.5);
		\draw[line width=2mm, white] (7.5,0) -- (7.5,1.5);
		\draw (7.5,0) -- (7.5,1.5);
		\foreach \x/\c in {0/{1}, 1.5/{2}, 3/{3}, 4.5/{4}, 6/{5}}
		{
			\draw[fill=white] (\x,0) circle (0.4);
			\draw (\x,0) node {$A[\c]$};
		}
		\draw[fill=white] (7.5,0) circle (0.4);
		\draw (7.5,0) node {$A[N]$};
	\end{tikzpicture}.
\end{center}
Contracting the two tensors $A$ and $B$ vertically, and combining the two horizontal lines into a new horizontal line ranging from $1, 2, \dots, D_1 D_2$:
\begin{center}
	\begin{tikzpicture}[scale=0.75]
		\draw (-4,0.5) -- (-4,-1.5);
		\draw[line width = 0.5mm] (-5,-0.5) -- (-3,-0.5);
		\draw[fill=white] (-4,-0.5) circle (0.4);
		\draw (-4,-0.5) node[scale=0.75] {$C[n]$};
		\draw (-5.25,-0.5) node {$\gamma$};
		\draw (-2.75,-0.5) node {$\delta$};
		\draw (-4,-1.75) node {$k$};
		\draw (-4,0.75) node {$j$};
		\draw (-2.0,-0.5) node {$=$};
		\draw (0,-1) -- (0,1);
		\draw (-1,0) -- (1,0);
		\draw (-1.25,0) node {$\alpha_1$};
		\draw (1.25,0) node {$\beta_1$};
		\draw (0,1.25) node {$j$};
		\draw (0,-2) -- (0,0);
		\draw (-1,-1) -- (1,-1);
		\draw[fill=white] (0,-1) circle (0.4);
		\draw (0,-1) node[scale=0.75] {$B[n]$};
		\draw (-1.25,-1) node {$\alpha_2$};
		\draw (1.25,-1) node {$\beta_2$};
		\draw (0,-2.25) node {$k$};
		\draw[fill=white] (0,0) circle (0.4);
		\draw (0,0) node[scale=0.75] {$A[n]$};
	\end{tikzpicture},
\end{center}
results in a new MPO $\mu\nu$ with bond dimension $D_1 D_2$:
\begin{center}
	\begin{tikzpicture}[scale=0.75, every node/.style={scale=0.75}]
		\draw[line width = 0.5mm] (-0.3,0) arc (-90:-270:0.5);
		\draw[line width = 0.5mm] (-0.3,1) -- (7.8,1);
		\draw[line width=2mm, white] (0,-1.5) -- (0,1.5);
		\draw (0,-1.5) -- (0,1.5);	
		\foreach \x in {1.5, 3.0, ..., 7}
		{
			\draw[line width = 0.5mm] (\x-1.5,0) -- (\x,0);
			\draw[line width=2mm, white] (\x,-1.5) -- (\x,1.5);
			\draw (\x,-1.5) -- (\x,1.5);
		}
		\draw[line width = 0.5mm] (6,0) -- (6.45,0);
		\draw (6.8,0) node {$\dots$};
		\draw[line width = 0.5mm] (7.05,0) -- (7.5,0);
		\draw[line width = 0.5mm] (7.8,0) arc (-90:90:0.5);
		\draw[line width=2mm, white] (7.5,-1.5) -- (7.5,1.5);
		\draw (7.5,-1.5) -- (7.5,1.5);
		\foreach \x/\c in {0/{1}, 1.5/{2}, 3/{3}, 4.5/{4}, 6/{5}}
		{
			\draw[fill=white] (\x,0) circle (0.4);
			\draw (\x,0) node {$C[\c]$};
		}
		\draw[fill=white] (7.5,0) circle (0.4);
		\draw (7.5,0) node {$C[N]$};
	\end{tikzpicture}.
\end{center}

The bond dimension $D$ is a \emph{refinement parameter} limiting the correlations occurring in the MPO/MPS representation of an operator/state: when the correlations have a finite range there is a finite $D$ capable of representing operator/state accurately. As the cost of tensor network computations is polynomial in $D$, the MPO/MPS ansatz is a basis for powerful numerical methods. The bond dimension of an MPS is directly connected to the \emph{entanglement between a bipartition} of the chain: when one computes the entanglement entropy of a contiguous collection $[j,k] = \{j,j+1,\dots,k \}$ of spins one may derive the bound
\begin{equation}
S(\rho_{[j,k]}) \le 2\log_2(D),
\end{equation}
on the entanglement between the region $[j,k]$ and the rest of the chain (see, e.g., \cite{fannesFinitelyCorrelatedStates1992,vidal_efficient_2003a,vidal_efficient_2003b} for an elaboration of this result amongst many others). Accordingly, if a quantum state has a large bipartite entanglement then a larger $D$ is required to represent it as an MPS, and hence the harder it is to approximate it numerically.

A central tool for tensor network manipulations is the \emph{singular value decomposition} (SVD), according to which, for all operators $T$ there exist unitaries $U$ and $V$ and a diagonal matrix $S$ with non-negative real numbers on the diagonal called \emph{singular values}, such that
\begin{equation}
	T = USV^\dag.
\end{equation}
This may be diagrammatically represented as follows:
\begin{center}
	\begin{tikzpicture}[scale=0.75]
		\draw (-1,0) -- (1,0);
		\draw[fill=white] circle (0.4);
		\draw (0,0) node {$T$};
		\draw (1.5,0) node {$=$};
		\draw (2,0) -- (6,0);
		\draw[fill=white] (3,0) circle (0.4);
		\draw (3,0) node {$U$};
		\draw[fill=white] (4,0) circle (0.4);
		\draw (4,0) node {$S$};
		\draw[fill=white] (5,0) circle (0.4);
		\draw (5,0) node {$V^\dag$};
	\end{tikzpicture}.
\end{center}

So far we have discussed tensor networks for finite collections of particles. A crucial advantage of the tensor-network formalism is that we can easily extend the MPO (as well as MPS) ansatz to apply to \emph{infinite}-sized systems; we simply allow the network to extend to infinity from either side:
\begin{center}
	\begin{tikzpicture}[scale=0.75, every node/.style={scale=0.75}]
		\draw (-1,0) node[left] {$\dots$} -- (0,0);
		\draw (7.5,0) -- (8.5,0) node[right] {$\dots$};
		\draw[line width=2mm, white] (0,-1.5) -- (0,1.5);
		\draw (0,-1.5) -- (0,1.5);	
		\foreach \x in {1.5, 3.0, ..., 7}
		{
			\draw (\x-1.5,0) -- (\x,0);
			\draw[line width=2mm, white] (\x,-1.5) -- (\x,1.5);
			\draw (\x,-1.5) -- (\x,1.5);
		}
		\draw (6,0) -- (6.45,0);
		\draw (6.8,0) node {$\dots$};
		\draw (7.05,0) -- (7.5,0);
		\draw[line width=2mm, white] (7.5,-1.5) -- (7.5,1.5);
		\draw (7.5,-1.5) -- (7.5,1.5);
		\foreach \x/\c in {0/{1}, 1.5/{2}, 3/{3}, 4.5/{4}, 6/{5}}
		{
			\draw[fill=white] (\x,0) circle (0.4);
			\draw (\x,0) node {$A[\c]$};
		}
		\draw[fill=white] (7.5,0) circle (0.4);
		\draw (7.5,0) node {$A[N]$};
	\end{tikzpicture}.
\end{center}
In order to work with such networks it is expedient to assume \emph{translation invariance} (TI), which is imposed by assuming the tensor does not vary from site to site, so for each $n$: $A[n]=A$. With this simple assumption it becomes possible to contract and evaluate infinite MPO (iMPO).

A key primitive operation for manipulations involving iMPOs is the \emph{trace}. Diagrammatically the trace $\tr(\mu)$ of an iMPO may be obtained by connecting the vertical legs:
\begin{center}
	\begin{tikzpicture}[scale=0.75, every node/.style={scale=0.75}]
		\draw (-1,0) node[left] {$\dots$} -- (0,0);
		\draw (7.5,0) -- (8.5,0) node[right] {$\dots$};
		\draw[line width=2mm, white] (0,-1.5) -- (0,1.5);	
		\draw (0,-0.5) -- (0,0.5) to[out=90,in=180] (0+0.3,0.8) to[out=0,in=90] (0+0.6,0.5) -- (0+0.6,-0.5) to[out=-90,in=0] (0+0.3,-.8) to[out=180,in=-90] (0,-.5);
		\foreach \x in {1.5, 3.0, ..., 7}
		{
			\draw (\x,-0.5) -- (\x,0.5) to[out=90,in=180] (\x+0.3,0.8) to[out=0,in=90] (\x+0.6,0.5) -- (\x+0.6,-0.5) to[out=-90,in=0] (\x+0.3,-.8) to[out=180,in=-90] (\x,-.5);
			\draw[line width=2mm, white] (\x-1.5,0) -- (\x,0);
			\draw (\x-1.5,0) -- (\x,0);
		}
		\draw[line width=2mm, white] (6,0) -- (7.5,0);
		\draw (6,0) -- (7.5,0);
		\draw (7.5,-0.5) -- (7.5,0.5) to[out=90,in=180] (7.5+0.3,0.8) to[out=0,in=90] (7.5+0.6,0.5) -- (7.5+0.6,-0.5) to[out=-90,in=0] (7.5+0.3,-.8) to[out=180,in=-90] (7.5,-.5);
		\draw (7.05,0) -- (7.5,0);
		\foreach \x/\c in {0/{1}, 1.5/{2}, 3/{3}, 4.5/{4}, 6/{5}}
		{
			\draw[fill=white] (\x,0) circle (0.4);
			\draw (\x,0) node {$A$};
		}
		\draw[fill=white] (7.5,0) circle (0.4);
		\draw (7.5,0) node {$A$};
		\draw[line width=2mm, white] (7.9,0) -- (8.5,0);
		\draw (7.9,0) -- (8.5,0);
	\end{tikzpicture}.
\end{center}
Define
\begin{center}
	\begin{tikzpicture}[scale=0.75,every node/.style={scale=0.75}]
		\begin{scope}[shift={(-5,0)}]
			\draw (-1.5,0) -- (1.5,0);
			\draw (2.25,0) node[right] {$=$};
			\draw[fill=white] (0,0) circle (0.4);
			\draw (0,0) node {$E$};
		\end{scope}
		\draw (0,-0.5) -- (0,0.5) to[out=90, in=90,looseness=2] (0.75,0.5) to[out=-90, in=90] (0.75,-0.5) to[out=-90, in=-90,looseness=2] (0,-0.5);
		\draw[line width = 1.5mm, white] (-1.5,0) -- (1.5,0);
		\draw (-1.5,0) -- (1.5,0);
		\draw[fill=white] (0,0) circle (0.4);
		\draw (0,0) node {${A}$};
	\end{tikzpicture}.
\end{center}
In this way we obtain for the trace $\tr(\mu)$ a tensor network involving the infinite product of so-called \emph{transfer matrices} $E$:
\begin{center}
	\begin{tikzpicture}[scale=0.75, every node/.style={scale=0.75}]
		\draw (-1,0) node[left] {$\dots$} -- (0,0);
		\draw (7.5,0) -- (8.5,0) node[right] {$\dots$};
		\draw[line width=2mm, white] (0,-1.5) -- (0,1.5);	
		\foreach \x in {1.5, 3.0, ..., 7}
		{
			\draw[line width=2mm, white] (\x-1.5,0) -- (\x,0);
			\draw (\x-1.5,0) -- (\x,0);
		}
		\draw[line width=2mm, white] (6,0) -- (7.5,0);
		\draw (6,0) -- (7.5,0);
		\foreach \x/\c in {0/{1}, 1.5/{2}, 3/{3}, 4.5/{4}, 6/{5}}
		{
			\draw[fill=white] (\x,0) circle (0.4);
			\draw (\x,0) node {$E$};
		}
		\draw[fill=white] (7.5,0) circle (0.4);
		\draw (7.5,0) node {$E$};
		\draw[line width=2mm, white] (7.9,0) -- (8.5,0);
		\draw (7.9,0) -- (8.5,0);
	\end{tikzpicture}
\end{center}
or, in equations,
\begin{equation}
	\tr(\mu) = \lim_{n\rightarrow \infty}\tr(E^{n}).
\end{equation}
To calculate this expression we note that when $E$ is diagonalizable (does not have to be Hermitian) we can decompose it: $E = \sum_{j} \lambda_j |r_j) (l_j|$, where $|r_j)$, $(l_j|$ are respectively right and left eigenvectors of $E$ which can be normalized that $(l_i|r_j)=\delta_{ij}$. This decomposition in diagrammes looks like
\begin{center}
	\begin{tikzpicture}[scale=0.7, every node/.style={scale=0.75}]
		\begin{scope}[shift={(4,0.75)}]
			\draw (-1,0) -- (1,0);
			\draw[fill=white] (0,0) circle (0.428);
			\draw (0,0) node {$E$};
		\end{scope}
		\begin{scope}[shift={(8.25,0.75)}]
			\draw (-2,-0.2) node[scale=10/8] {$\displaystyle =\sum_{j} \lambda_j$};
			\draw (-1,0) -- (0,0);
			\draw[fill=white] (0,0) circle (0.428);
			\draw (0,0) node {$r_j$};
			\draw (1.5,0) -- (2.25,0);
			\draw[fill=white] (1.25,0) circle (0.428);
			\draw (1.25,0) node {$l_j$};
		\end{scope}
	\end{tikzpicture},
\end{center}
and give us dominant contribution of $E^{n}$, determined by the (here assumed unique) leading eigenvalue $\lambda_1$:
\begin{equation} \label{eq:TM eig1}
	E^{n} \sim  \lambda_1^n |r_1) (l_1|,
\end{equation}
and $\tr(E^{n}) \sim  \lambda_1^n$.
Thus, the limit in an expression such as
\begin{equation}
	\lim_{n\rightarrow\infty}\frac{\tr(E^n)}{\lambda_1^n}
\end{equation}
exists and is equal to $1$.

\subsection{MPO representation of $\rho_\varphi$ and $\rho'_\varphi$} \label{subsec:mpo rho}
In this subsection we exploit the tensor-network representation to write $\rho_\varphi$ and its derivative $\rho'_\varphi$ as MPOs. The initial assumption we make to obtain our representation is that $\rho_0$ admits an efficient representation, with some finite bond dimension $D_{\rho_0}$, as an MPO:
\begin{equation}
	\rho_0 = \sum_{\mathbf{j},\mathbf{k}}\tr\left(R_0[1]^{j_1}_{k_1}R_0[2]^{j_2}_{k_2} \dots R_0[N]^{j_N}_{k_N}\right) |\mathbf{j}\rangle\langle\mathbf{k}|,
\end{equation}
where for brevity we have introduced $\mathbf{j} = \{j_1,\dots,j_N\}$. Exploiting channel/state duality to vectorize $\rho_0$ we obtain the MPS representation:
\begin{center}
	\begin{tikzpicture}[scale=0.75]
		\draw (-2,0) node {$|\rho_0\rangle =$};
		\draw (-0.3,0) arc (90:270:0.5);
		\draw (-0.3,-1) -- (7.8,-1);
		\draw[line width=0.5mm] (0,0) -- (0,1.5);	
		\foreach \x in {1.5, 3.0, ..., 7}
		{
			\draw (\x-1.5,0) -- (\x,0);
			\draw[line width=0.5mm] (\x,0) -- (\x,1.5);
		}
		\draw (6,0) -- (6.45,0);
		\draw (6.8,0) node {$\dots$};
		\draw (7.05,0) -- (7.5,0);
		\draw (7.8,0) arc (90:-90:0.5);
		\draw[line width=2mm, white] (7.5,0) -- (7.5,1.5);
		\draw[line width=0.5mm] (7.5,0) -- (7.5,1.5);
		\foreach \x/\c in {0/{1}, 1.5/{2}, 3/{3}, 4.5/{4}, 6/{5}}
		{
			\draw[fill=white] (\x,0) circle (0.4);
			\draw (\x,0) node[scale=0.65] {$R_0[\c]$};
		}
		\draw[fill=white] (7.5,0) circle (0.4);
		\draw (7.5,0) node[scale=0.65] {$R_0[N]$};
	\end{tikzpicture}.
\end{center}
Our first goal is to find the tensor network representation of $\rho_\varphi = \Lambda_\varphi(\rho_0)$. We achieve this in two steps. First we exploit the local structure of the noise channel, Eq.~(\ref{eq:localnoisechannel}), to build a tensor network for $\Lambda|\rho_0\rangle$---the result of this construction is not, in general, in MPS form. Then we exploit the singular value decomposition to put the resulting tensor network back into MPS form. Finally, we apply the local unitary parameter imprinting and return to the original MPO form.

For definiteness, we focus on the situation when the $\Lambda$ operator can be described by a subsequent action of singe-particle $\Lambda^{[n]}$ and two-particle terms $\Lambda^{[n,n+1]}$---which in the following are denoting by $Y$ and $X$, respectively---while retaining translation invariance. Physically this is the case when single and two-particle evolution terms commute and no particle is distinguished---generalizations to more complex situations are tedious but straightforward. In the tensor network representation the action of the $\Lambda$ operator therefore takes the following form:
\begin{center}
	\begin{tikzpicture}[scale=0.75]
		\draw (-2,1.55) node {$\Lambda|\rho_0\rangle =$};
		\draw (-0.3,0) arc (90:270:0.5);
		\draw (-0.3,-1) -- (7.8,-1);
		\draw[line width=0.5mm] (0,0) -- (0,0.8);	
		\foreach \x in {1.5, 3.0, ..., 7}
		{
			\draw (\x-1.5,0) -- (\x,0);
			\draw[line width=0.5mm] (\x,0) -- (\x,0.8);
		}
		\draw (6,0) -- (6.45,0);
		\draw (6.8,0) node {$\dots$};
		\draw (7.05,0) -- (7.5,0);
		\draw (7.8,0) arc (90:-90:0.5);
		\draw[line width=2mm, white] (7.5,0) -- (7.5,1);
		\draw[line width=0.5mm] (7.5,0) -- (7.5,0.8);
		\foreach \x/\c in {0/{1}, 1.5/{2}, 3/{3}, 4.5/{4}, 6/{5}}
		{
			\draw[fill=white] (\x,0) circle (0.4);
			\draw (\x,0) node[scale=0.65] {$R_0[\c]$};
		}
		\draw[fill=white] (7.5,0) circle (0.4);
		\draw (7.5,0) node[scale=0.65] {$R_0[N]$};
		\draw[rounded corners, fill=white, rotate around={45:(0.75,1.7)}] (-0.5,1.4) rectangle (2.0,2.0);
 		\draw[rounded corners, fill=white, rotate around={45:(3.75,1.7)}] (2.5,1.4) rectangle (5.,2.0);
		\draw[rounded corners, fill=white, rotate around={45:(6.75,1.7)}] (5.5,1.4) rectangle (8.,2.0);
		\draw (0.75, 1.7) node {$X$};
		\draw (3.75, 1.7) node {$X$};
		\fill[white,rotate around={45:(6.75,1.7)} ] (6.25,1.3) rectangle (7.25,2.1);
		\draw (6.8,1.7) node {$\dots$};
		\foreach \x in {0, 1.5, 3.0, ..., 8.5}
		{
			\draw[line width=0.5mm] (\x,1.4) -- (\x,2.0);
		}
		\draw[rounded corners, fill=white, rotate around={45:(2.25,1.7)}] (1.0,1.4) rectangle (3.5,2.0);
		\draw[rounded corners, fill=white, rotate around={45:(5.25,1.7)}] (4.0,1.4) rectangle (6.5, 2.0);
		\draw (2.25, 1.7) node {$X$};
		\draw (5.25, 1.7) node {$X$};
		\begin{scope}
    		\clip (-0.5,0) rectangle (0.5,3);
    		\draw[rounded corners, fill=white, rotate around={45:(-0.75,1.7)}] (-1.,1.4) rectangle (0.5,2.0);
		\end{scope}
		\begin{scope}
    		\clip (7,0) rectangle (8,3);
    		\draw[rounded corners, fill=white, rotate around={45:(8.25,1.7)}] (7,1.4) rectangle (8.5,2.0);
		\end{scope}
		\foreach \x in {0, 1.5, 3.0, ..., 8.5}
		{
			\draw[line width=0.5mm] (\x,2.7) -- (\x,3);
		}
		\foreach \x in {0, 1.5, 3.0, ..., 8.5}
		{
			\draw[line width=0.5mm] (\x,3.6) -- (\x,4.1);
		}
		\foreach \x in {0, 1.5, ..., 8.5}
		{
			\draw[fill=white] (\x,3.3) circle (0.4);
			\draw (\x,3.3) node {$Y$};
		}
	\end{tikzpicture},
\end{center}
where we place $X$ operators in a skewed orientation in order to maintain a manifestly translation invariant model.

This state is no longer in MPS form. The operator $X$ is defined with respect to a product basis $|\alpha\rangle = |j,k\rangle$ for the doubled legs and it acts on two neighbouring subsystems $|\alpha\rangle$ and $|\beta\rangle$, i.e., it is the tensor
\begin{equation}
	X = \sum_{\alpha',\alpha,\beta',\beta} X_{\alpha',\alpha,\beta',\beta} |\alpha'\rangle\langle\alpha|\otimes|\beta'\rangle\langle\beta|.
\end{equation}
By vectorizing each of the subsystems we can express it as a matrix:
\begin{equation}
	x = \sum_{(\alpha',\alpha),(\beta',\beta)} x_{(\alpha',\alpha),(\beta',\beta)} |(\alpha',\alpha)\rangle\langle(\beta',\beta)|.
\end{equation}
Applying the SVD to this matrix gives us the representation
\begin{equation}
	x_{(\alpha',\alpha),(\beta',\beta)} = \sum_{\gamma,\gamma'} U_{(\alpha',\alpha),\gamma} S_{\gamma,\gamma'} \left[V^\dag\right]_{\gamma', (\beta',\beta)},
\end{equation}
where $S_{\gamma,\gamma'} = s_\gamma \delta_{\gamma,\gamma'}$ with singular values $s_\gamma$ and $U$ and $V$ are unitary operators. Graphically this becomes:
\begin{center}
	\begin{tikzpicture}[scale=0.75]
		\draw[line width=0.5mm] (0.3,0) arc (0:-90:0.3);
		\draw[line width=0.5mm] (0.3,0.6) arc (0:90:0.3);
		\draw[line width=0.5mm] (1.8,0) arc (-180:-90:0.3);
		\draw[line width=0.5mm] (1.8,0.6) arc (180:90:0.3);
		\draw (4,0.3) -- (6,0.3);
		\draw[line width=0.5mm] (4,0) arc (0:-90:0.3);
		\draw[line width=0.5mm] (4,0.6) arc (0:90:0.3);
		\draw[line width=0.5mm] (6,0) arc (-180:-90:0.3);
		\draw[line width=0.5mm] (6,0.6) arc (180:90:0.3);
		\draw (2.9,0.3) node {$=$};
		\draw (-0.3,-0.3) node {$\alpha$};
		\draw (-0.3,0.9) node {$\alpha'$};
		\draw (2.4,-0.3) node {$\beta$};
		\draw (2.4,0.9) node {$\beta'$};
		\draw (3.4,-0.3) node {$\alpha$};
		\draw (3.4,0.9) node {$\alpha'$};
		\draw (6.7,-0.3) node {$\beta$};
		\draw (6.7,0.9) node {$\beta'$};
		\draw[fill=white] (4,0.3) circle (0.4);
		\draw (4,0.3) node {$U$};
		\draw[fill=white] (5,0.3) circle (0.4);
		\draw (5,0.3) node {$S$};
		\draw[fill=white] (6,0.3) circle (0.4);
		\draw (6,0.3) node {$V^\dag$};
		\draw[rounded corners, fill=white] (0,-0.1) rectangle (2.1,0.7);
		\draw (1.05, 0.3) node {$X$};
	\end{tikzpicture}.
\end{center}
(Here we regard the two vertical legs of $X$ on the left as a doubled leg acting on a single virtual system and the two vertical legs on the right as acting on a second virtual system: in this way one can think of $X$ as a simple matrix acting on a virtual system and we can then apply the SVD.)

We apply the SVD to each $X$ operator and absorb $\sqrt{S}$ into the $U$ tensor from the right (respectively, into the $V^\dag$ tensor from the left). Let $D^{(2)}$ (the upper index indicates two-particle nature of the noise) be the number of non-zero (or more practically non-negligible) singular values $s_\gamma$. Introducing $T=U \sqrt{S} $, $W= \sqrt{S} V^\dag$ results in the following tensor network,
\begin{center}
	\begin{tikzpicture}[scale=0.75]
		\draw (6,1.1) -- (6.5,1.47);
		\draw (7.0,1.83) -- (7.5,2.2);
		\draw (7.5,1.1) -- (8.5,1.6);
		\draw (0,2.2) -- (-1,1.6);
		\foreach \x in {0, 1.5, 3.0, ..., 8.5}
		{
			\draw[line width=0.5mm] (\x,0.3) -- (\x,0.8);
		}
		\fill[white] (6.45,0.6) rectangle (7.05,1.5);
		\draw (6.8,1.65) node {$\dots$};
		\foreach \x in {0, 1.5, 3.0, ..., 8.5}
		{
			\draw[line width=0.5mm] (\x,1.4) -- (\x,1.9);
			\draw[line width=0.5mm] (\x,1.4) -- (\x,1.9);
		}
		\draw (0,1.1) -- (1.5,2.2);
		\draw (3,1.1) -- (4.5,2.2);
	      \draw (1.5,1.1) -- (3.0,2.2);
		\draw (4.5,1.1) -- (6.,2.2);
		\foreach \x in {0, 1.5, 3.0, ..., 8.5}
		{
			\draw[line width=0.5mm] (\x,2.5) -- (\x,3);
			\draw[fill=white] (\x,1.1) circle (0.4);
		}
		\foreach \x in {0, 1.5, 3.0, ..., 8.5}
		{
			\draw[fill=white] (\x,2.2) circle (0.4);
		}
		\foreach \x in {0, 3, ..., 8.5}
		{
			\draw (\x, 1.1) node {$T$};
			\draw (\x, 2.2) node {$W$};
		}
		\foreach \x in {1.5, 4.5, ..., 8.5}
		{
			\draw (\x, 1.1) node {$T$};
			\draw (\x, 2.2) node {$W$};
		}
	\end{tikzpicture},
\end{center}
in place of the layer of $X$s.

The final step to obtain an MPO representation for $\rho_\varphi$ is to combine the tensors $R_0[n]$, $T$, $W$, $Y$ and a tensor $Z= e^{-\ii h\varphi}\otimes (e^{\ii h\varphi})^\mathrm{T}$---which represents the unitary phase encoding process---into a single new MPS tensor $R_\varphi[n]$:
\begin{center}
	\begin{tikzpicture}[scale=0.75]
		\draw (-1.7,2.25) -- (-4.3,2.25);
		\draw (-1.7,2.15) -- (-4.3,2.15);
		\draw (0.3,0) -- (2.7,0);
		\draw[line width=0.5mm] (1.5,0.3) -- (1.5,0.8);
		\draw[line width=0.5mm] (1.5,1.4) -- (1.5,1.9);
		\draw[line width=0.5mm] (1.5,1.4) -- (1.5,1.9);
		\draw (0.3,2.2) -- (1.5,2.2);
		\draw[line width=0.5mm] (-3,2.5) -- (-3,3.5);
		\draw[line width=0.5mm] (1.5,2.5) -- (1.5,3);
		\draw (1.8,1.1) -- (2.7,1.1);
		\draw[line width=0.5mm] (1.5,2.5) -- (1.5,3);
		\draw[line width=0.5mm] (1.5,3.6) -- (1.5,4.1);
		\draw[line width=0.5mm] (1.5,4.7) -- (1.5,5.2);

		\draw[fill=white] (-3,2.2) circle (0.4);
		\draw (-3,2.2) node[scale=0.65] {$R_\varphi[n]$};
		\draw (-0.7,2.2) node {$=$};
		\draw[fill=white] (1.5,0) circle (0.4);
		\draw (1.5, 0) node[scale=0.65] {$R_0[n]$};
		\draw[fill=white] (1.5,1.1) circle (0.4);
		\draw[fill=white] (1.5,2.2) circle (0.4);
		\draw (1.5, 2.2) node {$W$};
		\draw (1.5, 1.1) node {$T$};
		\draw[fill=white] (1.5,3.3) circle (0.4);
		\draw (1.5, 3.3) node {$Y$};
		\draw[fill=white] (1.5,4.4) circle (0.4);
		\draw (1.5, 4.4) node {$Z$};
	\end{tikzpicture}.
\end{center}
The doubled horizontal legs can be then combined into thicker horizontal legs to yield the MPS representation:
\begin{center}
	\begin{tikzpicture}[scale=0.75]
		\draw (-2,0) node {$|\rho_\varphi\rangle =$};
		\draw[line width=0.5mm] (-0.3,0) arc (90:270:0.5);
		\draw[line width=0.5mm] (-0.3,-1) -- (7.8,-1);
		\draw[line width=0.5mm] (0,0) -- (0,1.5);	
		\foreach \x in {1.5, 3.0, ..., 7}
		{
			\draw[line width=0.5mm] (\x-1.5,0) -- (\x,0);
			\draw[line width=0.5mm] (\x,0) -- (\x,1.5);
		}
		\draw[line width=0.5mm] (6,0) -- (6.45,0);
		\draw (6.8,0) node {$\dots$};
		\draw[line width=0.5mm] (7.05,0) -- (7.5,0);
		\draw[line width=0.5mm] (7.8,0) arc (90:-90:0.5);
		\draw[line width=2mm, white] (7.5,0) -- (7.5,1.5);
		\draw[line width=0.5mm] (7.5,0) -- (7.5,1.5);
		\foreach \x/\c in {0/{1}, 1.5/{2}, 3/{3}, 4.5/{4}, 6/{5}}
		{
			\draw[fill=white] (\x,0) circle (0.4);
			\draw (\x,0) node[scale=0.6] {$R_\varphi[\c]$};
		}
		\draw[fill=white] (7.5,0) circle (0.4);
		\draw (7.5,0) node[scale=0.6] {$R_\varphi[N]$};
	\end{tikzpicture}.
\end{center}
This representation can be put into MPO form by splitting the vertical legs back into the original two legs and bending:
\begin{center}
	\begin{tikzpicture}[scale=0.75]
		\draw (-2,0) node {$\rho_\varphi =$};
		\draw[line width = 0.5mm] (-0.3,0) arc (-90:-270:0.5);
		\draw[line width = 0.5mm] (-0.3,1) -- (7.8,1);
		\draw[line width=2mm, white] (0,-1.5) -- (0,1.5);
		\draw (0,-1.5) -- (0,1.5);	
		\foreach \x in {1.5, 3.0, ..., 7}
		{
			\draw[line width = 0.5mm] (\x-1.5,0) -- (\x,0);
			\draw[line width=2mm, white] (\x,-1.5) -- (\x,1.5);
			\draw (\x,-1.5) -- (\x,1.5);
		}
		\draw[line width = 0.5mm] (6,0) -- (6.45,0);
		\draw (6.8,0) node {$\dots$};
		\draw[line width = 0.5mm] (7.05,0) -- (7.5,0);
		\draw[line width = 0.5mm] (7.8,0) arc (-90:90:0.5);
		\draw[line width=2mm, white] (7.5,-1.5) -- (7.5,1.5);
		\draw (7.5,-1.5) -- (7.5,1.5);
		\foreach \x/\c in {0/{1}, 1.5/{2}, 3/{3}, 4.5/{4}, 6/{5}}
		{
			\draw[fill=white] (\x,0) circle (0.4);
			\draw (\x,0) node[scale=0.6] {$R_\varphi[\c]$};
		}
		\draw[fill=white] (7.5,0) circle (0.4);
		\draw (7.5,0) node[scale=0.6] {$R_\varphi[N]$};
	\end{tikzpicture}.
\end{center}
As a result we end up with an MPO with the bond dimension $D_{\rho} = D_{\rho_0} D^{(2)}$. The generalization of this derivation beyond the case of nearest-neighbour correlations will lead to an MPO representation of the density matrix $\rho_\varphi$ with bond dimension $D_\rho = D_{\rho_0} D_r$. Here $D_r = \prod_{k=2}^r D^{(k)} $ represents the contribution to the effective bond dimension of the output state resulting from the action of the correlated noise, where $D^{(k)}$ is the number of non-zero singular values that will appear when considering the $k$-particle noise term (the upper bound on $D^{(k)}$ is $d^{2(k-1)}$).

We finally move on to the task of writing the  $\rho'_\varphi$ operator as an MPO. It is possible to do this efficiently thanks to the fact that this operator is given as a commutator of $\rho_\varphi$ with $H$, where $H$ is a sum of local Hamiltonians
\begin{equation}
	\rho'_\varphi = \sum_{n=1}^N  \ii \left[\rho_\varphi,h^{[n]}\right].
\end{equation}
As a result we can regard $\rho'_\varphi$ as a sum of $N$ MPOs where each of them represents the original $\rho_\varphi$ MPO modified by an action of the Hamiltonian $h$ on consecutive particles. In what follows we assume that the basis $\ket{j}$ associated with the physical indices $j$ is chosen to be the eigenbasis of the local Hamiltonian $h$, $h= \sum_{j} \epsilon_j \ket{j}\bra{j}$, where $\epsilon_j$ are the corresponding eigenvalues. With this choice of local basis, the MPO representation of $\rho'_\varphi$ can be easily written at the  cost of doubling the bond dimension ($D_{\rho'} = 2 D_{\rho_0} D_r$):
\begin{equation} \label{eq:rhoprimempo}
	\rho'_\varphi = \sum_{\mathbf{j},\mathbf{k}}\tr\left(R'[1]^{j_1}_{k_1}R'[2]^{j_2}_{k_2} \dots R'[N]^{j_N}_{k_N}\right) |\mathbf{j}\rangle\langle\mathbf{k}|,
\end{equation}
where $R'[n]^{j_n}_{k_n}$ is equal to
\begin{equation} \label{eq:rhop mpo}
	\begin{cases}
		\begin{pmatrix} \ii (\epsilon_{k_1}-\epsilon_{j_1}) & 1\\ 0 & 0 \end{pmatrix}
		\otimes R_\varphi[1]^{j_1}_{k_1} & \text{for } n = 1,\\
		\begin{pmatrix} 1 & 0\\ \ii (\epsilon_{k_n}-\epsilon_{j_n}) & 1 \end{pmatrix}
		\otimes R_\varphi[n]^{j_n}_{k_n} & \text{for } n \in [2,\dots,N-1],\\
		\begin{pmatrix} 1 & 0\\ \ii (\epsilon_{k_N}-\epsilon_{j_N}) & 0 \end{pmatrix}
		\otimes R_\varphi[N]^{j_N}_{k_N} & \text{for } n = N.
	\end{cases}
\end{equation}
The $2\times 2$ matrices that appear in the above construction are responsible for the increase of the bond dimension but guarantee that the effect of trace in \eqref{eq:rhoprimempo} is equivalent to that resulting from the sum of $\rho_\varphi$ MPO acted upon consecutively by the commutator of the local Hamiltonians corresponding to different particles.

\subsection{Optimization of the QFI} \label{subsec:mpo finite}
As indicated in Sec.~\ref{subsec:qmetrology intro}, maximization of our FoM $F(\rho_0,L)$  leading to the maximal possible QFI for a given metrological model is a two-step iterative process. In the first part of this subsection we show how to maximize the FoM over a Hermitian operator $L$ with fixed $\rho_0$, and in the second part we focus on the maximization over the input state $\rho_0$ with fixed $L$.

We search for the optimal $L$ in the form of an MPO
\begin{equation}
	L = \sum_{\mathbf{j},\mathbf{k}}\tr\left(S[1]^{j_1}_{k_1}S[2]^{j_2}_{k_2} \dots S[N]^{j_N}_{k_N}\right) |\mathbf{j}\rangle\langle\mathbf{k}|,
\end{equation}
with a finite bond dimension $D_L$. Without loss of generality, each $S[n]^{j_n}_{k_n}$ is assumed Hermitian in its physical indices,
\begin{equation} \label{Hgauge}
    S[n]^{j_n}_{k_n} = \overline{S[n]^{k_n}_{j_n}},
\end{equation}
to ensure that $L$ is Hermitian. We call this condition a \emph{Hermitian gauge}.

The bond dimension $D_L$ for $L$ is expected to be small for weakly correlated noise models. Indeed, in the limit of an uncorrelated product state $\rho_\varphi = \varrho_\varphi^{\otimes N}$, $L$ is a sum of local operators, $L = \sum_{n=1}^N l^{[n]}$. Here $l^{[n]}$ is the SLD for the single-particle problem applied to particle $n$. In a similar way as for $\rho_\varphi^\prime$ presented above, the sum can be represented by an MPO with a bond dimension $2$. Therefore, $D_L=2$ is the limiting value for uncorrelated noise models.

The FoM to be maximized is
\begin{center}
	\begin{tikzpicture}[scale=0.5, every node/.style={scale=0.6}]
		\draw (-2,-0.5) node {\LARGE $2$};
		\draw (10,-0.5) node {\LARGE $-$};
		\draw (-1,0) to[out=180,in=180,looseness=2] (-1,0.8) -- (8,0.8) to[out=0,in=0,looseness=2] (8,0);
		\draw (-1,-2) to[out=180,in=180,looseness=2] (-1,-1.2) -- (8,-1.2) to[out=0,in=0,looseness=2] (8,-2);
		\foreach \x in {0, 2, 4, 7}
		{
			\draw[line width=1.5mm, white] (\x,-3) -- (\x,1) to[out=90, in=90,looseness=2] (\x+0.75,1) to[out=-90, in=90,looseness=2] (\x+0.75,-3) to[out=-90, in=-90,looseness=2] (\x+0,-3);
			\draw (\x,-3) -- (\x,1) to[out=90, in=90,looseness=2] (\x+0.75,1) to[out=-90, in=90,looseness=2] (\x+0.75,-3) to[out=-90, in=-90,looseness=2] (\x+0,-3);
		}
		\draw[line width=1.5mm, white] (-1,0) -- (5,0);
		\draw (-1,0) -- (5,0);
		\draw (5.5,0) node {$\dots$};
		\draw[line width=1.5mm, white] (6,0) -- (8,0);
		\draw (6,0) -- (8,0);

		\draw[line width=1.5mm, white] (-1,-2) -- (5,-2);
		\draw (-1,-2) -- (5,-2);
		\draw[line width=1.5mm, white] (6,-2) -- (8,-2);
		\draw (6,-2) -- (8,-2);

		\draw (5.5,-2) node {$\dots$};

		\foreach \x/\c in {0/{1}, 2/{2}, 4/{3}, 7/{N}}
		{
			\draw[fill=white] (\x,0) circle (0.6);
			\draw (\x,0) node {$R'[\c]$};
			\draw[fill=white] (\x,-2) circle (0.6);
			\draw (\x,-2) node {$S[\c]$};
		}

		\begin{scope}[shift={(5,-6)}]
		\draw (-1,0) to[out=180,in=180,looseness=2] (-1,0.8) -- (8,0.8) to[out=0,in=0,looseness=2] (8,0);
		\draw (-1,-2) to[out=180,in=180,looseness=2] (-1,-1.2) -- (8,-1.2) to[out=0,in=0,looseness=2] (8,-2);
		\draw (-1,-4) to[out=180,in=180,looseness=2] (-1,-3.2) -- (8,-3.2) to[out=0,in=0,looseness=2] (8,-4);
		\foreach \x in {0, 2, 4, 7}
		{
			\draw[line width=1.5mm, white] (\x,-5) -- (\x,1) to[out=90, in=90,looseness=2] (\x+0.75,1) to[out=-90, in=90,looseness=2] (\x+0.75,-5) to[out=-90, in=-90,looseness=2] (\x+0,-5);
			\draw (\x,-5) -- (\x,1) to[out=90, in=90,looseness=2] (\x+0.75,1) to[out=-90, in=90,looseness=2] (\x+0.75,-5) to[out=-90, in=-90,looseness=2] (\x+0,-5);
		}
		\draw[line width=1.5mm, white] (-1,0) -- (5,0);
		\draw (-1,0) -- (5,0);
		\draw (5.5,0) node {$\dots$};
		\draw[line width=1.5mm, white] (6,0) -- (8,0);
		\draw (6,0) -- (8,0);

		\draw[line width=1.5mm, white] (-1,-2) -- (5,-2);
		\draw (-1,-2) -- (5,-2);
		\draw[line width=1.5mm, white] (6,-2) -- (8,-2);
		\draw (6,-2) -- (8,-2);

		\draw[line width=1.5mm, white] (-1,-4) -- (5,-4);
		\draw (-1,-4) -- (5,-4);
		\draw[line width=1.5mm, white] (6,-4) -- (8,-4);
		\draw (6,-4) -- (8,-4);

		\draw (5.5,-2) node {$\dots$};
		\draw (5.5,-4) node {$\dots$};
		\foreach \x/\c in {0/{1}, 2/{2}, 4/{3}, 7/{N}}
		{
			\draw[fill=white] (\x,0) circle (0.6);
			\draw (\x,0) node {$S[\c]$};
			\draw[fill=white] (\x,-2) circle (0.6);
			\draw (\x,-2) node {$R_{\varphi}[\c]$};
			\draw[fill=white] (\x,-4) circle (0.6);
			\draw (\x,-4) node {$S[\c]$};
		}
		\end{scope}
	\end{tikzpicture}.
\end{center}
Maximization of the FoM over $L$ is equivalent to a joint maximization over each tensor $S[n]$. We relax this optimization problem by iterating an optimization loop. In the loop we first find the optimal $S[1]$, then $S[2]$, and so on up to $S[N]$, after which we go back to $S[1]$. The optimization over each $S[n]$ is performed with all other tensors fixed. The loop is repeated until the FoM converges.

After fixing the other tensors, the FoM becomes quadratic in $S[n]$ and the optimal $S[n]$ is found as a solution to a linear equation. For definiteness, to explain the procedure, we focus on the generic example of $S[2]$. After vectorizing $S[2] \rightarrow |S[2]\rangle$,
\begin{center}
	\begin{tikzpicture}[scale=0.75, every node/.style={scale=1}]
		\draw (0,-1) -- (0,1);
		\draw (-1,0) -- (1,0);
		\draw[fill=white] (0,0) circle (0.4);
		\draw (0,0) node[scale=0.85] {$S[2]$};
		\draw (-1.25,0) node {$\gamma$};
		\draw (1.25,0) node {$\delta$};
		\draw (0,-1.25) node {$k_2$};
		\draw (0,1.25) node {$j_2$};
		\draw[->] (1.75,0) -- (2.5,0);
		\draw (4,-1.25) -- (4,1) to[in = 90, out=90] (5,1) -- (5,-0.5) to[in = 90, out=270] (4.1,-1) -- (4.1,-1.25);
		\draw (4,0) -- (3.25,0) to[in = 180, out=180] (3.25,-0.6) to[in = 90, out=0] (3.95,-1) -- (3.95,-1.25);
		\draw (4,0) -- (4.75,0) to[in = 0, out= 0] (4.75,-0.6) to[in = 90, out=180] (4.05,-1) -- (4.05,-1.25);
		\draw[fill=white] (4,0) circle (0.4);
		\draw (4,0) node[scale=0.85] {$S[2]$};
		\draw[line width=2pt] (8,0) -- (8,-1);
		\draw[fill=white] (8,0) circle (0.4);
		\draw[->] (5.75,0) -- (6.5,0);
		\draw (8,0) node[scale=0.85] {$S[2]$};
		\draw (8,-1.25) node {$\alpha$};
	\end{tikzpicture},
\end{center}
we can represent the $S[2]$-FoM as
\begin{center}
	\begin{tikzpicture}[scale=0.5, every node/.style={scale=0.6}]
		\draw[->] (-1,-3.5) -- (-1,-5);
		\draw (-1,-5) node[below] {\LARGE $|b\rangle$};
		\draw (-2.5,-0.5) node {\LARGE $2$};
		\draw (10,-0.5) node {\LARGE $-$};
		\draw (-1,0) to[out=180,in=180,looseness=2] (-1,0.8) -- (8,0.8) to[out=0,in=0,looseness=2] (8,0);
		\draw (-1,-2) to[out=180,in=180,looseness=2] (-1,-1.2) -- (8,-1.2) to[out=0,in=0,looseness=2] (8,-2);

		\draw[line width=1.5mm, white] (2,-5) -- (2,1) to[out=90, in=90,looseness=2] (2+0.75,1) -- (2+0.75,0) to[out=-90, in=90,looseness=2] (2.05,-3) -- (2+0.05,-5) to[out=-90, in=-90,looseness=2] (2+0,-5);
		\draw (2,-5) -- (2,1) to[out=90, in=90,looseness=2] (2+0.75,1) -- (2+0.75,0) to[out=-90, in=90,looseness=2] (2.05,-3) -- (2+0.05,-5) to[out=-90, in=-90,looseness=2] (2+0,-5);

		\foreach \x in {0, 4, 7}
		{
			\draw[line width=1.5mm, white] (\x,-3) -- (\x,1) to[out=90, in=90,looseness=2] (\x+0.75,1) to[out=-90, in=90,looseness=2] (\x+0.75,-3) to[out=-90, in=-90,looseness=2] (\x+0,-3);
			\draw (\x,-3) -- (\x,1) to[out=90, in=90,looseness=2] (\x+0.75,1) to[out=-90, in=90,looseness=2] (\x+0.75,-3) to[out=-90, in=-90,looseness=2] (\x+0,-3);
		}
		\draw[line width=1.5mm, white] (-1,0) -- (5,0);
		\draw (-1,0) -- (5,0);
		\draw (5.5,0) node {$\dots$};
		\draw[line width=1.5mm, white] (6,0) -- (8,0);
		\draw (6,0) -- (8,0);

		\draw[line width=1.5mm, white] (-1,-2) -- (1,-2);
		\draw[line width=1.5mm, white] (3,-2) -- (5,-2);
		\draw (-1,-2) -- (1,-2) to[out=0,in=90] (1.95,-4) -- (1.95,-5) -- (2.1,-5) -- (2.1,-4) to[out=90,in=180] (3,-2) -- (5,-2);

		\draw[line width=1.5mm, white] (6,-2) -- (8,-2);
		\draw (6,-2) -- (8,-2);

		\draw (5.5,-2) node {$\dots$};
		
		\foreach \x/\c in {0/{1}, 2/{2}, 4/{3}, 7/{N}}
		{
			\draw[fill=white] (\x,0) circle (0.6);
			\draw (\x,0) node {$R'[\c]$};
		}
		\foreach \x/\c in {0/{1}, 4/{3}, 7/{N}}
		{
			\draw[fill=white] (\x,-2) circle (0.6);
			\draw (\x,-2) node {$S[\c]$};
		}

		\draw[fill=white] (2,-5) circle (0.6);
		\draw (2,-5) node {$S[2]$};

		\draw[rounded corners, fill=none,dashed] (-1.75,-4) rectangle (8.75,2.0);

		\draw (2,-3.5) node[right] {\LARGE $\alpha$};

		\begin{scope}[shift={(5,-9)}]
		\draw[->] (8,1.5) -- (8,3);
		\draw (8,3) node[above] {\LARGE $A$};
		\draw (-1,0) to[out=180,in=180,looseness=2] (-1,0.8) -- (8,0.8) to[out=0,in=0,looseness=2] (8,0);
		\draw (-1,-2) to[out=180,in=180,looseness=2] (-1,-1.2) -- (8,-1.2) to[out=0,in=0,looseness=2] (8,-2);
		\draw (-1,-4) to[out=180,in=180,looseness=2] (-1,-3.2) -- (8,-3.2) to[out=0,in=0,looseness=2] (8,-4);
		
		\draw[line width=1.5mm, white] (2,-7) -- (2,3) to[out=90, in=90,looseness=2] (2+0.05,3) -- (2.05,1) to[out=-90, in=90,looseness=2] (2+0.75,-1.5) -- (2+0.75,-2.5) to[out=-90,in=90,looseness=2] (2.05,-5) -- (2.05,-6) to[out=-90, in=-90,looseness=2] (2+0.05,-7);

		\draw (2,-7) -- (2,3) to[out=90, in=90,looseness=2] (2+0.05,3) -- (2.05,1) to[out=-90, in=90,looseness=2] (2+0.75,-1.5) -- (2+0.75,-2.5) to[out=-90,in=90,looseness=2] (2.05,-5) -- (2.05,-6) to[out=-90, in=-90,looseness=2] (2+0.05,-7);

		\foreach \x in {0, 4, 7}
		{
			\draw[line width=1.5mm, white] (\x,-5) -- (\x,1) to[out=90, in=90,looseness=2] (\x+0.75,1) to[out=-90, in=90,looseness=2] (\x+0.75,-5) to[out=-90, in=-90,looseness=2] (\x+0,-5);
			\draw (\x,-5) -- (\x,1) to[out=90, in=90,looseness=2] (\x+0.75,1) to[out=-90, in=90,looseness=2] (\x+0.75,-5) to[out=-90, in=-90,looseness=2] (\x+0,-5);
		}
		\draw[line width=1.5mm, white] (-1,0) -- (1,0);
		\draw[line width=1.5mm, white] (3,0) -- (5,0);
		\draw (-1,0) -- (1,0) to[out=0,in=-90] (1.95,2) -- (1.95,3) -- (2.1,3) -- (2.1,2) to[out=-90,in=180] (3,0) -- (5,0);

		\draw (5.5,0) node {$\dots$};
		\draw[line width=1.5mm, white] (6,0) -- (8,0);
		\draw (6,0) -- (8,0);

		\draw[line width=1.5mm, white] (-1,-2) -- (5,-2);
		\draw (-1,-2) -- (5,-2);

		\draw[line width=1.5mm, white] (6,-2) -- (8,-2);
		\draw (6,-2) -- (8,-2);

		\draw[line width=1.5mm, white] (-1,-4) -- (1,-4);
		\draw[line width=1.5mm, white] (3,-4) -- (5,-4);

		\draw (-1,-4) -- (1,-4) to[out=0,in=90] (1.95,-5) -- (1.95,-7) -- (2.1,-7) -- (2.1,-5) to[out=90,in=180] (3,-4) -- (5,-4);

		\draw[line width=1.5mm, white] (6,-4) -- (8,-4);
		\draw (6,-4) -- (8,-4);

		\draw (5.5,-2) node {$\dots$};
		\draw (5.5,-4) node {$\dots$};
		\foreach \x/\c in {0/{1}, 4/{3}, 7/{N}}
		{
			\draw[fill=white] (\x,0) circle (0.6);
			\draw (\x,0) node {$S[\c]$};
			\draw[fill=white] (\x,-2) circle (0.6);
			\draw (\x,-2) node {$R_{\varphi}[\c]$};
			\draw[fill=white] (\x,-4) circle (0.6);
			\draw (\x,-4) node {$S[\c]$};
		}
		\draw[fill=white] (2,3) circle (0.6);
		\draw (2,3) node {$S[2]$};
		\draw[fill=white] (2,-2) circle (0.6);
		\draw (2,-2) node {$R_{\varphi}[2]$};
		\draw[fill=white] (2,-7) circle (0.6);
		\draw (2,-7) node {$S[2]$};
		\draw[rounded corners, fill=none,dashed] (-1.75,-6) rectangle (8.75,2.0);
		\draw (2,-5.5) node[right] {\LARGE $\beta$};
		\draw (2,1.5) node[right] {\LARGE $\alpha$};
		\end{scope}
	\end{tikzpicture},
\end{center}
which can be written in a compact way as
\begin{equation}
	F(\rho_0,L) = 2\sum_\alpha b_\alpha S[2]_\alpha - \sum_{\alpha\beta} S[2]_\alpha A_{\alpha\beta} S[2]_\beta.
\end{equation}
Here $b_\alpha$ are the elements of the vector $|b\rangle$, and $A_{\alpha,\beta}$ are the elements of the matrix $A$. Both $|b\rangle$ and $A$ describe the entire tensor network complementing the distinguished vector $\ket{S[2]}$ in the two respective terms of the $S[2]$-FoM. After taking a derivative with respect to $S[2]_\alpha$, we obtain a linear equation for the extremum:
\begin{equation} \label{eq:condLfin}
	\frac{1}{2}\left(A+A^\mathrm{T}\right) |S[2]\rangle = |b\rangle.
\end{equation}
The $d^2D_L^2 \times d^2D_L^2$ matrix $\widetilde{A} = \frac{1}{2}\left(A+A^\mathrm{T}\right)$ typically has a non-zero kernel and the linear equation does not have a unique solution. We use the Moore-Penrose pseudo-inverse, $\widetilde{A}^+$, to obtain a solution $|S[2]\rangle=\widetilde{A}^+|b\rangle$ that does not contain any zero modes of $\widetilde{A}$.

If the linear equation was non-singular, then its exact solution would satisfy the Hermitian gauge (\ref{Hgauge}). For the typical singular case, using an SVD of $\widetilde{A}$ to construct its pseudo-inverse, we have to truncate singular values falling below a small but finite cut-off, set by $\kappa$ multiplied by the highest singular value. As the cut-off solution $|S[2]\rangle$ need not satisfy the Hermitian gauge condition exactly, we filter out its small anti-Hermitian part with the substitution:
\begin{equation}
	S[2]^{j_2}_{k_2} \rightarrow \frac{1}{2} \left(S[2]^{j_2}_{k_2}+\overline{S[2]^{k_2}_{j_2}}\right).
\end{equation}
From experience, this substitution can improve numerical stability but is not necessary when all initial $S[n]$ are in the Hermitian gauge (\ref{Hgauge}) and $\kappa$ is large enough to suppress the anti-Hermitian part of the solution. However, with too large a cut-off the final optimized $L$ does not reach the maximal possible value of the QFI. Therefore, we adjust $\kappa$ to obtain the highest QFI achievable without compromising the stability.

Now we move on to the maximization of the FoM over the input state $\rho_0$ for a fixed $L$. We start by rewriting $F(\rho_0,L)$ as
\begin{align}
	F(\rho_0,L) &= 2\tr\left(\rho'_\varphi L \right) - \tr\left(\rho_\varphi L^2\right) = \nonumber\\
	&= 2\tr\left(\ii \left[\Lambda_\varphi(\rho_0),H\right] L \right) - \tr\left(\Lambda_\varphi(\rho_0)  L^2\right)  \nonumber\\
	&= 2\tr\left(\rho_0 \ii \left[H,\Lambda^*_\varphi(L)\right]\right) - \tr\left(\rho_0\Lambda^*_\varphi(L^2)\right),
\end{align}
where by $\Lambda^*_\varphi(\cdot)$ we denote the channel which is dual to $\Lambda_\varphi(\cdot)$ (the evolution written in the Heisenberg picture). We can rewrite this as
\begin{align} \label{Fnew}
	F(\rho_0,L) &= \tr\left[\rho_0 (2 L'^{*}_{\varphi}- L^*_{2,\varphi})\right],
\end{align}
where we introduce $L^*_{\varphi} = \Lambda^*_{\varphi}(L)$, $L'^{*}_{\varphi} = \frac{\dd L^*_{\varphi}}{\dd \varphi} = \ii \left[H, L^*_{\varphi}\right]$ and $L^*_{2,\varphi} = \Lambda^*_{\varphi}(L^2)$. By analogy with the construction of the MPO representation for $\rho_\varphi = \Lambda_\varphi(\rho_0)$ and $\rho'_\varphi = \ii \left[\rho_\varphi,H\right]$ in Sec.~\ref{subsec:mpo rho}, we can easily construct the MPO representation of $L^*_{2,\varphi}$ and $L'^{*}_{\varphi}$ from the known MPO form of $L$. The tensors determining the MPO form of $L^*_{2,\varphi}$ and $L'^{*}_{\varphi}$ are denoted by $S_2[n]$ and $S'[n]$, respectively, and their respective bond dimensions are $D_{L_2} = D_L^2 D_r$ and $D_{L'} = 2 D_L D_r$.

The quantity $F(\rho_0,L)$ in (\ref{Fnew}) is maximal when $\rho_0$ is a projection on the eigenvector associated with the maximal eigenvalue of the Hermitian operator $2L'^{*}_{\varphi}-L^{*}_{2,\varphi}$. Hence, without loss of generality, we can assume a pure input state $\rho_0 = |\psi \rangle\!\langle \psi|$ with $|\psi \rangle$ being an MPS with bond dimension $D_\psi$:
\begin{equation}
	|\psi \rangle = \sum_{\mathbf{j}}\tr\left(P[1]^{j_1}P[2]^{j_2} \dots P[N]^{j_N}\right) |\mathbf{j}\rangle.
\end{equation}
The input state $\rho_0$ has bond dimension $D_{\rho_0} = D_\psi^2$ and its MPO tensors are $R_0[n]^{j_n}_{k_n} = P[n]^{j_n} \otimes \overline{P[n]^{k_n}}$.

The maximization of $F(\rho_0,L)$ over the input state $|\psi\rangle$ is equivalent to the variational optimization for the ground state of a many-body ``Hamiltonian'' $L^*_{2,\varphi}-2L'^*_{\varphi}$, a problem widely discussed in the many-body physics MPS literature \cite{White1992,Schollwock2011,bridgemanHandwavingInterpretiveDance2017}. After reinterpreting our problem as a variational minimization of the ``energy''
\begin{equation} \label{energy}
	-F(\rho_0,L) = \frac{\langle \psi|L^*_{2,\varphi}-2L'^*_{\varphi}|\psi \rangle}{\langle \psi|\psi \rangle},
\end{equation}
we proceed iteratively in a similar way as in the case of the maximization of $F(\rho_0,L)$ over $L$.

For example, in order to find the minimum over $P[2]$, we begin by vectorizing the tensor $P[2] \rightarrow |P[2]\rangle$ and expressing the ``energy'' (\ref{energy}) as a Rayleigh quotient
\begin{equation} \label{FN}
	-F(\rho_0,L) = \frac{\langle P[2]|\mathcal{F}|P[2] \rangle}{\langle P[2]|\mathcal{N}|P[2] \rangle}.
\end{equation}
Here the $d D_\psi^2 \times d D_\psi^2$ matrices are
\begin{center}
	\begin{tikzpicture}[scale=0.5, every node/.style={scale=0.6}]
		\draw (-1,0) to[out=180,in=180,looseness=2] (-1,0.8) -- (8,0.8) to[out=0,in=0,looseness=2] (8,0);
		\draw (-1,-2) to[out=180,in=180,looseness=2] (-1,-1.2) -- (8,-1.2) to[out=0,in=0,looseness=2] (8,-2);
		\draw (-1,-4) to[out=180,in=180,looseness=2] (-1,-3.2) -- (8,-3.2) to[out=0,in=0,looseness=2] (8,-4);
		

		\draw[line width=1.5mm, white] (2,-6) -- (2,2);
		\draw (2,-6) -- (2,2);
		

		\foreach \x in {0, 4, 7}
		{
			\draw[line width=1.5mm, white] (\x,-4) -- (\x,0);
			\draw (\x,-4) -- (\x,0);
		}
		\draw[line width=1.5mm, white] (-1,0) -- (1,0);
		\draw[line width=1.5mm, white] (3,0) -- (5,0);
		\draw (-1,0) -- (1,0) to[out=0,in=-90] (1.95,2) -- (1.95,2);
		\draw (2.05,2) -- (2.05,2) to[out=-90,in=180] (3,0) -- (5,0);
		\draw (2,2) node[above] {\LARGE $\alpha$};

		\draw (5.5,0) node {$\dots$};
		\draw[line width=1.5mm, white] (6,0) -- (8,0);
		\draw (6,0) -- (8,0);

		\draw[line width=1.5mm, white] (-1,-2) -- (5,-2);
		\draw (-1,-2) -- (5,-2);

		\draw[line width=1.5mm, white] (6,-2) -- (8,-2);
		\draw (6,-2) -- (8,-2);

		\draw[line width=1.5mm, white] (-1,-4) -- (1,-4);
		\draw[line width=1.5mm, white] (3,-4) -- (5,-4);

		\draw (-1,-4) -- (1,-4) to[out=0,in=90] (1.95,-5) -- (1.95,-6);
		\draw (2.05,-6) -- (2.05,-5) to[out=90,in=180] (3,-4) -- (5,-4);

		\draw[line width=1.5mm, white] (6,-4) -- (8,-4);
		\draw (6,-4) -- (8,-4);
		\draw (2,-6) node[below] {\LARGE $\beta$};

		\draw (5.5,-2) node {$\dots$};
		\draw (5.5,-4) node {$\dots$};
		\foreach \x/\c in {0/{1}, 4/{3}, 7/{N}}
		{
			\draw[fill=white] (\x,0) circle (0.6);
			\draw (\x,0) node {$P[\c]$};
			\draw[fill=white] (\x,-2) circle (0.6);
			\draw (\x,-2) node {$S_{2}[\c]$};
			\draw[fill=white] (\x,-4) circle (0.6);
			\draw (\x,-4) node {$\overline{P[\c]}$};
		}
		\draw[fill=white] (2,-2) circle (0.6);
		\draw (2,-2) node {$S_2[2]$};
		\draw (-2,-2) node[left] {\LARGE $\mathcal{F}_{\alpha\beta}=$};
		\draw (9,-2) node[right] {\LARGE $-$};
		\begin{scope}[shift={(5,-9)}]
			\draw (-1,0) to[out=180,in=180,looseness=2] (-1,0.8) -- (8,0.8) to[out=0,in=0,looseness=2] (8,0);
			\draw (-1,-2) to[out=180,in=180,looseness=2] (-1,-1.2) -- (8,-1.2) to[out=0,in=0,looseness=2] (8,-2);
			\draw (-1,-4) to[out=180,in=180,looseness=2] (-1,-3.2) -- (8,-3.2) to[out=0,in=0,looseness=2] (8,-4);
			

			\draw[line width=1.5mm, white] (2,-6) -- (2,2);
			\draw (2,-6) -- (2,2);
			

			\foreach \x in {0, 4, 7}
			{
				\draw[line width=1.5mm, white] (\x,-4) -- (\x,0);
				\draw (\x,-4) -- (\x,0);
			}
			\draw[line width=1.5mm, white] (-1,0) -- (1,0);
			\draw[line width=1.5mm, white] (3,0) -- (5,0);
			\draw (-1,0) -- (1,0) to[out=0,in=-90] (1.95,2) -- (1.95,2);
			\draw (2.05,2) -- (2.05,2) to[out=-90,in=180] (3,0) -- (5,0);
			\draw (2,2) node[above] {\LARGE $\alpha$};

			\draw (5.5,0) node {$\dots$};
			\draw[line width=1.5mm, white] (6,0) -- (8,0);
			\draw (6,0) -- (8,0);

			\draw[line width=1.5mm, white] (-1,-2) -- (5,-2);
			\draw (-1,-2) -- (5,-2);

			\draw[line width=1.5mm, white] (6,-2) -- (8,-2);
			\draw (6,-2) -- (8,-2);

			\draw[line width=1.5mm, white] (-1,-4) -- (1,-4);
			\draw[line width=1.5mm, white] (3,-4) -- (5,-4);

			\draw (-1,-4) -- (1,-4) to[out=0,in=90] (1.95,-5) -- (1.95,-6);
			\draw (2.05,-6) -- (2.05,-5) to[out=90,in=180] (3,-4) -- (5,-4);

			\draw[line width=1.5mm, white] (6,-4) -- (8,-4);
			\draw (6,-4) -- (8,-4);
			\draw (2,-6) node[below] {\LARGE $\beta$};

			\draw (5.5,-2) node {$\dots$};
			\draw (5.5,-4) node {$\dots$};
			\foreach \x/\c in {0/{1}, 4/{3}, 7/{N}}
			{
				\draw[fill=white] (\x,0) circle (0.6);
				\draw (\x,0) node {$P[\c]$};
				\draw[fill=white] (\x,-2) circle (0.6);
				\draw (\x,-2) node {$S'[\c]$};
				\draw[fill=white] (\x,-4) circle (0.6);
				\draw (\x,-4) node {$\overline{P[\c]}$};
			}
			\draw[fill=white] (2,-2) circle (0.6);
			\draw (2,-2) node {$S'[2]$};
			\draw (-2,-2) node {\LARGE $2$};
		\end{scope}
	\end{tikzpicture}
\end{center}
and
\begin{center}
	\begin{tikzpicture}[scale=0.5, every node/.style={scale=0.6}]
		\draw (-1,0) to[out=180,in=180,looseness=2] (-1,0.8) -- (8,0.8) to[out=0,in=0,looseness=2] (8,0);
		\draw (-1,-2) to[out=180,in=180,looseness=2] (-1,-1.2) -- (8,-1.2) to[out=0,in=0,looseness=2] (8,-2);

		\draw[line width=1.5mm, white] (2,-4) -- (2,2);
		\draw (2,-4) -- (2,2);

		\foreach \x in {0, 4, 7}
		{
			\draw[line width=1.5mm, white] (\x,-2) -- (\x,0);
			\draw (\x,-2) -- (\x,0);
		}
		\draw[line width=1.5mm, white] (-1,0) -- (1,0);
		\draw[line width=1.5mm, white] (3,0) -- (5,0);
		\draw (-1,0) -- (1,0) to[out=0,in=-90] (1.95,2) -- (1.95,2);
		\draw (2.05,2) -- (2.05,2) to[out=-90,in=180] (3,0) -- (5,0);
		\draw (2,2) node[above] {\LARGE $\alpha$};

		\draw (5.5,0) node {$\dots$};
		\draw[line width=1.5mm, white] (6,0) -- (8,0);
		\draw (6,0) -- (8,0);

		\draw[line width=1.5mm, white] (-1,-4) -- (1,-4);
		\draw[line width=1.5mm, white] (3,-4) -- (5,-4);

		\draw (-1,-2) -- (1,-2) to[out=0,in=90] (1.95,-3) -- (1.95,-4);
		\draw (2.05,-4) -- (2.05,-3) to[out=90,in=180] (3,-2) -- (5,-2);

		\draw[line width=1.5mm, white] (6,-2) -- (8,-2);
		\draw (6,-2) -- (8,-2);
		\draw (2,-4) node[below] {\LARGE $\beta$};

		\draw (5.5,-2) node {$\dots$};
		\foreach \x/\c in {0/{1}, 4/{3}, 7/{N}}
		{
			\draw[fill=white] (\x,0) circle (0.6);
			\draw (\x,0) node {$P[\c]$};
			\draw[fill=white] (\x,-2) circle (0.6);
			\draw (\x,-2) node {$\overline{P[\c]}$};
		}
		\draw (-2,-1) node[left] {\LARGE $\mathcal{N}_{\alpha\beta}=$};
	\end{tikzpicture}.
\end{center}
After taking the derivative of (\ref{FN}) we obtain the condition for the extremum:
\begin{equation} \label{eq:condP}
	\mathcal{F} |P[2]\rangle = -F(\rho_0,L)~ \mathcal{N}|P[2]\rangle,
\end{equation}
which is a generalized eigenvalue problem with eigenvalue $-F(\rho_0,L)$. By multiplying it with a pseudo-inverse of matrix $\mathcal{N}$ we bring it into the form of ordinary eigenvalue problem, for which we obtain the lowest eigenvalue and its corresponding eigenvector using the Lanczos algorithm.

Modification of the entire tensor-network framework to calculate the maximal QFI for systems with open boundary conditions (OBC) poses no problem and, for systems which are not sensitive to boundary conditions, is even advisable. In OBC the MPS representing $|\psi \rangle$ can be brought to a canonical form, where the matrix $\mathcal{N}$ becomes an identity and Eq.~\eqref{eq:condP} reduces to a standard eigenvalue problem. There is no need to pseudo-invert $\mathcal{N}$.

To summarize the procedure: one needs to iteratively determine the optimal $L$ for a given $\rho_0$ and then the optimal $\rho_0$ for a given $L$, until one observes convergence of the final result, e.g.\ the FoM does not change more than, say, $0.1\%$ after a fixed number of steps. From our numerical experience this happens very rapidly, typically after $5$ iterations of the $\rho_0$ and $L$ optimization steps.

While running the algorithm, one has to choose the bond-dimensions for the input state, $D_{\psi}$, as well as for the SLD, $D_L$, over which the optimization is performed. As in all tensor-network algorithms, keeping the bond-dimension as low as possible is essential for their efficiency. In our calculations, we typically started with a product input state, $D_\psi=1$, and optimized for $L$ with a minimal non-trivial $D_L=2$. Then we increased one of the bond dimensions, either $D_\psi$ or $D_L$, each time repeating the optimization procedure, until we found that the QFI did not change when increasing the $D$'s more than by, e.g., $1\%$ and hence assumed that relative error of our method is around $1\%$.

\subsection{Asymptotic limit} \label{subsec:mpo inf}
The previous subsection describes an algorithm that functions for a system with a finite number of particles $N$. For quantum metrological problems in the presence of decoherence, it is the generic situation that the optimal quantum enhancement thanks to the use of entanglement leads asymptotically (for large $N$) to an improvement by a constant factor over product-state strategies. Even though the finite-system MPO approach allows us to achieve values of $N$ that are inaccessible via exact full-Hilbert space computation, it may sometimes be not enough to reach the asymptotic limit and determine the quantum enhancement coefficient with the desired precision. For this reason, we would like to have a procedure that allows us to go \emph{directly} to the infinite-particle limit, calculate the maximal achievable QFI per particle and, as a result, determine the maximal quantum enhancement coefficient.

For this purpose, we exploit the infinite MPO/MPS (iMPO/iMPS) approach, see e.g.\ \cite{TopoCincioVidal}. We assume that all tensors are translationally invariant (TI). Then we take the limit of infinite $N$. Technically this limit is most natural in the case of PBCs, where it is enough to notice that, for any TI transfer matrix $E$, the spectral decomposition of $E^N$ is dominated by the leading eigenvalue and eigenvector of $E$. This is why in the following discussion we proceed with PBCs. In the OBC case, which is arguably more natural in some metrological contexts, one has in principle to consider the boundary conditions at infinity. However, $E^k$ applied to a boundary vector gives the leading eigenvector of $E$ when $k$ becomes longer than the finite correlation range (just as in the Lanczos algorithm). Therefore, in the bulk (i.e., far from the boundaries), all equations the TI tensors have to satisfy become the same as for the PBC.

When the input state $|\psi \rangle$ is TI then the final state $\rho_\varphi$ is TI as well. There is a problem, however, with the operator $\rho'_\varphi$. Its construction as an MPO in Eq.~\eqref{eq:rhop mpo} is \emph{not} TI. Because of this, instead of calculating the derivative exactly, we approximate it by a difference of two TI iMPOs:
\begin{equation} \label{eq:rho_phi_epsilon}
	\rho'_\varphi = \frac{\rho_{\varphi+\varepsilon}-\rho_\varphi}{\varepsilon}
\end{equation}
with infinitesimal parameter $\varepsilon$. Motivated by the defining equation for the SLD (\ref{eq:sld}), we can consider a similar expansion of the operator $L$:
\begin{equation} \label{eq:tildeL}
	L = \frac{\widetilde{L}-\mathds{1}}{\varepsilon}.
\end{equation}
Here $\widetilde{L}$ and $\openone$ are solutions of Eq. (\ref{eq:sld}) when $\rho^\prime_\varphi$ is replaced by, respectively, $\rho_{\varphi+\varepsilon}$ and $\rho_{\varphi}$. We search for the optimal $\widetilde{L}$ which is better suited for the TI formalism than the original operator $L$. Let us denote by
\begin{equation}
	f(\rho_0,\widetilde{L}) = \frac{1}{N} F(\rho_0,L)
\end{equation}
the QFI per particle that we want to maximize:
\begin{equation}
	f(\rho_0,\widetilde{L}) = \frac{1}{N \varepsilon^2} \left[2\tr\left(\rho_{\varphi+\varepsilon} \widetilde{L} \right) - \tr\left(\rho_\varphi \widetilde{L}^2\right)-1\right].
\end{equation}
A TI iMPO is defined by only one tensor which we assign respectively as: $|\psi\rangle \rightarrow P$, $\rho_0 \rightarrow R_0$, $\rho_\varphi \rightarrow R_\varphi$, $\widetilde{L} \rightarrow \widetilde{S}$, $\widetilde{L}^*_\varphi \rightarrow \widetilde{S}_\varphi$, $\widetilde{L}^*_{2,\varphi} \rightarrow \widetilde{S}_2$.

Optimization of a tensor-network consisting of identical tensors $A$ is a highly nonlinear problem in $A$ and one might think that an approach similar to the one used in the previous subsection is not applicable here. Fortunately, an efficient method for the problem was developed in \cite{Corboz2016}. The main idea is to find the optimal tensor $A_\mathrm{new}$ at one site, treating all other tensors $A$ as fixed, and then rather then replacing all tensors by $A_\mathrm{new}$ perform a flexible substitution,
\begin{equation}
	A \to A_\mathrm{new} \sin\left(\lambda \pi\right) - A \cos\left(\lambda \pi\right),
\end{equation}
with a mixing angle $\lambda$. The angle is optimized to yield the best possible FoM.

We now apply this approach to our two-step iterative procedure. First we need to find the optimal $\widetilde{L}$ which is equivalent to the determination of the optimal local tensor $\widetilde{S}$.

As explained in Sec.~\ref{subsec:mpo review}, the trace of an operator represented as an iMPO is defined by its transfer matrix, so we start by introducing transfer matrices $E_1$ and $E_2$ associated with, respectively, $\rho_{\varphi+\varepsilon} \widetilde{L}$ and $\rho_\varphi \widetilde{L}^2$:
\begin{center}
	\begin{tikzpicture}[scale=0.7, every node/.style={scale=0.75}]
		\draw (0,-0.5) -- (0,2) to[out=90, in=90,looseness=2] (0.75,2) to[out=-90, in=90] (0.75,-0.5) to[out=-90, in=-90,looseness=2] (0,-0.5);
		\draw[line width = 1.5mm, white] (-1.5,0) -- (1.5,0);
		\draw (-1.5,0) -- (1.5,0);
		\draw[line width = 1.5mm, white] (-1.5,1.5) -- (1.5,1.5);
		\draw (-1.5,1.5) -- (1.5,1.5);
		\draw[fill=white] (0,0) circle (0.428);
		\draw (0,0) node {$\widetilde{S}$};
		\draw[fill=white] (0,1.5) circle (0.428);
		\draw (0,1.5) node {$R_{\varphi+\varepsilon}$};
		
		\draw[->] (1.5,0.75) -- (2.5,0.75);
		\begin{scope}[shift={(4,0.75)}]
		\draw[line width = 0.5mm] (-1,0) -- (1,0);
		\draw[fill=white] (0,0) circle (0.428);
		\draw (0,0) node {$E_1$};
		\end{scope}
		\begin{scope}[shift={(8.25,0.75)}]
		\draw (-2,-0.2) node[scale=0.9*4/3] {$\displaystyle =\sum_{j} \lambda_{1,j}$};
		\draw[line width = 0.5mm] (-1,0) -- (0,0);
		\draw[fill=white] (0,0) circle (0.428);
		\draw (0,0) node {$r_{1,j}$};
		\draw[line width = 0.5mm] (1.5,0) -- (2.25,0);
		\draw[fill=white] (1.25,0) circle (0.428);
		\draw (1.25,0) node {$l_{1,j}$};
		\end{scope}
	\end{tikzpicture}
\end{center}
\begin{center}
	\begin{tikzpicture}[scale=0.75, every node/.style={scale=0.8}]
		\draw[->] (0,0) -- (1,0);
		\begin{scope}[shift={(4,0)}]
		\draw (-1,-1) to[out=0,in=-90] (0,-0.5) -- (0,0.5) to[out=90,in=0] (-1,1);
		\draw (2.25,-1) to[out=180,in=-90] (1.25,-0.5) -- (1.25,0.5) to[out=90,in=180] (2.25,1);
		\draw (-2,-0.2) node[scale=0.9*10/8] {$\displaystyle \sum_{j} \lambda_{1,j}$};
		\draw[fill=white] (0,0) circle (0.4);
		\draw (0,0) node {$r_{1,j}$};
		\draw[fill=white] (1.25,0) circle (0.4);
		\draw (1.25,0) node {$l_{1,j}$};
		\end{scope}
	\end{tikzpicture},
\end{center}
\begin{center}
	\begin{tikzpicture}[scale=0.725, every node/.style={scale=0.8}]
		\begin{scope}[shift={(0,-0.75)}]
		\draw (0,-0.5) -- (0,3.5) to[out=90, in=90,looseness=2] (0.75,3.5) to[out=-90, in=90] (0.75,-0.5) to[out=-90, in=-90,looseness=2] (0,-0.5);
		\draw[line width = 1.5mm, white] (-1.5,3) -- (1.5,3);
		\draw (-1.5,3) -- (1.5,3);
		\draw[line width = 1.5mm, white] (-1.5,0) -- (1.5,0);
		\draw (-1.5,0) -- (1.5,0);
		\draw[line width = 1.5mm, white] (-1.5,1.5) -- (1.5,1.5);
		\draw (-1.5,1.5) -- (1.5,1.5);
		\draw[fill=white] (0,0) circle (0.41);
		\draw (0,0) node {$\widetilde{S}$};
		\draw[fill=white] (0,1.5) circle (0.41);
		\draw (0,1.5) node {$R_{\varphi}$};
		\draw[fill=white] (0,3) circle (0.41);
		\draw (0,3) node {$\widetilde{S}$};
		\end{scope}
		\draw[->] (1.75,0.75) -- (2.75,0.75);
		\begin{scope}[shift={(4,0.75)}]
		\draw[line width = 0.5mm] (-1,0) -- (1,0);
		\draw[fill=white] (0,0) circle (0.41);
		\draw (0,0) node {$E_2$};
		\end{scope}
		\begin{scope}[shift={(8.25,0.75)}]
		\draw (-2,-0.2) node[scale=0.9*10/8] {$\displaystyle =\sum_{j} \lambda_{2,j}$};
		\draw[line width = 0.5mm] (-1,0) -- (0,0);
		\draw[fill=white] (0,0) circle (0.41);
		\draw (0,0) node {$r_{2,j}$};
		\draw[line width = 0.5mm] (1.5,0) -- (2.25,0);
		\draw[fill=white] (1.25,0) circle (0.41);
		\draw (1.25,0) node {$l_{2,j}$};
		\end{scope}
	\end{tikzpicture}
\end{center}
\begin{center}
	\begin{tikzpicture}[scale=0.725, every node/.style={scale=0.8}]
		\draw[->] (0,0) -- (1,0);
		\begin{scope}[shift={(4,0)}]
		\draw (-1,-1) to[out=0,in=-90] (0,-0.5) -- (0,0.5) to[out=90,in=0] (-1,1);
		\draw (2.25,-1) to[out=180,in=-90] (1.25,-0.5) -- (1.25,0.5) to[out=90,in=180] (2.25,1);
		\draw (-1,0) -- (0,0);
		\draw (1.25,0) -- (2.25,0);
		\draw (-2,-0.2) node[scale=0.9*10/8] {$\displaystyle \sum_{j} \lambda_{2,j}$};
		\draw[fill=white] (0,0) circle (0.41);
		\draw (0,0) node {$r_{2,j}$};
		\draw[fill=white] (1.25,0) circle (0.41);
		\draw (1.25,0) node {$l_{2,j}$};
		\end{scope}
	\end{tikzpicture},
\end{center}
which allow us to write $\tr\left(\rho_{\varphi+\varepsilon} \widetilde{L} \right) = \tr E_1^N$ and $\tr\left(\rho_\varphi \widetilde{L}^2\right) = \tr E_2^N$. Now using the fact that $E_i^N$ is determined by its leading eigenvalue (see Eq.~\eqref{eq:TM eig1}) we can write $\tr E_i^N=\lambda_{i,1}^{N-1} (l_{i,1}|E_i|r_{i,1})$ and express $f(\rho_0, \widetilde{L})$ as:
\begin{center}
	\begin{tikzpicture}[scale=0.55, every node/.style={scale=0.7}]
		\draw (0,-0.75) -- (0,2.75) to[out=90, in=90,looseness=2] (0.75,2.75) to[out=-90, in=90] (0.75,-0.75) to[out=-90, in=-90,looseness=2] (0,-0.75);
		\draw[line width = 1.5mm, white] (-1.5,0) -- (1.5,0);
		\draw (-1.5,0) -- (1.5,0) to[out=0,in=-90] (2,0.5) -- (2,1.5) to[out=90,in=0] (1.5,2);
		\draw[line width = 1.5mm, white] (-1.5,2) -- (1.5,2);
		\draw (-1.5,0) to[out=180,in=-90] (-2,0.5) -- (-2,1.5) to[out=90,in=180] (-1.5,2)  -- (1.5,2);
		\draw[fill=white] (0,0) circle (0.54);
		\draw (0,0) node {$\widetilde{S}$};
		\draw[fill=white] (0,2) circle (0.54);
		\draw (0,2) node {$R_{\varphi+\varepsilon}$};
		
		\draw[fill=white] (2,1) circle (0.54);
		\draw (2,1) node {$r_{1,1}$};
		
		\draw[fill=white] (-2,1) circle (0.54);
		\draw (-2,1) node {$l_{1,1}$};
		\draw (-2.5,1) node[left] {$\displaystyle \frac{2\lambda_{1,1}^{N-1}}{N\varepsilon^2}$};
		\begin{scope}[shift={(7.5,1)}]
			\draw (0,-2.25) -- (0,2.25) to[out=90, in=90,looseness=2] (0.75,2.25) to[out=-90, in=90] (0.75,-2.25) to[out=-90, in=-90,looseness=2] (0,-2.25);
			
			\draw[line width = 1.5mm, white] (-1.5,-1.5) -- (1.5,-1.5);
			\draw[line width = 1.5mm, white] (-1.5,0) -- (1.5,0);
			\draw (-1.5,-1.5) -- (1.5,-1.5) to[out=0,in=-90] (2,-1) -- (2,1) to[out=90,in=0] (1.5,1.5);
			\draw[line width = 1.5mm, white] (-1.5,1.5) -- (1.5,1.5);
			\draw (-1.5,-1.5) to[out=180,in=-90] (-2,-1) -- (-2,1) to[out=90,in=180] (-1.5,1.5)  -- (1.5,1.5);
			\draw (-1.5,0) -- (1.5,0);
			
			\draw[fill=white] (0,-1.5) circle (0.54);
			\draw (0,-1.5) node {$\widetilde{S}$};
			\draw[fill=white] (0,0) circle (0.54);
			\draw (0,0) node {$R_{\varphi}$};
			\draw[fill=white] (0,1.5) circle (0.54);
			\draw (0,1.5) node {$\widetilde{S}$};
			
			\draw[fill=white] (2,0) circle (0.54);
			\draw (2,0) node {$r_{2,1}$};
			
			\draw[fill=white] (-2,0) circle (0.54);
			\draw (-2,0) node {$l_{2,1}$};
			\draw (-2.5,0) node[left] {$-\displaystyle \frac{\lambda_{2,1}^{N-1}}{N\varepsilon^2}$};
			\draw (2.5,0) node[right] {$-\displaystyle \frac{1}{N\varepsilon^2}$};
		\end{scope}
	\end{tikzpicture},
\end{center}
which after vectorizing $\widetilde{S}$ is equal to
\begin{center}
	\begin{tikzpicture}[scale=0.5, every node/.style={scale=0.7}]
		\draw (-2,-1) node {$\frac{2\lambda_{1,1}^{N-1}}{N\varepsilon^2}$};

		\draw (2,-5) -- (2,1) to[out=90, in=90,looseness=2] (2+0.75,1) -- (2+0.75,0) to[out=-90, in=90,looseness=2] (2.05,-3) -- (2+0.05,-5) to[out=-90, in=-90,looseness=2] (2+0,-5);

		\draw[line width=1.5mm, white] (-1,0) -- (5,0);
		\draw (0,-1) to[out=90,in=180] (1,0) -- (3,0) to[out=0,in=90] (4,-1);

		\draw[line width=1.5mm, white] (-1,-2) -- (1,-2);
		\draw[line width=1.5mm, white] (3,-2) -- (5,-2);
		\draw (0,-1) to[out=-90,in=180] (1,-2) to[out=0,in=90] (1.95,-4) -- (1.95,-5) -- (2.1,-5) -- (2.1,-4) to[out=90,in=180] (3,-2) to[out=0,in=-90] (4,-1);

		\draw[fill=white] (2,-5) circle (0.6);
		\draw (2,-5) node {$\widetilde{S}$};

		\draw[fill=white] (2,0) circle (0.6);
		\draw (2,0) node {$R_{\varphi+\varepsilon}$};

		\draw[fill=white] (0,-1) circle (0.6);
		\draw (0,-1) node {$l_{1,1}$};

		\draw[fill=white] (4,-1) circle (0.6);
		\draw (4,-1) node {$r_{1,1}$};

		\draw[rounded corners, fill=none,dashed] (-1,-4) rectangle (5,2.0);

		\draw[->] (-0.5,1.5) -- (-0.5,2.5);
		\draw (-0.5,2.5) node[above] {\LARGE $|b\rangle$};

		\draw (2,-3.5) node[right] {\LARGE $\alpha$};

		\begin{scope}[shift={(8.25,1)}]
		\draw (-2.25,-2) node {$-\frac{\lambda_{2,1}^{N-1}}{N\varepsilon^2}$};
		\draw (6,-2) node {$-\frac{1}{N\varepsilon^2}$};

		\draw (2,-7) -- (2,3) to[out=90, in=90,looseness=2] (2+0.05,3) -- (2.05,1) to[out=-90, in=90,looseness=2] (2+0.75,-1.5) -- (2+0.75,-2.5) to[out=-90,in=90,looseness=2] (2.05,-5) -- (2.05,-6) to[out=-90, in=-90,looseness=2] (2+0.05,-7);

		\draw[line width=1.5mm, white] (-1,0) -- (1,0);
		\draw[line width=1.5mm, white] (3,0) -- (5,0);
		\draw (0,-2) to[out=90,in=180] (1,0) to[out=0,in=-90] (1.95,2) -- (1.95,3) -- (2.1,3) -- (2.1,2) to[out=-90,in=180] (3,0) to[out=0,in=90] (4,-2);

		\draw[line width=1.5mm, white] (0,-2) -- (4,-2);
		\draw (0,-2) -- (4,-2);

		\draw[line width=1.5mm, white] (-1,-4) -- (1,-4);
		\draw[line width=1.5mm, white] (3,-4) -- (5,-4);

		\draw (0,-2) to[out=-90,in=180] (1,-4) to[out=0,in=90] (1.95,-5) -- (1.95,-7) -- (2.1,-7) -- (2.1,-5) to[out=90,in=180] (3,-4) to[out=0,in=-90] (4,-2);

		\draw[fill=white] (2,3) circle (0.6);
		\draw (2,3) node {$\widetilde{S}$};
		\draw[fill=white] (2,-2) circle (0.6);
		\draw (2,-2) node {$R_{\varphi}$};
		\draw[fill=white] (2,-7) circle (0.6);
		\draw (2,-7) node {$\widetilde{S}$};

		\draw[fill=white] (0,-2) circle (0.6);
		\draw (0,-2) node {$l_{2,1}$};

		\draw[fill=white] (4,-2) circle (0.6);
		\draw (4,-2) node {$r_{2,1}$};

		\draw[rounded corners, fill=none,dashed] (-1,-6) rectangle (5,2.0);

		\draw (2,-5.5) node[right] {\LARGE $\beta$};
		\draw (2,1.5) node[right] {\LARGE $\alpha$};

		\draw[->] (-0.5,1.5) -- (-0.5,2.5);
		\draw (-0.5,2.5) node[above] {\LARGE $A$};
		\end{scope}
	\end{tikzpicture},
\end{center}
or when written in as equation:
\begin{equation}
	f(\rho_0,\widetilde{L})  = \frac{\lambda_{1,1}^{N-1} \sum_\alpha b_\alpha \widetilde{S}_\alpha - \lambda_{2,1}^{N-1} \sum_{\alpha\beta} \widetilde{S}_\alpha A_{\alpha\beta} \widetilde{S}_\beta -1}{N\varepsilon^2}.
\end{equation}
The condition for an extremum reads:
\begin{equation} \label{eq:condLinf}
	\frac{1}{2} \lambda_{1,1}^{N-1} \left(A+A^\mathrm{T}\right) |\widetilde{S}\rangle = \lambda_{2,1}^{N-1} |b\rangle.
\end{equation}

For $N\to\infty$ the powers of the eigenvalues may seem to pose a problem. Fortunately, however, this problem can be circumvented. For a given $\widetilde{L}$ we can calculate the associated value of our FoM per particle:
\begin{equation} \label{eq:FoM inf val}
	f(\rho_0, \widetilde{L}) = \frac{1}{N\varepsilon^2} \left(2\lambda_{1,1}^N-\lambda_{2,1}^N-1 \right),
\end{equation}
but going back to the roots (Eq.~\eqref{eq:qfimax}) we see that FoM per particle should have form $f(\rho_0, \widetilde{L}) = 2f_1-f_2$ where $f_1$ and $f_2$ are of the same order of magnitude as the asymptotic limit of the QFI per particle. Assuming that our calculations are in the regime of $N\rightarrow\infty$, $\varepsilon \rightarrow 0$, and $N \varepsilon^2 \rightarrow 0$, and remembering binomial expansion
\begin{equation}
	(1+\varepsilon^2 f_i)^N = 1+N\varepsilon^2 f_i+O[(N\varepsilon^2)^2],
\end{equation}
it is to be expected that the highest eigenvalues of the transfer matrices have the form:
\begin{equation}
	\lambda_{1,1} = 1+\varepsilon^2 f_1, \quad
	\lambda_{2,1} = 1+\varepsilon^2 f_2,
\end{equation}
which after inserting into Eq.~\eqref{eq:FoM inf val} and using binomial expansion to the first order give us exactly $f(\rho_0, \widetilde{L}) = 2f_1-f_2$. Note that, it means that we can calculate value of FoM per particle in a simple way:
\begin{equation}
	f(\rho_0, \widetilde{L}) \approx \frac{1}{\varepsilon^2} \left(2\lambda_{1,1}-\lambda_{2,1}-1 \right).
\end{equation}
It is clear now that for the purpose of solving Eq.~\eqref{eq:condLinf} we can approximate $\lambda_{1,1}^{N-1}$ and $\lambda_{2,1}^{N-1}$ by ones and, hence, bring the condition for the optimal $\widetilde{S}$ to a simpler form:
\begin{equation}
	\frac{1}{2} \left(A+A^\mathrm{T}\right) |\widetilde{S}\rangle = |b\rangle.
\end{equation}
This equation is solved with a pseudo-inverse and its anti-Hermitian part is filtered out at every iteration step.

The SLD is always traceless in any unitary parameter estimation problem---see Eq.~\eqref{eq:L}. Note also that $\bra{\lambda} \rho'_\varphi \ket{\lambda}=\ii \bra{\lambda} [H,\rho_\varphi] \ket{\lambda} = 0$ for any $|\lambda\rangle$ which is an eigenstate of $\rho_\varphi$. We can ensure that solution $\widetilde{S}$ has proper normalization using the condition $\tr L=0$ or, equivalently $\tr \widetilde{L} = \tr \mathds{1} = d^N$ which in the language of transfer matrices means that the highest eigenvalue of the transfer matrix,
\begin{center}
	\begin{tikzpicture}[scale=0.75]
		\draw (0,-0.5) -- (0,0.5) to[out=90, in=90,looseness=2] (0.75,0.5) to[out=-90, in=90] (0.75,-0.5) to[out=-90, in=-90,looseness=2] (0,-0.5);
		\draw[line width = 1.5mm, white] (-1.5,0) -- (1.5,0);
		\draw (-1.5,0) -- (1.5,0);
		\draw[fill=white] (0,0) circle (0.4);
		\draw (0,0) node {$\widetilde{S}$};
	\end{tikzpicture},
\end{center}
has to be equal to $d$.

Now we turn to the second part of our optimization procedure, namely the variational minimization over the input state. This step does not introduce any qualitatively new challenges, so we only briefly discuss it for completeness. As for the $\rho'_\varphi$, we approximate the exact derivative of $L'^*_\varphi$ by its discrete version:
\begin{equation}
	L'^*_{\varphi} = \frac{L^*_{\varphi+\varepsilon}-L^*_{\varphi}}{\varepsilon}.
\end{equation}
After the expansion $L = (\widetilde{L}-\mathds{1})/\varepsilon$, our task becomes equivalent to minimization of the ``energy density'':
\begin{equation} \label{eq:varinf}
	-f(\rho_0, \widetilde{L})  = \frac{\langle \psi|\widetilde{L}^*_{2,\varphi}-2\widetilde{L}^*_{\varphi+\varepsilon} + \mathds{1}|\psi \rangle}{N \varepsilon^2 \langle \psi|\psi \rangle}
\end{equation}
over $\ket{\psi}$. Using transfer matrices $E_3$, $E_4$ and $E_5$ associated with, respectively, $\langle \psi|\widetilde{L}^*_{2,\varphi}|\psi \rangle$, $\langle \psi|\widetilde{L}^*_{\varphi+\varepsilon}|\psi \rangle$ and $\langle \psi|\psi \rangle$,
\begin{center}
	\begin{tikzpicture}[scale=0.75, every node/.style={scale=0.8}]
		\begin{scope}[shift={(0,0)}]
			\draw (0,0) -- (0,3);
			\draw[line width = 1.5mm, white] (-1.5,3) -- (1.5,3);
			\draw (-1.5,3) -- (1.5,3);
			\draw[line width = 1.5mm, white] (-1.5,0) -- (1.5,0);
			\draw (-1.5,0) -- (1.5,0);
			\draw[line width = 1.5mm, white] (-1.5,1.5) -- (1.5,1.5);
			\draw (-1.5,1.5) -- (1.5,1.5);
			\draw[fill=white] (0,0) circle (0.4);
			\draw (0,0) node {$\overline{P}$};
			\draw[fill=white] (0,1.5) circle (0.4);
			\draw (0,1.5) node {$\widetilde{S}_2$};
			\draw[fill=white] (0,3) circle (0.4);
			\draw (0,3) node {${P}$};
		\end{scope}
		\begin{scope}[shift={(6,1.5)}]
			\draw (-1,-1.5) to[out=0,in=-90] (0,-0.5) -- (0,0.5) to[out=90,in=0] (-1,1.5);
			\draw (2.25,-1.5) to[out=180,in=-90] (1.25,-0.5) -- (1.25,0.5) to[out=90,in=180] (2.25,1.5);
			\draw (-1,0) -- (0,0);
			\draw (1.25,0) -- (2.25,0);
			\draw (-2.5,-0.2) node[scale=0.9*10/8] {$=\displaystyle \sum_{j} \lambda_{3,j}$};
			\draw[fill=white] (0,0) circle (0.4);
			\draw (0,0) node {$r_{3,j}$};
			\draw[fill=white] (1.25,0) circle (0.4);
			\draw (1.25,0) node {$l_{3,j}$};
		\end{scope}
	\end{tikzpicture}
\end{center}
\begin{center}
	\begin{tikzpicture}[scale=0.75, every node/.style={scale=0.8}]
		\begin{scope}[shift={(0,0)}]
			\draw (0,0) -- (0,3);
			\draw[line width = 1.5mm, white] (-1.5,3) -- (1.5,3);
			\draw (-1.5,3) -- (1.5,3);
			\draw[line width = 1.5mm, white] (-1.5,0) -- (1.5,0);
			\draw (-1.5,0) -- (1.5,0);
			\draw[line width = 1.5mm, white] (-1.5,1.5) -- (1.5,1.5);
			\draw (-1.5,1.5) -- (1.5,1.5);
			\draw[fill=white] (0,0) circle (0.4);
			\draw (0,0) node {$\overline{P}$};
			\draw[fill=white] (0,1.5) circle (0.4);
			\draw (0,1.5) node[scale=0.9] {$\widetilde{S}_{\varphi+\varepsilon}$};
			\draw[fill=white] (0,3) circle (0.4);
			\draw (0,3) node {${P}$};
		\end{scope}
		\begin{scope}[shift={(6,1.5)}]
			\draw (-1,-1.5) to[out=0,in=-90] (0,-0.5) -- (0,0.5) to[out=90,in=0] (-1,1.5);
			\draw (2.25,-1.5) to[out=180,in=-90] (1.25,-0.5) -- (1.25,0.5) to[out=90,in=180] (2.25,1.5);
			\draw (-1,0) -- (0,0);
			\draw (1.25,0) -- (2.25,0);
			\draw (-2.5,-0.2) node[scale=0.9*10/8] {$=\displaystyle \sum_{j} \lambda_{4,j}$};
			\draw[fill=white] (0,0) circle (0.4);
			\draw (0,0) node {$r_{4,j}$};
			\draw[fill=white] (1.25,0) circle (0.4);
			\draw (1.25,0) node {$l_{4,j}$};
		\end{scope}
	\end{tikzpicture}
\end{center}
\begin{center}
	\begin{tikzpicture}[scale=0.75, every node/.style={scale=0.8}]
		\begin{scope}[shift={(0,0)}]
			\draw (0,0.75) -- (0,2.25);
			\draw[line width = 1.5mm, white] (-1.5,2.25) -- (1.5,2.25);
			\draw (-1.5,2.25) -- (1.5,2.25);
			\draw[line width = 1.5mm, white] (-1.5,0.75) -- (1.5,0.75);
			\draw (-1.5,0.75) -- (1.5,0.75);
			\draw[fill=white] (0,0.75) circle (0.4);
			\draw (0,0.75) node {$\overline{P}$};
			\draw[fill=white] (0,2.25) circle (0.4);
			\draw (0,2.25) node {${P}$};
		\end{scope}
		\begin{scope}[shift={(6,1.5)}]
			\draw (-1,-1.5) to[out=0,in=-90] (0,-0.5) -- (0,0.5) to[out=90,in=0] (-1,1.5);
			\draw (2.25,-1.5) to[out=180,in=-90] (1.25,-0.5) -- (1.25,0.5) to[out=90,in=180] (2.25,1.5);
			\draw (-2.5,-0.2) node[scale=0.9*10/8] {$=\displaystyle \sum_{j} \lambda_{5,j}$};
			\draw[fill=white] (0,0) circle (0.4);
			\draw (0,0) node {$r_{5,j}$};
			\draw[fill=white] (1.25,0) circle (0.4);
			\draw (1.25,0) node {$l_{5,j}$};
		\end{scope}
	\end{tikzpicture}
\end{center}
we can rewrite Eq.~\eqref{eq:varinf} in a diagrammatic form:
\begin{widetext}
	\begin{center}
		\begin{tikzpicture}[scale=0.6, every node/.style={scale=0.7}]
			\draw (-12,-2.5) node[left] {$-f(\rho_0, \widetilde{L})=$};
			\draw (-12,-2.5) -- (10,-2.5);
			\begin{scope}[shift={(0,-5.5)}]
				\draw (0,2) -- (0,0);
				\draw[line width = 1.5mm, white] (-1.5,0) -- (1.5,0);
				\draw (-1.5,0) -- (1.5,0) to[out=0,in=-90] (2,0.5) -- (2,1.5) to[out=90,in=0] (1.5,2);
				\draw[line width = 1.5mm, white] (-1.5,2) -- (1.5,2);
				\draw (-1.5,0) to[out=180,in=-90] (-2,0.5) -- (-2,1.5) to[out=90,in=180] (-1.5,2)  -- (1.5,2);
				\draw[fill=white] (0,0) circle (0.5);
				\draw (0,0) node {$\overline{P}$};
				\draw[fill=white] (0,2) circle (0.5);
				\draw (0,2) node {$P$};
	
				\draw[fill=white] (2,1) circle (0.5);
				\draw (2,1) node {$r_{5,1}$};
	
				\draw[fill=white] (-2,1) circle (0.5);
				\draw (-2,1) node {$l_{5,1}$};
				\draw (-2.5,1) node[left] {$\displaystyle N\varepsilon^2$};
			\end{scope}
	
			\begin{scope}[shift={(7,-1)}]
				\draw (0,2) -- (0,0);
				\draw[line width = 1.5mm, white] (-1.5,0) -- (1.5,0);
				\draw (-1.5,0) -- (1.5,0) to[out=0,in=-90] (2,0.5) -- (2,1.5) to[out=90,in=0] (1.5,2);
				\draw[line width = 1.5mm, white] (-1.5,2) -- (1.5,2);
				\draw (-1.5,0) to[out=180,in=-90] (-2,0.5) -- (-2,1.5) to[out=90,in=180] (-1.5,2)  -- (1.5,2);
				\draw[fill=white] (0,0) circle (0.5);
				\draw (0,0) node {$\overline{P}$};
				\draw[fill=white] (0,2) circle (0.5);
				\draw (0,2) node {$P$};
	
				\draw[fill=white] (2,1) circle (0.5);
				\draw (2,1) node {$r_{5,1}$};
	
				\draw[fill=white] (-2,1) circle (0.5);
				\draw (-2,1) node {$l_{5,1}$};
				\draw (-2.5,1) node[left] {$+\lambda_{5,1}^{N-1}$};
			\end{scope}

			\begin{scope}[shift={(0,0)}]
				\draw (0,-2) -- (0,2);
				\draw[line width = 1.5mm, white] (-1.5,-1.5) -- (1.5,-1.5);
				\draw[line width = 1.5mm, white] (-1.5,0) -- (1.5,0);
				\draw (-1.5,-1.5) -- (1.5,-1.5) to[out=0,in=-90] (2,-1) -- (2,1) to[out=90,in=0] (1.5,1.5);
				\draw[line width = 1.5mm, white] (-1.5,1.5) -- (1.5,1.5);
				\draw (-1.5,-1.5) to[out=180,in=-90] (-2,-1) -- (-2,1) to[out=90,in=180] (-1.5,1.5)  -- (1.5,1.5);
				\draw (-1.5,0) -- (1.5,0);
	
				\draw[fill=white] (0,-1.5) circle (0.5);
				\draw (0,-1.5) node {$\overline{P}$};
				\draw[fill=white] (0,0) circle (0.5);
				\draw (0,0) node {$\widetilde{S}_{\varphi+\varepsilon}$};
				\draw[fill=white] (0,1.5) circle (0.5);
				\draw (0,1.5) node {$P$};
	
				\draw[fill=white] (2,0) circle (0.5);
				\draw (2,0) node {$r_{4,1}$};
	
				\draw[fill=white] (-2,0) circle (0.5);
				\draw (-2,0) node {$l_{4,1}$};
				\draw (-2.5,0) node[left] {$-2\lambda_{4,1}^{N-1}$};
			\end{scope}
	
			\begin{scope}[shift={(-7.5,0)}]
				\draw (0,-2) -- (0,2);
				\draw[line width = 1.5mm, white] (-1.5,-1.5) -- (1.5,-1.5);
				\draw[line width = 1.5mm, white] (-1.5,0) -- (1.5,0);
				\draw (-1.5,-1.5) -- (1.5,-1.5) to[out=0,in=-90] (2,-1) -- (2,1) to[out=90,in=0] (1.5,1.5);
				\draw[line width = 1.5mm, white] (-1.5,1.5) -- (1.5,1.5);
				\draw (-1.5,-1.5) to[out=180,in=-90] (-2,-1) -- (-2,1) to[out=90,in=180] (-1.5,1.5)  -- (1.5,1.5);
				\draw (-1.5,0) -- (1.5,0);
	
				\draw[fill=white] (0,-1.5) circle (0.5);
				\draw (0,-1.5) node {$\overline{P}$};
				\draw[fill=white] (0,0) circle (0.5);
				\draw (0,0) node {$\widetilde{S}_{2}$};
				\draw[fill=white] (0,1.5) circle (0.5);
				\draw (0,1.5) node {$P$};
	
				\draw[fill=white] (2,0) circle (0.5);
				\draw (2,0) node {$r_{3,1}$};
	
				\draw[fill=white] (-2,0) circle (0.5);
				\draw (-2,0) node {$l_{3,1}$};
				\draw (-2.5,0) node[left] {$\lambda_{3,1}^{N-1}$};
			\end{scope}
		\end{tikzpicture}.
	\end{center}
\end{widetext}
As previously, we expect that $\lambda_{i,1} = 1+\varepsilon^2 f_i$ and for the purpose of finding the optimal tensor $P$, we can approximate $\lambda_{3,1}^{N-1}$ and $\lambda_{4,1}^{N-1}$ by ones. After taking the derivative we obtain the condition for the extremum:
\begin{equation} \label{eq:condPinf}
	\mathcal{F} |P\rangle = g \mathcal{N}|P\rangle,
\end{equation}
where $g = -f(\rho_0, \widetilde{L}) N\varepsilon^2-\lambda_{5,1}^{N-1}$ is a generalised eigenvalue, whereas the matrices $\mathcal{F}$ and $\mathcal{N}$ are defined as:
\begin{center}
	\begin{tikzpicture}[scale=0.7, every node/.style={scale=0.75}]
		\begin{scope}[shift={(0,0)}]
			\draw (0,-2.5) -- (0,2.5);

			\draw (-2,0) to[out=90,in=-90,looseness=1.5] (-0.05,2) -- (-0.05,2.5);
			\draw (2,0) to[out=90,in=-90,looseness=1.5] (0.05,2) -- (0.05,2.5);

			\draw (-2,0) to[out=-90,in=90,looseness=1.5] (-0.05,-2) -- (-0.05,-2.5);
			\draw (2,0) to[out=-90,in=90,looseness=1.5] (0.05,-2) -- (0.05,-2.5);

			\draw (-1.5,0) -- (1.5,0);

			\draw[fill=white] (0,0) circle (0.428);
			\draw (0,0) node {$\widetilde{S}_2$};

			\draw[fill=white] (2,0) circle (0.428);
			\draw (2,0) node {$r_{3,1}$};

			\draw[fill=white] (-2,0) circle (0.428);
			\draw (-2,0) node {$l_{3,1}$};

			\draw (0,2.5) node[right] {$\alpha$};
			\draw (0,-2.5) node[right] {$\beta$};
			\draw (-2.5,0) node[left] {$\mathcal{F}_{\alpha\beta}=$};
		\end{scope}
		\begin{scope}[shift={(6,0)}]
			\draw (0,-2.5) -- (0,2.5);

			\draw (-2,0) to[out=90,in=-90,looseness=1.5] (-0.05,2) -- (-0.05,2.5);
			\draw (2,0) to[out=90,in=-90,looseness=1.5] (0.05,2) -- (0.05,2.5);

			\draw (-2,0) to[out=-90,in=90,looseness=1.5] (-0.05,-2) -- (-0.05,-2.5);
			\draw (2,0) to[out=-90,in=90,looseness=1.5] (0.05,-2) -- (0.05,-2.5);

			\draw (-1.5,0) -- (1.5,0);

			\draw[fill=white] (0,0) circle (0.428);
			\draw (0,0) node {$\widetilde{S}_{\varphi+\varepsilon}$};

			\draw[fill=white] (2,0) circle (0.428);
			\draw (2,0) node {$r_{4,1}$};

			\draw[fill=white] (-2,0) circle (0.428);
			\draw (-2,0) node {$l_{4,1}$};

			\draw (0,2.5) node[right] {$\alpha$};
			\draw (0,-2.5) node[right] {$\beta$};
			\draw (-2.5,0) node[left] {$-2$}; 		
		\end{scope}
	\end{tikzpicture},
\end{center}
\begin{center}
	\begin{tikzpicture}[scale=0.7, every node/.style={scale=0.75}]
		\begin{scope}[shift={(0,0)}]
			\draw (0,-2.5) -- (0,2.5);

			\draw (-2,0) to[out=90,in=-90,looseness=1.5] (-0.05,2) -- (-0.05,2.5);
			\draw (2,0) to[out=90,in=-90,looseness=1.5] (0.05,2) -- (0.05,2.5);

			\draw (-2,0) to[out=-90,in=90,looseness=1.5] (-0.05,-2) -- (-0.05,-2.5);
			\draw (2,0) to[out=-90,in=90,looseness=1.5] (0.05,-2) -- (0.05,-2.5);

			\draw[fill=white] (2,0) circle (0.428);
			\draw (2,0) node {$r_{5,1}$};

			\draw[fill=white] (-2,0) circle (0.428);
			\draw (-2,0) node {$l_{5,1}$};

			\draw (0,2.5) node[right] {$\alpha$};
			\draw (0,-2.5) node[right] {$\beta$};
			\draw (-2.5,0) node[left] {$\mathcal{N}_{\alpha\beta}=$};
		\end{scope}
		\begin{scope}[shift={(6,0)}]
			\draw (0,-2.5) -- (0,2.5);

			\draw (-2,0) to[out=90,in=-90,looseness=1.5] (-0.05,2) -- (-0.05,2.5);
			\draw (2,0) to[out=90,in=-90,looseness=1.5] (0.05,2) -- (0.05,2.5);

			\draw (-2,0) to[out=-90,in=90,looseness=1.5] (-0.05,-2) -- (-0.05,-2.5);
			\draw (2,0) to[out=-90,in=90,looseness=1.5] (0.05,-2) -- (0.05,-2.5);

			\draw[fill=white] (2,0) circle (0.428);
			\draw (2,0) node {$r_{5,1}$};

			\draw[fill=white] (-2,0) circle (0.428);
			\draw (-2,0) node {$l_{5,1}$};

			\draw (0,2.5) node[right] {$\alpha$};
			\draw (0,-2.5) node[right] {$\beta$};
			\draw (-2.7,0) node[left] {$=$};
			\draw (-1,0) node {$\otimes$};
			\draw (1,0) node {$\otimes$};
			\draw (0,0) node[right] {$\mathds{1}$};
			\draw[rounded corners, fill=none,dashed] (-2.7,-1) rectangle (2.7,1);
		\end{scope}
	\end{tikzpicture}.
\end{center}
The matrix $\mathcal{N}$ is a tensor product of three matrices, $r_{5,1}$, the identity, and $l_{5,1}$, hence its pseudo-inverse $\mathcal{N}^+$ can be obtained as a tensor product of (pseudo-)inverses of the smaller matrices. Applying $\mathcal{N}^+$ to Eq.~\eqref{eq:condPinf} we bring it into the form of a standard eigenvalue problem. We solve this eigenproblem with respect to the smallest eigenvalue and its corresponding eigenvector $|\psi\rangle$ and require that $|\psi\rangle$ is normalized so that $\lambda_{5,1} = 1$. Then we calculate the asymptotic value of the QFI per particle:
\begin{multline}
	-f(\rho_0,\widetilde{L}) = \frac{1}{N\varepsilon^2} \left(\lambda_{3,1}^N-2\lambda_{4,1}^N+1 \right) \approx \\
	\approx f_3-2f_4 = \frac{1}{\varepsilon^2} \left(\lambda_{3,1}-2\lambda_{4,1}+1 \right).
\end{multline}

Just as for the finite $N$, iterating the $\widetilde{L}$ and $|\psi\rangle$ optimization steps leads to the optimal solution with the maximal QFI per particle.

While performing the numerics one should choose $\varepsilon$ to be small but not too small as  too small values may lead to numerical instabilities. Our general strategy in obtaining numerical results reported in the next section, was to lower the value of $\varepsilon$ until we observed no noticeable change in the obtained results, while still remaining in the regime where algorithm was stable. In all the examples we studied in this paper this approach resulted in the  choice of  $\varepsilon \approx 10^{-3}-10^{-4}$ (instabilities started to appear for $\varepsilon < 10^{-6}$). Notice that in the asymptotic iMPO approach described above we required $N \varepsilon^2$ to be small---on the order of the precision we expect from the numerical results. In other words setting the precision requirements to $10^{-2}$ this implies that $N \varepsilon^2 \approx 10^{-2}$ and hence $N \approx 10^4-10^6$. What this physically means is that in our setup the QFI per particle does not change in any noticeable way for larger $N$ and hence the asymptotic behaviour in the actual $N \rightarrow \infty$ limit
may be inferred from this results.

\section{Applications} \label{sec:example}
In this section we present three applications of our framework. The examples were chosen in a way so as to highlight the possibility of applying the framework to a variety of qualitatively different physical problems and therefore demonstrate versatility of the approach. The first example -- of magnetic field sensing with locally correlated magnetic field fluctuations from Sec~\ref{subsec:example dephase} -- falls into the general metrological model structure as outlined in Sec.~\ref{sec:qmetrology}, and should be regarded as a typical representative of the models that can be dealt with efficiently using the MPO framework. In this case it is the standard QFI FoM that is being employed in the optimization process. However, in order to demonstrate that the utility of MPO based approach to metrology goes beyond the QFI optimization tasks, in Sec.~\ref{subsec:example allan} we show how the framework can be adapted to calculate the fundamental bound on the achievable Allan variance in the atomic clock stabilization problem. In this case not only the physical setup is different (temporal LO noise correlations affecting the atoms, rather than spatial noise correlations between the probes), but also the employed FoM. Instead of the QFI the QAVAR is optimized---a quantity in fact more closely related with the Bayesian variance rather than the QFI. Finally, in Sec.~\ref{subsec:example fidelity}, the third example demonstrates that the utility of the techniques of calculating QFI for mixed state via the MPO formalism developed in this paper, goes beyond the realm of metrological applications and can be directly applied to exact calculation of fidelity susceptibility in many-body thermal states---a task deemed too hard for all the state-of-the art methods.

\subsection{Magnetic field sensing with locally correlated noise} \label{subsec:example dephase}
Consider $N$ particles each with spin $s$ (in principle this might be an effective spin of some number of ``sub-particles'') which interact for a fixed time $t$ with an external magnetic field $B$ (assumed to be in the $z$ direction) whose strength fluctuates. The fluctuations induce an effective dephasing process on the particles. We will go beyond the standard model of independent field fluctuations resulting in independent dephasing of atoms \cite{Huelga1997, Escher2011}, by taking into account correlations between field fluctuations at the nearest neighbour particle sites. This will lead us to a model where we will be able to study the impact of correlations (and anti-correlations) in the effective dephasing process on the metrological potential of the system.

For a moment let us assume that the field $B$ is fixed. The Hamiltonian corresponding to the dynamics of an $n$-th spin in a static magnetic field is $- g  S^{[n]}_z B$, where $S^{[n]}_z$ is the $z$ spin component of the $n$-th particle and $g$ is the gyromagnetic ratio of the particle.  In order to stay consistent with the abstract notation introduced in Sec.~\ref{subsec:qmetrology local}, we will identify $\varphi = g B t$, $h^{[n]} =  S_z^{[n]}/\hbar$. The uncertainty of estimation of $B$ will be related with the uncertainty of a standard phase estimation problem via a simple proportionality relation $\Delta \widetilde{B} = \Delta \widetilde{\varphi}/(g t)$.

Now, taking into account the presence of fluctuations let us write the magnetic field at site $n$ as $B^{[n]}(t) = B + \delta B^{[n]}(t)$. Here we assume that fluctuations are Gaussian and have no relevant temporal correlations (white noise). The corresponding variance as well as the nearest-neighbour correlation functions of the fluctuating field read: $\langle \delta B^{[n]}(t) \delta B^{[n]}(t') \rangle  = \sigma^2 \delta(t-t')$, $\langle \delta B^{[n]}(t) \delta B^{[n+1]}(t')  \rangle  = \chi \delta(t-t')$, where $\chi$ represents the strength of correlations and may be both positive and negative (anti-correlation)---for simplicity, we assume periodic boundary conditions, so in fact also the particles $N$ and $1$ are correlated.

Let $|\mathbf{j}\rangle = |j_1, j_2, \dots, j_N\rangle$, $j_n \in \{-s,\dots,s\}$ be the eigenbasis with well defined eigenvalues of local Hamiltonians $h^{[n]} = S^{[n]}_z/\hbar$ operators equal to $j_n$: $h^{[n]} \ket{\mathbf{j}} = j_n \ket{\mathbf{j}}$. Using this basis we can easily write the evolution of the density matrix under the action of noise by applying the standard technique of the cumulant expansion \cite{vankampen1981}:
\begin{equation} \label{eq:lambdarho0}
	\Lambda(\rho_0) =  \sum_{\mathbf{j},\mathbf{k}} \langle \mathbf{j}|\rho_0|\mathbf{k}\rangle e^{-\frac12 (\mathbf{j} - \mathbf{k})^\mathrm{T} C (\mathbf{j} - \mathbf{k})} |\mathbf{j}\rangle\langle\mathbf{k}|,
\end{equation}
where $\mathbf{j}, \mathbf{k}$ are column vectors $\mathbf{j} = (j_1, j_2,\dots, j_N)^\mathrm{T}$, and $C$ is the correlation matrix
\begin{equation}
	C = \begin{pmatrix}
			c_1    & c_2 & 0   & \dots & c_2    \\
			c_2    & c_1 & c_2 &        &        \\
			0      & c_2 & c_1 &        &        \\
			\vdots &     &     & \ddots & \vdots \\
			c_2    &     &     & \dots & c_1    \\
		\end{pmatrix},
\end{equation}
where $c_1 = \sigma^2 g^2 t$, $c_2 = \chi g^2  t$. It is straightforward to extend the discussion here to deal with the more general longer range correlations covering up to $r$ neighbouring particles, in which case the matrix $C$ will be $2r+1$ diagonal.

For a better physical insight, let us provide the corresponding master equation in the form of Eq.~\ref{eq:master}. One may obtain it by simply differentiating \eqref{eq:lambdarho0} over $t$. As a result we get the master equation with single particle $L^{[n]} = \sqrt{\gamma_1} h^{[n]}$ and two particle $L^{[n,n+1]} = \sqrt{|\gamma_2|}(h^{[n]} + \t{sgn}(\gamma_2)h^{[n+1]})$ dephasing operators.  The dephasing rates  $\gamma_1$, $\gamma_2$ are related with field fluctuation properties as follows: $\gamma_1 = (\sigma^2 - 2 |\chi|) g^2$, $\gamma_2 = \chi g^2$---note that $\sigma^2 \geq 2 |\chi| $ by virtue of positivity of the correlation matrix $C$ so the rate $\gamma_1$ is always positive.

In order to write the evolution manifestly in the MPO formalism, we will replace $|\mathbf{j}\rangle\langle \mathbf{k}| \rightarrow |\mathbf{j}\rangle |\mathbf{k}\rangle$ which forms a basis for the vectorized input density matrix $|\rho_0\rangle$. The action of $\Lambda$ on $\rho_0$ is identical to the action of the operator $e^\Gamma$ on $|\rho_0\rangle$, i.e., $|\Lambda(\rho_0)\rangle = e^\Gamma|\rho_0\rangle$, where
\begin{equation}
	\Gamma = -\frac{c_1}{2}\sum_{n=1}^N \Upsilon^{[n]} - c_2\sum_{n=1}^N \Xi^{[n,n+1]},
\end{equation}
with
\begin{equation}
	\Upsilon^{[n]} = \left(h^{[n]} \otimes \openone -\openone \otimes h^{[n]} \right)^2,
\end{equation}
and
\begin{equation}
	\Xi^{[n,n+1]} = \left(h^{[n]} \otimes \openone -\openone \otimes h^{[n]}\right)\left(h^{[n+1]} \otimes \openone -\openone \otimes h^{[n+1]}\right).
\end{equation}
Note that the $\Upsilon^{[n]}$ and $\Xi^{[n,n+1]}$ mutually commute with each other, so that
\begin{equation}
	e^\Gamma = \prod_{n=1}^N e^{-\frac{c_1}{2}\Upsilon^{[n]}} e^{- c_2 \Xi^{[n,n+1]}}.
\end{equation}
Denoting $Y^{[n]} = e^{-\frac{c_1}{2}\Upsilon^{[n]}}$ and $X^{[n,n+1]} = e^{- c_2 \Xi^{[n,n+1]}}$ we finally obtain
\begin{equation} \label{eq:egamma}
	e^\Gamma = \prod_{n=1}^N Y^{[n]}X^{[n,n+1]},
\end{equation}
which is the form of evolution the same as discussed in Sec.~\ref{subsec:mpo rho} guaranteeing efficient MPO description.

After evolution through quantum channel $\Lambda$, the phase is imprinted in our state through unitary evolution according to Eq.~\eqref{eq:rhophi} with local Hamiltonians $h^{[n]}$---in the notation of Sec.~\ref{sec:mpo} this is represented by the action of $\prod_{n=1}^N Z^{[n]}$, where $Z^{[n]} = e^{-\ii h^{[n]}\varphi}\otimes (e^{\ii h^{[n]}\varphi})^\mathrm{T}$.  Written in the basis $|\mathbf{j}\rangle$, $\rho'_\varphi = \ii \left[\rho_\varphi,H\right]$ reads:
\begin{equation}
	\rho'_\varphi = \sum_{\mathbf{j},\mathbf{k}} \langle \mathbf{j}|\rho_\varphi|\mathbf{k}\rangle
	\ii \sum_n \left(k_n-j_n\right) |\mathbf{j}\rangle\langle\mathbf{k}|.
\end{equation}

\begin{figure*}
	\centering
	\includegraphics[width=\textwidth]{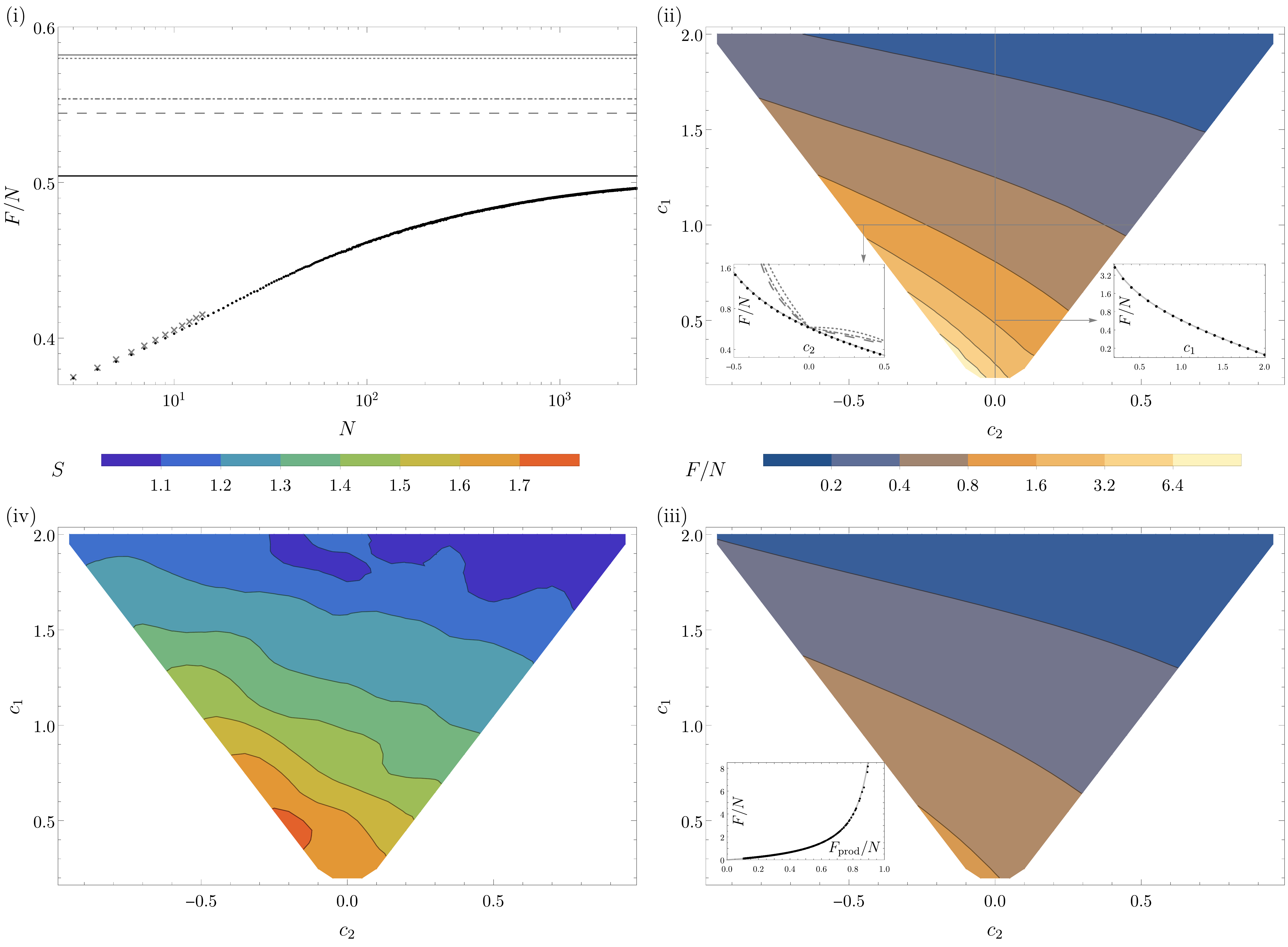}
	\caption{(i) Comparison of the QFI per particle for a magnetic field sensing problem in presence of locally correlated dephasing as a function of the number of spins in a chain $N$ (for dephasing noise model with local noise parameter $c_1 = 1$ and correlation parameter $c_2 = 0.1$) calculated using the finite MPO approach (black dots) with asymptotic value obtained using iMPO approach (black  solid line).  Grey crosses indicate results obtained via the standard full Hilbert space description.  Grey lines show state-of-the-art bounds on $\t{QFI}/N$ obtained for decomposition of the dynamics into effectively independent channels  $\Lambda_1$ (dotted), $\Lambda_2$ (dash-dotted), $\Lambda_3$ (dashed).
For comparison, the solid gray line corresponds to the bound obtained when all correlations are neglected and only local dephasing noise is taken into account.
	(ii) Asymptotic value (obtained using iMPOs) of QFI per particle for dephasing type noise in function of local $c_1$ and between nearest neighbours $c_2$ noise parameters. Black equipotential lines are in logarithmic scale. Left inset shows a slice of the main plot along $c_1=1$ and presents the results obtained using the iMPO approach (black dots) compared with the exact asymptotic result for a weakly squeezed state strategy (see the text) $F/N=e^{-c_1}/\left(1-e^{-c_1}+2e^{-c_1}\sinh(c_2)\right)$  (light grey line) and the state-of-the-art bounds on QFI/N obtained for channel $\Lambda_1$ (grey dotted line), $\Lambda_2$ (grey dash-dotted line), $\Lambda_3$ (grey dashed line). Right inset shows a slice of the main plot along $c_2=0$ and presents result obtained using iMPO approach (black dots) compared with the known exact result for strictly local noise $F/N=\eta^2/(1-\eta^2)$ with $\eta=e^{-c_1/2}$ (light grey line).
	(iii) Asymptotic value (obtained using the iMPOs) of QFI per particle for product input states ($D_\psi=1$) as a function of $c_1$, $c_2$ noise parameters. Inset shows the dependence of the optimal QFI per particle as a function of the corresponding product state QFI per particle $F_\mathrm{prod}/N$, revealing the functional dependence known from the uncorrelated dephasing case $\frac{F}{N}=\frac{F_\mathrm{prod}}{N}/\left(1-\frac{F_\mathrm{prod}}{N}\right)$.
	(iv) Entanglement of the optimal iMPO state quantified via the Von Neumann entropy (in bits) of the reduced density matrix (obtained by tracing out half of the spin chain) of the optimal state (obtained using iMPO approach) as a function of $c_1$ and $c_2$ noise parameters.}
	\label{fig:plot dephase}
\end{figure*}

Now we are ready to make use of the methods described in Sec.~\ref{sec:mpo} in order to find the optimal probe states and the corresponding maximal QFI. All the numerical results that follow correspond to the $N$ spin $1/2$ particle case, $h^{[n]} = \sigma_z^{[n]}/2$. We perform numerical calculations using MATLAB software with the help of ncon() function \cite{NCON} for tensor contractions. We should stress that all the optimization algorithms are written from scratch with the metrological context in mind, since standard optimization procedures utilized by many-body physics community are not easily adapted to the optimization tasks that we face.

First, in Fig.~\ref{fig:plot dephase}(i) we present a comparison of results of the QFI optimization procedure for exemplary dephasing parameters $c_1=1$, $c_2=0.1$ (correlated noise) obtained using the finite number of particles $N$ MPO approach and the asymptotic value of the QFI per particle obtained using the iMPO approach---note that we plot $F/N$ as we expect asymptotic linear scaling of the QFI and hence convergence of $F/N$ to a fixed value. The results obtained via the two approaches are in very good agreement. When we plot QFI versus $N$ and fit to the data for large $N$ a straight line we obtain a slope of $0.500$. It compares well with the asymptotic limit obtained directly from the iMPO method as $0.504$. This is a numerical confirmation that indeed the iMPO approach which, as described in Sec.~\ref{subsec:mpo inf} is much more conceptually involved, yields correct results. This is a highly relevant observation as it is numerically much more efficient to obtain asymptotic properties of the QFI using directly the iMPO approach rather than performing finite $N$ computations and extrapolating them to $N \rightarrow \infty$. While performing the optimization we have set the error tolerance on the level of $1\%$, which resulted in the maximal bond dimension required in the numerical procedure to be only $D_\psi = 4$ for the input state and $D_L = 2$ for the SLD. This demonstrates how efficient the MPO description is in this case. Crosses overlaid on the plot indicate the regime where direct calculations using the standard full Hilbert space description were possible on the same high performance PC on which the MPO algorithms were run. This regime is clearly very far from the one were we observe the convergence to the asymptotic linear scaling of the QFI with $N$, and is only accessible numerically using the MPO based methods. In Fig.~\ref{fig:plot dephase}(ii) we present the contour plot depicting the asymptotic value of the QFI per particle as a function of noise parameters and contrast with the achievable QFI when using product states, see Fig.~\ref{fig:plot dephase}(iii). In order to appreciate the amount of entanglement that is present in the optimal states, in Fig.~\ref{fig:plot dephase}(iv) we provide a contour plot of the results of calculation of the von Neuman entropy for the iMPO \cite{Cirac2011} corresponding to the reduced density matrix of the system when half of the particles is traced out (optimal QFI can be achieved by many different states so some fluctuations of entropy are to be expected). We see a clear relation, between the amount of entanglement and the increase in sensing precision.

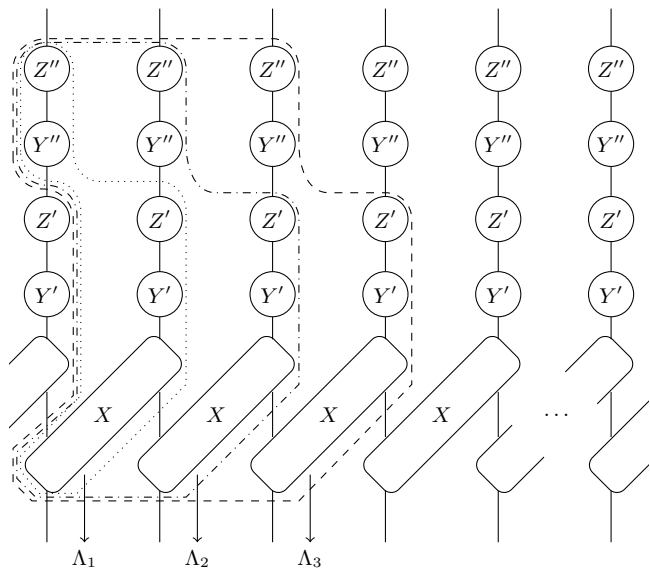
\begin{figure}
	\begin{center}
		\begin{tikzpicture}[scale=1, every node/.style={scale=0.9}]
			\draw (0,0) -- (0,0.8);	
			\foreach \x in {1.5, 3.0, ..., 7}
			{
				\draw (\x,0) -- (\x,0.8);
			}
	
			\draw[line width=2mm, white] (7.5,0) -- (7.5,1);
			\draw (7.5,0) -- (7.5,0.8);
	
			\draw[rounded corners, fill=white, rotate around={45:(0.75,1.7)}] (-0.5,1.4) rectangle (2.0,2.0);
	 		\draw[rounded corners, fill=white, rotate around={45:(3.75,1.7)}] (2.5,1.4) rectangle (5.,2.0);
			\draw[rounded corners, fill=white, rotate around={45:(6.75,1.7)}] (5.5,1.4) rectangle (8.,2.0);
			\draw (0.75, 1.7) node {$X$};
			\draw (3.75, 1.7) node {$X$};
			\fill[white,rotate around={45:(6.75,1.7)} ] (6.25,1.3) rectangle (7.25,2.1);
			\draw (6.8,1.7) node {$\dots$};
			\foreach \x in {0, 1.5, 3.0, ..., 8.5}
			{
				\draw (\x,1.4) -- (\x,2.0);
			}
			\draw[rounded corners, fill=white, rotate around={45:(2.25,1.7)}] (1.0,1.4) rectangle (3.5,2.0);
			\draw[rounded corners, fill=white, rotate around={45:(5.25,1.7)}] (4.0,1.4) rectangle (6.5, 2.0);
			\draw (2.25, 1.7) node {$X$};
			\draw (5.25, 1.7) node {$X$};
			\begin{scope}
	    		\clip (-0.5,0) rectangle (0.5,3);
	    		\draw[rounded corners, fill=white, rotate around={45:(-0.75,1.7)}] (-1.,1.4) rectangle (0.5,2.0);
			\end{scope}
			\begin{scope}
	    		\clip (7,0) rectangle (8,3);
	    		\draw[rounded corners, fill=white, rotate around={45:(8.25,1.7)}] (7,1.4) rectangle (8.5,2.0);
			\end{scope}
			\foreach \x in {0, 1.5, 3.0, ..., 8.5}
			{
				\draw (\x,2.7) -- (\x,3);
			}
			\foreach \x in {0, 1.5, 3.0, ..., 8.5}
			{
				\draw (\x,3.6) -- (\x,7.1);
			}
			\foreach \x in {0, 1.5, ..., 8.5}
			{
				\draw[fill=white] (\x,3.3) circle (0.3);
				\draw (\x,3.3) node {$Y^{\prime}$};
				\draw[fill=white] (\x,4.3) circle (0.3);
				\draw (\x,4.3) node {$Z^{\prime}$};
				\draw[fill=white] (\x,5.3) circle (0.3);
				\draw (\x,5.3) node {$Y^{\prime\prime}$};
				\draw[fill=white] (\x,6.3) circle (0.3);
				\draw (\x,6.3) node {$Z^{\prime\prime}$};
			}
			\draw[dotted] (-0.35,5.2) -- (-0.35,6.3) to[out=90,in=180] (0,6.65) to[out=0,in=90] (0.35,6.3) -- (0.35, 5.2) to[out=-90,in=180] (0.7,4.8) -- (1.5,4.8) to[out=0,in=90] (1.85,4.4) -- (1.85,2.1) -- (0.4,0.65) -- (0,0.65) -- (-0.1,0.65) -- (-0.35,0.95) -- (-0.35,1.15) -- (0.8-0.35,0.8+1.15) -- (0.8-0.35,1+1.15) -- (0.8-0.35,4.4) to[out=90,in=0] (0.8-0.7,4.8) to[out=180,in=-90] (-0.35,5.2);
			\draw[dash dot] (-0.4,5.15) -- (-0.4,6.25) to[out=90,in=180] (0,6.65) -- (1.5,6.65) to[out=0,in=90] (1.85,6.3) -- (1.85,5.3) to[out=-90,in=180] (2.25,4.65) -- (3,4.65) to[out=0,in=90] (3.35, 4.3) -- (3.35,2.1) -- (3.35-1.5,0.6) -- (-0.15,0.6) -- (-0.4,0.85) -- (-0.4,1.2) -- (0.4,0.8+1.1) -- (0.4,4.35) to[out=90,in=0] (0,4.75) to[out=180,in=-90] (-0.4,5.15);
			\draw[dashed] (-0.45,5.1) -- (-0.45,6.3) to[out=90,in=180] (0,6.7) -- (3,6.7) to[out=0,in=90] (3.35,6.3) -- (3.35,5.3) to[out=-90,in=180] (3.75,4.65) -- (4.5,4.65) to[out=0,in=90] (4.85, 4.3) -- (4.85,2.1) -- (4.85-1.5,0.55) -- (-0.2,0.55) -- (-0.45,0.8) -- (-0.45,1.25) -- (0.35,0.8+1.15) -- (0.35,4.3) to[out=90,in=0] (-0.05,4.7) to[out=180,in=-90] (-0.45,5.1);
			\draw[->] (0.5,0.9) -- (0.5,0);
			\draw (0.5,0) node[below] {$\Lambda_1$};
			\draw[->] (2,0.9) -- (2,0);
			\draw (2,0) node[below] {$\Lambda_2$};
			\draw[->] (3.5,0.9) -- (3.5,0);
			\draw (3.5,0) node[below] {$\Lambda_3$};
		\end{tikzpicture}.
	\end{center}
	\caption{Different ways of rewriting the locally correlated noise dynamics as a collection of independent quantum channels: $\Lambda_1$ (dotted), $\Lambda_2$ (dash-dotted), $\Lambda_3$ (dashed).}
	\label{fig:bounds}
\end{figure}

Let us now ask the question, whether some insight into the problem could have been gained by ingeniously adapting the state-of-the-art methods of deriving fundamental bounds in quantum metrology developed with uncorrelated noise models in mind \cite{Escher2011, Demkowicz2012, demkowicz2014using, demkowicz2017adaptive, zhou2018achieving}. These methods provide easily calculable asymptotic bounds, based on just the knowledge of the Kraus operators of elementary probe dynamics or alternatively noise jump operators appearing in the quantum master equation. Even though the noise in our problem is correlated we can formally divide the evolution as a collection of independent channels acting on two, three or more particles---a similar trick has been employed in a recent study of the impact of many-body effects in atomic interferometry \cite{Czajkowski2019}. In the scheme below we indicate the possible constructions. First we formally decompose local dephasing gates $Y$ as products $Y= Y^{\prime} Y^{\prime \prime}$ of local dephasing gates with corresponding $c_1^\prime$ and $c_1^{\prime \prime}$ such that $c_1 = c_1^\prime + c_1^{\prime\prime}$,  and the unitary encoding gate $Z$ as a product $Z= Z^{\prime}Z^{\prime\prime}$, where $Z^{\prime}$, $Z^{\prime\prime}$ are phase gates with corresponding phases $\varphi^\prime$, $\varphi^{\prime \prime}$, such that $\varphi = \varphi^\prime + \varphi^{\prime\prime}$. We can now unravel the total dynamics as effectively composed of $N$ independent channels $\Lambda_1$, or group the gates into $N/2$ larger channels $\Lambda_2$, or $N/3$ channel $\Lambda_3$, etc.---see Fig.~\ref{fig:bounds}.
We may now apply the bounds derived for uncorrelated noise models \cite{Demkowicz2012,demkowicz2014using} using either decomposition of the dynamics into $\Lambda_1$, $\Lambda_2$ or $\Lambda_3$ channels---we can group the operations in any way we please as in this case all elementary evolutions commute.  In order to obtain the tightest bound we numerically optimize the split of phases as well as noise contributions between gates $Y^\prime$, $Z^\prime$ and $Y^{\prime\prime}$, $Z^{\prime\prime}$, while making sure that the resulting $\Lambda_i$ is a legitimate quantum channel---note that the bare two qubit gate $X$ is not a proper quantum channel as it is not completely positive. The results obtained are depicted in the left inset of Fig.~\ref{fig:plot dephase}(ii). While the bounds are tight for the decorrelated noise model they are far from the actual achievable QFI in the correlated (or anti-correlated) noise regimes and as expected they improve when we increase the elementary channel size. Still, because this method scales badly with the elementary channel size we were forced to stop with $\Lambda_3$ channel. This demonstrates, that state-of-the-art methods developed with uncorrelated noise models in mind yield bounds that are far from satisfactory in case of correlated noise models.

As mentioned above, the use of the iMPO approach can greatly speed up process of calculating asymptotic value of QFI. In order to study the impact of noise correlations on the achievable QFI and the optimal states, for the rest of this section, we will therefore restrict ourselves to numerical results obtained using this approach. Using iMPO we have studied the optimal QFI for all noise parameters in the range $c_1 \in [0.2,2]$, and $c_2 \in (-c_1/2,c_1/2)$. We were able to impose the 0.3\% relative error on the obtained results, by going with state bond dimensions $D_\psi$ up to $15$ while keeping $D_{\widetilde{L}} = 1$. In interesting to note, that in the studied here cases higher $D_{\widetilde{L}}$ gave negligible gain in comparison to gain from increasing $D_\psi$---this is most probably related with the fact that local uncorrelated measurement are close to optimal, a fact often encountered in quantum metrology studies when the QFI is the only figure of merit to be optimised.

The main qualitative feature that clearly emerges from the Fig.~\ref{fig:plot dephase}(ii) is the decrease of the optimal QFI with the increase of correlated noise part parameter $c_2$. At the same time going into the anti-correlation regime (negative $c_2$) allows for a significant increase in the achievable QFI. This is to be expected as in the noise anti-correlation regime the noise operators responsible for correlations take the form of $L^{[n,n+1]} = \sqrt{|\gamma_2|}(h^{[n]} - h^{[n+1]})$ and are therefore linearly independent from the Hamiltonian operator  which is a sum of $h^{[n]}$. This is related with the fact that in the extreme case of perfect anti-correlations (no independent local dephasing noise component) it is possible to employ quantum error-correction inspired protocols in order to preserve the Heisenberg scaling \cite{layden2018spatial}. This is also reflected by the divergent behaviour of the state-of-the-art based bounds, which take into account all possible adaptive estimation strategies---note that there is no analogous divergence in the numerical results we obtain as we consider a parallel sensing strategy involving the most general entangled input states and most general measurement but no adaptive strategies. This again indicates, that if one aims at determining the potential of parallel entangled based strategies in presence of noise correlations, the state-of-the-art methods provide bounds which are very far from the actually achievable performance. The $c_2=0$ line on the Fig.~\ref{fig:plot dephase}(ii) corresponds to the strictly local dephasing model for which exact asymptotically saturable bound is known and reads $F/N = \eta^2/(1-\eta^2)$ \cite{Escher2011,Demkowicz2012}, where $\eta = e^{-c_1/2}$. Our numerical results obtained using iMPO agree perfectly with this formula (see right inset on Fig.~\ref{fig:plot dephase}(ii)).

In case of purely local dephasing it is known that in the limit of large number of particles the fundamental bound, $F/N = \eta^2/(1-\eta^2)$, can be saturated by protocols involving weakly spin squeezed states \cite{Orgikh2001, Escher2011}, e.g. one-axis twisted states \cite{MA2011}. Having obtained numerical optimal values of the QFI per particle using our MPO based methods, we want now to check whether the weakly spin squeezed strategy saturates the QFI in case of locally correlated noise model similarly as in the decorrelated noise scenario. For concreteness, consider the following one-axis squeezed state of $N$ particles:
\begin{equation}
	\ket{\psi} = e^{i \theta S_z^2} \ket{{+\tfrac{1}{2}}}^{\otimes N},
\end{equation}
where $\theta$ is the squeezing strength. We follow the standard protocol \cite{MA2011}, where the above state is rotated to the equator of the Bloch sphere so that the $\langle \vec{S} \rangle$ points in the $x$ direction, in a way that the direction in which the angular momentum has minimal variance is $y$. The state is then subject to locally correlated dephasing evolution and is rotated by an unknown angle $\varphi$. Assuming we operate around $\varphi \approx 0$, we measure the $S_y$ observable as this is the optimal choice in this case, from which value we infer the value of $\varphi$. Using the standard linear error propagation formula the resulting uncertainty of estimating the phase reads: $\Delta \widetilde{\varphi} =\sqrt{\Delta^2 S_y}/\left|\frac{\dd \langle S_y \rangle}{\dd \varphi}\right|$. In order to calculate the above quantity we move to the Heisenberg picture. Since we operate around $\varphi=0$, we can replace $\frac{\dd \langle S_y \rangle}{\dd \varphi} = \langle S_x \rangle$, and plug $\varphi=0$ everywhere. Under the locally correlated dephasing noise the relevant expectation values should be replaced according to the following rules:
\begin{align}
	&\langle s_x^{[n]} \rangle  \rightarrow \langle s_x^{[n]}  \rangle e^{-\frac{1}{2}c_1},
	   \langle s_y^{[n]} \rangle  \rightarrow \langle s_y^{[n]}  \rangle e^{-\frac{1}{2}c_1}  \\ \nonumber
	& \langle s_y^{[n]} s_y^{[n+2]} \rangle \rightarrow \langle s_y^{[n]} s_y^{[n+2]} \rangle  e^{-c_1} \\ \nonumber
	& \langle s_y^{[n]} s_y^{[n+1]} \rangle \rightarrow  \langle  s_y^{[n]} s_y^{[n+1]}   \cosh c_2+   s_x^{[n]} s_x^{[n+1]} \sinh c_2 \rangle  e^{-c_1}.
\end{align}
Taking the limit $N \rightarrow \infty$, $\theta \rightarrow 0$ in a way that $N \theta^2 \ll 1$ we obtain that the relevant expectation values on the squeezed state read \cite{MA2011} (assume $\hbar=1$): $\langle s_x^{[n]} \rangle  \rightarrow  \frac{1}{2} e^{-c_1}$, $\langle s_y^{[n]} \rangle  = 0$, $\langle s_y^{[n]} s_y^{[m]} \rangle  \rightarrow  - \frac{1}{4 (N-1)}$, $\langle s_x^{[n]} s_x^{[m]} \rangle  \rightarrow  \frac{1}{4}$ ($n \neq m$). As a result $\langle S_x\rangle = \frac{N}{2}e^{-c_1}$ while
\begin{multline}
	\Delta^2 S_y = \langle S_y^2 \rangle = \frac{1}{4} N   -  \frac{e^{-c_1}}{4(N-1)}(N-2)(N-1)+\\
	2(N-1) e^{-c_1}\left(\frac{1}{4}\sinh c_2  - \frac{1}{4(N-1)}\cosh c_2 \right).
\end{multline}
This leads to the final formula for the asymptotic precision $\Delta\widetilde{\varphi} = \sqrt{(1-e^{-c_1} +2 e^{-c_1} \sinh c_2)/(N e^{-c_1})}$ which can be related with the corresponding Fisher information per particle equal to:
\begin{equation}
	\frac{F}{N}  = \frac{e^{-c_1}}{1-e^{-c_1} +2 e^{-c_1} \sinh c_2}.
\end{equation}
We have checked that this formula agrees with our numerical results up to the desired accuracy (${<1\%}$), and the representative comparison of the numerical data and this formula is provided in the left inset of Fig.~\ref{fig:plot dephase}(ii). This implies that similarly as in the uncorrelated dephasing models, weakly spin-squeezed states are asymptotically optimal. Note that this does not imply that the optimal MPO states we have obtained in our numerical procedure are these kind of states. Quite contrary, our MPO approach favours states with low bond dimension and local correlations, while the above states due to their fully symmetric nature correlate all the particles with each other irrespectively of their distance. Had we considered a product state strategy, the only modification in the above reasoning would be a substitution $\langle s_y^{[n]} s_y^{[m]}\rangle = 0 \ (n \neq m)$, which would lead to the corresponding QFI per particle:
\begin{equation}
	\frac{F_{\t{prod}}}{N} = \frac{e^{-c_1}}{1+2 e^{-c_1} \sinh c_2},
\end{equation}
which also agrees perfectly with the numerical results we have presented in Fig.~\ref{fig:plot dephase}(iii).

Using the above formula, we may also go back to the original problem of magnetic field sensing.
Utilizing the relation $\varphi = g B t$ we get the corresponding magnetic field sensing precision:
\begin{equation}
\Delta \widetilde{B}_t =  \frac{1}{g t}\Delta \widetilde{\varphi} = \frac{1}{g t} \sqrt{\frac{1- e^{-\sigma^2 g^2 t} + 2 e^{-\sigma^2 g^2 t} \sinh(\chi g^2 t)}{N e^{-\sigma^2 g^2 t}}}.
\end{equation}
The above formula assumes a fixed interrogation  time $t$. We may generalize the considerations, and fix the \emph{total} interrogation time $T$
which we allow to split into $T/t$ independent interrogation steps.
The corresponding estimation uncertainty reads:
\begin{equation}
	\Delta \widetilde{B}_T = \frac{1}{\sqrt{T/t}}\Delta \widetilde{B}_t = \sqrt{\frac{1-e^{-\sigma^2 g^2 t} +2 e^{-\sigma^2 g^2 t} \sinh(\chi g^2 t)}{g^2 t T N e^{-\sigma^2 g^2 t}}},
\end{equation}
which when optimized over $t$ reaches the minimal value when $t \rightarrow 0$ and yields:
\begin{equation}
	\Delta \widetilde{B} = \sqrt{\frac{\sigma^2+2\chi}{T N}}.
\end{equation}

Based on the above results we can expect that weakly spin-squeezed states should also be optimal in case of a more general dephasing noise, provided the range of correlations $r$ is finite and we consider the asymptotic limit $N \rightarrow \infty$. In this case, following analogous calculations, we would arrive at the optimal magnetic field sensing precision of the form
\begin{equation}
	\Delta \widetilde{B} = \sqrt{\frac{\sigma^2+2\sum_{k=2}^r \chi_k}{T N}},
\end{equation}
where $\chi_k$ represent magnetic field correlations for particles at distance $k-1$:
$\langle \delta B^{[n]}(t) \delta B^{[n+k-1]}(t')\rangle = \chi_k \delta(t-t')$.
Comparing this result with the performance of the GHZ states for the same model, see \cite{Layden2019}, we notice that
there is the $\sqrt{e}$ factor improvement in performance of the optimally spin-squeezed states over the GHZ states familiar from uncorrelated dephasing considerations \cite{Huelga1997, Orgikh2001}.

\subsection{Atomic clock stabilization} \label{subsec:example allan}

When one follows the Bayesian rather than the frequentist line of reasoning and focuses on minimization of the Bayesian variance, an apparently similar computational problem arises as in the cases of optimization of the QFI \cite{Macieszczak2014, Demkowicz2015}. The goal is then to minimize:
\begin{equation}
	\left\langle  \Delta^2 \widetilde{\varphi} \right\rangle = \iint \dd \varphi \dd x \, p(x|\varphi) \left(\widetilde{\varphi}(x)-\varphi\right)^2,
\end{equation}
where $p(x|\varphi) = \tr({\rho}_{\varphi}{\Pi}_x)$ and $p(\varphi)$ is a prior distribution for the parameter to be estimated (for simplicity we assume that the prior is centered at $0$: $\int \dd \varphi \, p(\varphi) \varphi = 0$). The minimal achievable quadratic Bayesian cost for the problem, optimized over all measurements and estimators, reads \cite{Helstrom1976, Macieszczak2014}:
\begin{equation} \label{eq:bayesian}
	\left\langle  \Delta^2 \widetilde{\varphi} \right\rangle = \Delta^2_0  - \sup_{L} \left[2\tr\left(\underline{\rho}' L \right)-\tr\left(\underline{\rho} L^{2}\right)\right],
\end{equation}
where $\Delta^2_0$ is the variance of the prior distribution, $\underline{\rho} = \int \dd \varphi\, p(\varphi) \rho_\varphi$ is the output state averaged with the prior, while $\underline{\rho}' = \int \dd \varphi \, p(\varphi) \varphi \rho_{\varphi}$. It is clear from the above formula that the problem is computationally very similar to calculation of the QFI, as given in Eq.~\ref{eq:qfimax} (up to a replacement of the derivative of $\rho_\varphi$ with $\underline{\rho}^\prime$)

An important problem where the Bayesian line of reasoning is relevant, and which at the same time is very well suited for our tensor network framework is the atomic clock stabilization problem. A typical atomic clock operates in a feedback loop where the local oscillator (LO, e.g.\ laser) is stabilised to atomic reference frequency by periodically interrogating atoms (using radiation from the LO) and based on the measured response, the frequency of the LO is corrected \cite{Ludlow2015}. One of the main goals in the design of the clock interrogation scheme is to achieve the lowest instability typically quantified by the \emph{Allan variance} (AVAR) \cite{Allan1966,Riehle2004}:
\begin{equation} \label{eq:AVAR}
	\sigma^2(\tau) = \frac{1}{2\tau^2\omega_0^2} \left\langle \left(\int_{\tau}^{2\tau} \dd t \, \omega(t) - \int_{0}^{\tau} \dd t \, \omega(t) \right)^2 \right\rangle,
\end{equation}
where $\left\langle \cdot \right\rangle$  represents averaging over frequency fluctuations of the LO described by some stochastic process (which from the Bayesian estimation perspective plays the role of the prior distribution), $\tau$ denotes averaging time, $\omega_0$ atomic reference angular frequency and $\omega(t)$ time-dependent angular frequency of the LO.

Fixing the physical properties of the atoms the goal is to optimize their initial states states, interrogations times, measurements and feedback corrections in order to minimize the AVAR. Performing such a comprehensive optimization is not feasible. In \cite{Chabuda2016} a lower bound on the achievable AVAR was introduced the \emph{quantum Allan Variance} (QAVAR):
\begin{align}
	&\sigma_\mathrm{Q}^2(\tau) = \sigma_\mathrm{LO}^2(\tau) - \frac{1}{\omega_0^2} \sup_{\rho_0,L,T} F_\mathrm{A}(\tau;\rho_0,L,T), \\
	&F_\mathrm{A}(\tau;\rho_0,L,T) = \left[ 2\tr\left(\underline{\rho}' L \right)-\tr\left(\underline{\rho} L^{2}\right) \right]/2,
\end{align}
where $\sigma_\mathrm{LO}^2(\tau)$ is the AVAR of free running LO, $\frac{1}{\omega_0^2} F_\mathrm{A}(\tau;\rho_0,L,T)$ represents a correction to it from the feedback loop and $T$ is the interrogation time. We do not provide here explicit forms of the operators $\underline{\rho}$ and $\underline{\rho}'$ and refer the interested reader to \cite{Chabuda2016}, but just note that they are analogs of $\underline{\rho}$, $\underline{\rho}'$ as defined in the simple Bayesian estimation problem in Eq.~\eqref{eq:bayesian}. The important information is that, if the atoms with which the atomic clock interacts are described via states on some $d$ dimensional Hilbert space $\mathcal{H}$, then the $\underline{\rho}$ and $\underline{\rho}'$ objects act on a tensor space $\mathcal{H}^{\otimes N}$, where $N= 2(\tau/T)-1$ is the number of atomic cycles that need to be considered in order to calculate QAVAR. We assume here an idealized situation that one cycle of atomic clock operation lasts $T$---there is no dead time, and all the time is dedicated to interrogation of atoms by the LO. Typically we will think of the $d$ dimensional spaces of atomic states as being a fully symmetric subspace of $d-1$ two-level systems representing the relevant clock transition levels of the atoms---therefore we will use the notation for the states $\ket{\psi} = \sum_{k=0}^{d-1} a_k \ket{k}$, which correspond to the symmetric state, where $k$ atoms are in an excited state and $d-1-k$ atoms in the ground state. Interaction between the LO and the atomic sample has form of Ramsey interferometry and after a single interrogation step will effectively encode the phase in the above written state as $\ket{\psi}_T = \sum_{k=0}^{d-1} a_k e^{- i  k \int_0^T  \delta \omega(t) \dd t}\ket{k}$, where $\delta \omega(t)$ is the detuning of the LO frequency from the atomic reference frequency $\omega_0$.  The effect of LO fluctuations is equivalent to collective dephasing of atoms and hence the state will remain within the symmetric subspace---note that we are talking about the single Hilbert space $\mathcal{H}$ here which will be represented by single node in the MPO framework. The key feature from our perspective is the fact that the LO frequency fluctuations are temporally correlated, and hence the collective dephasing acting on atoms at different interrogation steps (in our representation different steps are formally represented as different product subsystems in the $\mathcal{H}^{\otimes N}$ space) will be correlated.

Assuming that LO fluctuations have finite correlations in time we can expect that QAVAR can be efficiently calculated using a tensor network in the form of a chain of length $N$ with $(d-1)$-dimensional physical indices on each site. Apart from a clear numerical efficiency advantages, the use of tensor networks approach also allows us to constrain the class of input states to be product (bond dimension $D_\psi=1$) which corresponds to the typical situation in which atomic samples in different time steps are independent of each other, as they prepared anew at the beginning of each interrogation step---note that we will still consider entanglement between physical atoms with which the LO interacts at a given interrogation time step.

LO fluctuations can be characterized by the autocorrelation function $R(t)$ which for the purpose of our example we choose to be a combination of Ornstein–Uhlenbeck (OU) process and white Gaussian frequency noise:
\begin{equation} \label{eq:acorr}
	R(t) = \alpha e^{-\gamma t} + \beta \delta(t),
\end{equation}
where we can interpret parameters $\alpha$, $\beta$ as strength of respectively OU process and white noise and $1/\gamma$ as OU correlation range. We choose $\alpha = 1 \, (\mathrm{rad}/\mathrm{s})^2$, $\beta = 0.1 \, (\mathrm{rad}/\mathrm{s})^2 \mathrm{s}$, $\gamma = 2 \, \mathrm{s}^{-1}$ for which noise correlations on the time scale that will correspond to the optimal interrogation time step $T$ (which in this case will happen to be around $1.5\,\mathrm{s}$) will be weak enough so that they will appreciable affect only the nearest-neighbour ``time-step subsystems''---studying noise with further correlations is possible but would require more computational resources because of the larger bond dimensions. For such a noise the AVAR of the free running LO reads:
\begin{equation}
	\sigma_\mathrm{LO}^2(\tau) = \frac{1}{\tau\omega_0^2} \left[ \frac{2\alpha}{\gamma} + \frac{\alpha}{\gamma^2\tau} \left(4e^{-\gamma\tau}-e^{-2\gamma\tau}-3\right) + \beta \right],
\end{equation}
and in the most interesting regime of large averaging times takes the form $\sigma_\mathrm{LO}^2(\tau) \simeq (2\alpha\gamma^{-1}+\beta)/(\tau\omega_0^2)$. We expect that in this limit QAVAR also takes the form $\sigma_\mathrm{Q}^2(\tau) \simeq c/(\tau\omega_0^2)$ with some constant $c$ which we will refer to as asymptotic coefficient.

\begin{figure*}
	\centering
	\includegraphics[width=2\columnwidth]{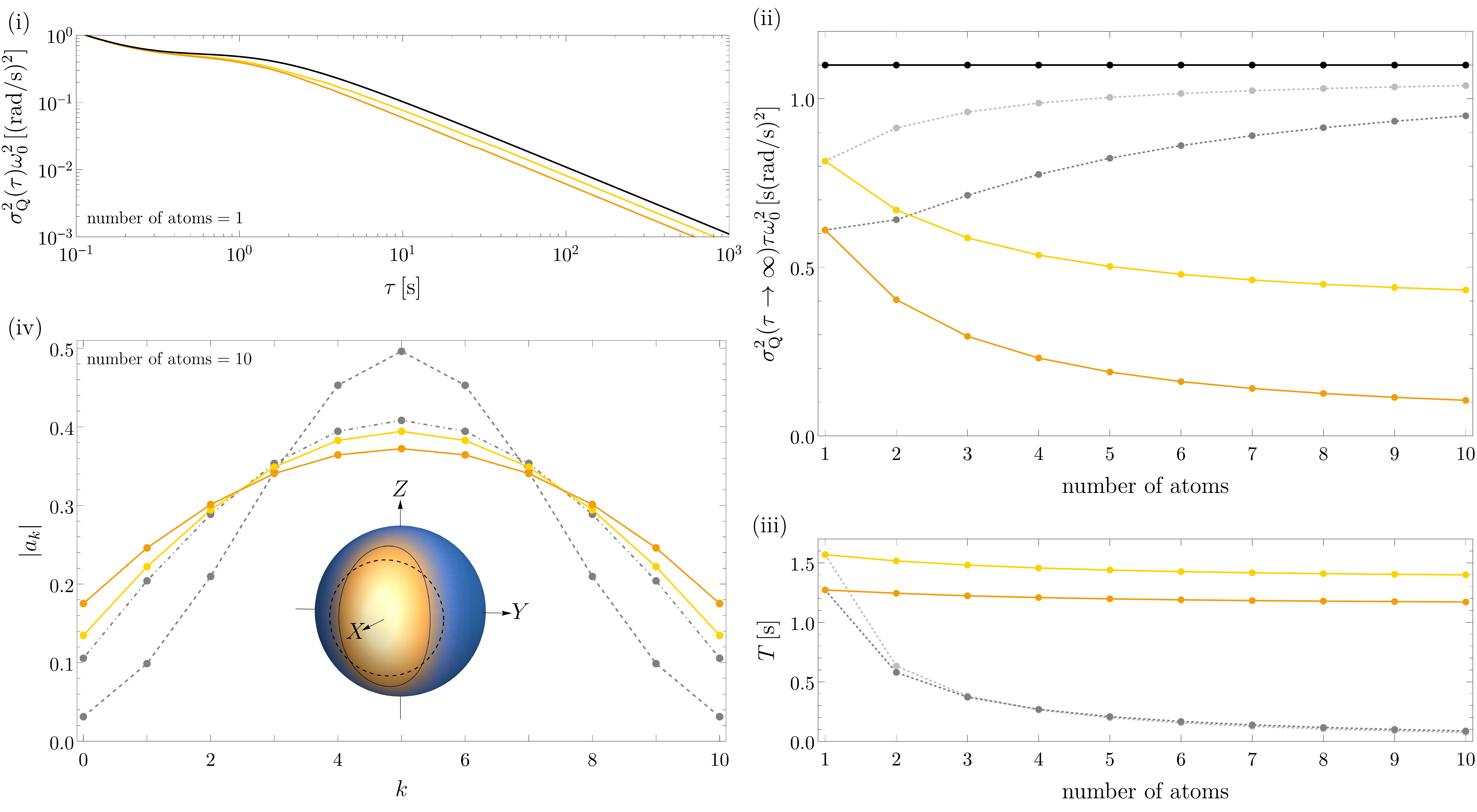}
	\caption{(i) QAVAR (times $\omega_0^2$) as a function of the averaging time $\tau$ for the atomic clock (based on one atom) with the LO noise which is strictly local (yellow line) or also includes the nearest neighbours correlations (orange line), plotted against the AVAR of uncorrected LO (black line).
	(ii) QAVAR asymptotic coefficient as a function of the number of atoms in the atomic clock with LO noise which is strictly local (yellow dots connected by solid line/light grey dots connected by dotted line) or also includes the nearest neighbours correlations (orange dots connected by solid line/grey dots connected by dotted line) for the optimal/NOON state, plotted against the AVAR asymptotic coefficient of uncorrected LO (black dots connected by solid line).
	(iii) Optimal interrogation times as a function of the number of atoms in the atomic clock with LO noise which is strictly local (yellow dots connected by solid line/light grey dots connected by dotted line) or also includes the nearest neighbours correlations (orange dots connected by solid line/grey dots connected by dotted line) for the optimal/NOON state.
	(iv) Absolute values of the probability amplitudes for the optimal states in the atomic clock (based on 10 atoms) with LO noise which is strictly local (yellow dots connected by solid line) or also includes the nearest neighbours correlations (orange dots connected by solid line), plotted against coherent spin state (CSS, grey dashed line) and the sine state \cite{Berry2000}---optimal in phase estimation in case of a completely unknown phase (grey dash-dotted line). In the centre Husimi Q distribution on the Bloch sphere for the optimal state (for the case with the nearest neighbours LO noise correlations) with marked equipotential lines where quasiprobability is equal to 0.1 for CSS (black dashed line) and for this optimal state (black solid line) which shows that the state is squeezed.}
	\label{fig:plot allan}
\end{figure*}
These calculations have been attempted in \cite{Chabuda2016} using the full Hilbert space description, but were not capable of approaching the regime where the character of the scaling of the QAVAR and the coefficient could be unambiguously read out. The tensor network framework, proposed in this paper, allows us to calculate QAVAR in previously inaccessible regime of large $\tau$. In Fig.~\ref{fig:plot allan}(i) we present the exemplary results for an atomic clock operating on one two-level atom which also shows that completely neglecting noise correlations in analysing the clock performance is unjustified. We see that the QAVAR curve flatten for $\tau \gtrsim 50\,\mathrm{s}$, which when taking into account the optimal interrogation times which in this case approaches $\sim1.3\,\mathrm{s}$, implies that for calculations of the QAVAR we would need to consider $\sim80$ interrogation steps and hence if full Hilbert space description was used for this purpose would require $2^{80}$ dimensional space---clearly an impossible task. Similarly as in the previous example, we may directly access the asymptotic behaviour  ($\tau \rightarrow \infty$) of the QAVAR function with the help of the iMPO approach.  Following this approach we calculate the QAVAR asymptotic coefficient $c$ and the corresponding optimal interrogation time $T$ as a function of the number of atoms in the clock, see Fig.~\ref{fig:plot allan}(ii-iii). From this figures we see that the differences in QAVAR between cases with strictly local noise and when nearest neighbour noise correlations are included only grow with the increasing number of atoms. This implies that noise correlations play an important role in the accurate analysis of clock performance. Note that, the definition \eqref{eq:AVAR} of the AVAR leads to an intrinsically not translationally invariant MPO for $\underline{\rho}'$ in the expression for QAVAR. Because of this when implementing the iMPO approach we approximate AVAR by an asymptotically equivalent expression:
\begin{equation}
	\sigma^2(\tau) \simeq \frac{1}{\tau^2\omega_0^2} \left\langle \left(\int_{0}^{\tau} \dd t \, \omega(t) \right)^2 \right\rangle,
\end{equation}
which coincide with previous definition for $\tau$ much larger then the noise correlation length (which is exactly the regime in which iMPO approach operates). In order to keep relative errors below $1\%$ the numerical calculations required bond dimension $D_L = 4$ for finite $N$ case and $D_{\widetilde{L}} = 1$ in the case of the iMPO approach.

We confront the results (which are optimized over the input state) with values obtained using a NOON/GHZ states as an input, $\ket{\psi} = (\ket{0} + \ket{d-1})/\sqrt{2}$. The NOON states are highly prone to dephasing noise, and hence the optimal interrogation times will be necessary reduced compared to the optimal (more robust states). This is visible in the Fig.~\ref{fig:plot allan}(ii-iii), where we see that even though the optimal $T$ for the NOON state scales down (in fact as $(d-1)^{-2}$), the noise quickly destroys any gain from the feedback loop to AVAR of a free running LO. This is a manifestation of a generic poor performance of the NOON/GHZ states in realistic (noisy) scenarios with increasing particle number $N$ \cite{Huelga1997, Escher2011, Demkowicz2012}.

Our framework allow us also to easily study the optimal input states. In Fig.~\ref{fig:plot allan}(iv) we plot the absolute values of the probability amplitudes $a_k$ for the optimal states (for LO noise which is strictly local or also includes the nearest neighbours correlations) alongside $a_k$ for coherent spin state (CSS) and the sine state \cite{Berry2000} (which is the optimal input state for Bayesian estimation of phase with a flat prior distribution). We see that our optimal states are nonclassical, which we can quantify by calculating their spin-squeezing parameter $\xi^2 = 2\Delta^2 J_y/\langle J_x \rangle$ \cite{MA2011} which is $1$ for CSS, $0.522$ for sine state, $0.456$ and $0.378$ for the optimal states for LO noise which is respectively strictly local or also includes the nearest neighbours correlations (squeezing in this last case can be also observed from Husimi Q distribution on Bloch sphere in the centre of the Fig.~\ref{fig:plot allan}(iv)).

\subsection{Fidelity susceptibility calculations for many-body thermal states} \label{subsec:example fidelity}
\begin{figure}
	\centering
	\includegraphics[width=\columnwidth]{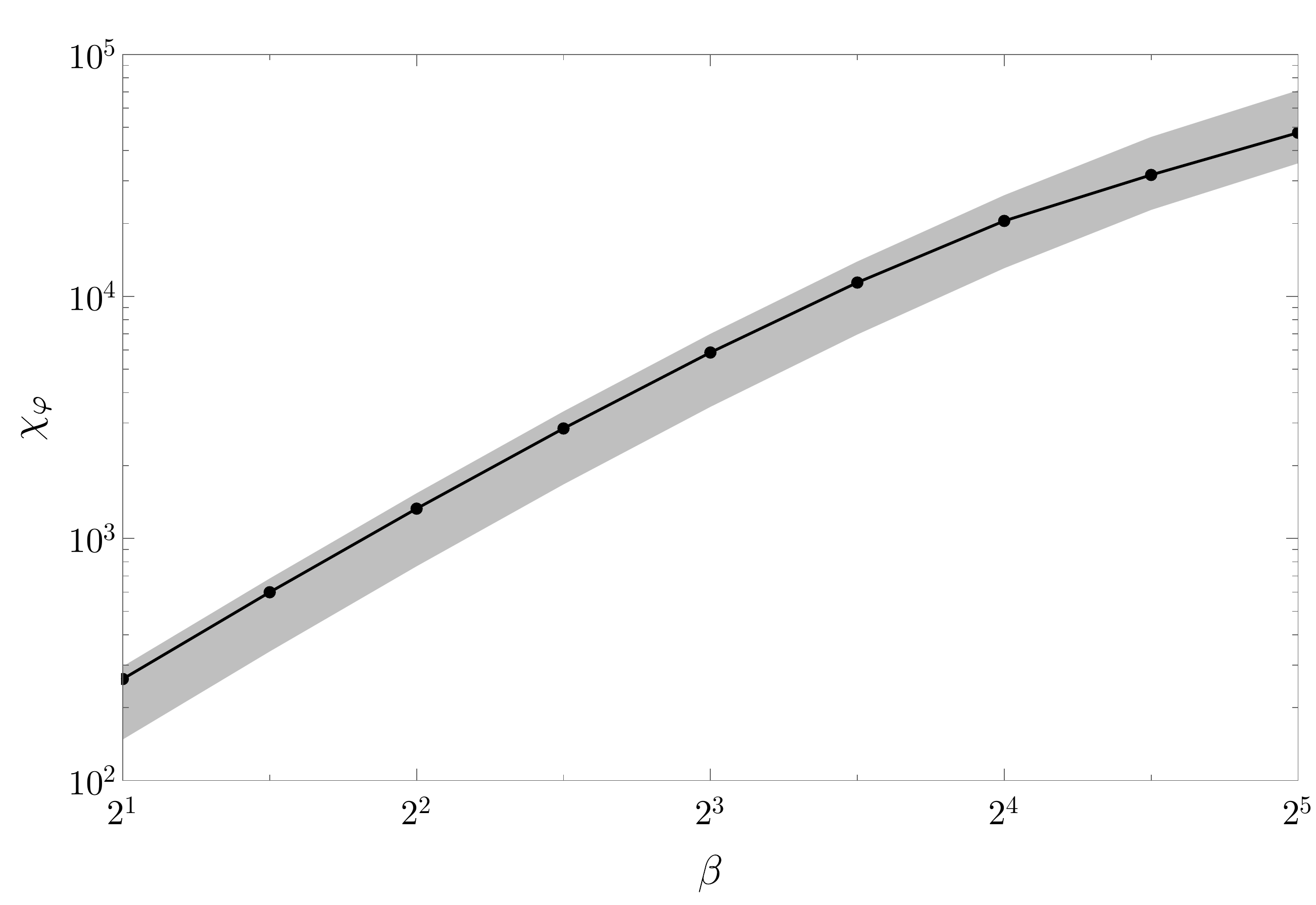}
	\caption{Exact fidelity susceptibility  for a thermal many-body state (dots connected by the line) at the critical point in function of dimensionless inverse temperature $\beta$ in the XX model (\ref{XX}) with $64$ spins. The shaded band shows the bounds (\ref{chibounds}). As predicted in \cite{Albuquerque,Sirker}, the exact value tends to the upper/lower bound for high/low temperatures.}
	\label{fig:plot fidelity}
\end{figure}

In condensed matter context, fidelity $\mathcal{F}(\varphi,\varphi+\varepsilon) = \tr\sqrt{\sqrt{\rho_\varphi}\rho_{\varphi+\varepsilon}\sqrt{\rho_\varphi}}$ between many-body states $\rho_\varphi$ and $\rho_{\varphi+\varepsilon}$, that differ by a small variation of a parameter $\varphi$ in a Hamiltonian, is a mean to identify the location $\varphi_c$ of a phase transition \cite{Zanardi,DamskiRams}. This is where the fidelity susceptibility $\chi_\varphi$, defined by $\mathcal{F} (\varphi,\varphi+\varepsilon) \approx 1-\frac12\chi_\varphi\varepsilon^2$, has a maximum indicating a fundamental change in the state of the system. This concept was employed in Ref. \cite{Rams2018} to evaluate the usefulness of a quantum phase transition, that happens at zero temperature, for precise sensing of the parameter $\varphi$ in a realistic system at a finite temperature. QFI defines a metric in the space of quantum states (the Bures metric)\cite{Braunstein1994} and is directly related with the fidelity susceptibility, namely $F=4\chi_\varphi$.

Unlike at zero temperature, the fidelity between a thermal many-body states represented by MPOs is not tractable in general. This is why a quasi-fidelity was employed $\widetilde{\mathcal{F}}(\varphi,\varphi+\varepsilon) = \sqrt{ \tr \sqrt{\rho_{\varphi}} \sqrt{\rho_{\varphi+\varepsilon}} }$ defining a quasi-susceptibility, $\widetilde{\mathcal{F}}(\varphi,\varphi+\varepsilon) \approx 1-\frac12\widetilde{\chi}_\varphi\varepsilon^2$, that provides bounds for the exact fidelity susceptibility \cite{Albuquerque,Sirker}:
\begin{equation} \label{chibounds}
	\widetilde{\chi}_\varphi\leq\chi_\varphi\leq2\widetilde{\chi}_\varphi.
\end{equation}
The Hamiltonian considered in Ref. \cite{Rams2018} was the spin-$\frac{1}{2}$ XX model
\begin{equation}
	H = -\sum_{n=1}^{N-1} \left( \sigma_x^{[n]}\sigma_x^{[n+1]} + \sigma_y^{[n]}\sigma_y^{[n+1]} \right) + \varphi \sum_{n=1}^N \sigma_x^{[n]},
	\label{XX}
\end{equation}
with a quantum critical point at $\varphi_c=0$. Taking the MPOs studied in Ref.~\cite{Rams2018}, we bypass the tractability problem employing the part of our scheme with $\rho'_\varphi=\left(\rho_{\varphi+\varepsilon}-\rho_{\varphi}\right)/\varepsilon$ to calculate the QFI $F=4\chi_\varphi$. This exact susceptibility for the chain with $64$ spins is shown in Fig.~\ref{fig:plot fidelity} together with the upper and lower bounds (\ref{chibounds}). The accuracy of the fidelity susceptibility is limited by the finite bond dimension $D_L$ as well the finite parameter difference $\varepsilon$. Nevertheless, we obtain satisfying results with relative error around $1\%$ for $\varepsilon=10^{-4}$ and $D_L = 4$.

This example demonstrates that the scheme for calculating fidelity susceptibility is a useful byproduct of our general algorithm. Beyond the present metrological context, it paves a way to generalize the zero-temperature fidelity approach to detecting quantum phase transitions \cite{Zanardi,DamskiRams}---by now standard in condensed matter physics---to phase transitions in quantum many-body systems at finite temperature. Their thermal states can be represented either by MPO, when DMRG on a cylinder is employed \cite{WeichselbaumPhysRevX}, or its two-dimensional generalization on an infinite lattice (iPEPO) \cite{CzarnikHubbard,CzarnikXiD}.

\section{Conclusions} \label{sec:conclusions}
We have provided a comprehensive framework for optimization of quantum metrological protocols using the MPO/MPS formalism. The potential to deal effectively with correlated noise models as well as directly access the asymptotic $N\rightarrow \infty$ is what makes this framework unique. We also expect that this framework may also be adapted to deal with even more challenging metrological problems including noisy multiparameter estimation \cite{Ragy2016, Szczykulska2016, Baumgratz2016}, waveform estimation \cite{Tsang2011, Berry2013} or the study of the effectiveness of adaptive metrological protocols including quantum error correction based schemes \cite{Arrad2014, Dur2014, zhou2018achieving, layden2018spatial, Gorecki2019}. We also expect that this numerical framework may be crucial for understanding better metrological models with temporally correlated noise especially of non-Markovian nature \cite{Chin2012}, where effective tools to find the optimal metrological protocols in such cases are yet to be developed.

\acknowledgments
We would like to thank Marek M. Rams, David Layden, Maciej Lewenstein, Shi-Ju Ran, Piet O. Schmidt and Ian D. Leroux for fruitful discussions. We are also indebt to Marek M. Rams for sharing with us the data from Ref.~\cite{Rams2018}.

KCh and RDD  acknowledge support from the  National Science Center (Poland) grant No. 2016/22/E/ST2/00559. Work of JD was funded by NCN together with European Union through QuantERA ERA NET program 2017/25/Z/ST2/03028. TJO was supported, in part, by the DFG through SFB 1227 (DQmat), the RTG 1991, and the cluster of excellence EXC 2123 QuantumFrontiers.

%

\end{document}